\documentclass[AMA,STIX1COL]{WileyNJD-v2}

\usepackage{moreverb}
\usepackage{geometry}
\usepackage{graphicx}
\usepackage[textwidth=8em,textsize=small]{todonotes}
\usepackage{caption}
\usepackage{subcaption}
\usepackage{color}
\usepackage{tikz}
\usepackage{float}
\usepackage{setspace}
\usepackage{nicefrac}
\usepackage{enumitem}
\usepackage{mathtools, nccmath}
\usepackage{threeparttable}
\usepackage{setspace}

\articletype{Research Article}%

\received{<day> <Month>, <year>}
\revised{<day> <Month>, <year>}
\accepted{<day> <Month>, <year>}

\begin{document}

\title{Causal Covariate Selection for the Imputation-based Regression Calibration Method for Exposure Measurement Error Bias Correction}

\author[1]{Wenze Tang*}

\author[2,3]{Donna Spiegelman}

\author[4]{Yujie Wu}

\author[1,4]{Molin Wang*}

\authormark{Tang et al.}

\address[1]{\orgdiv{Department of Epidemiology}, \orgname{Harvard School of Public Health}, \orgaddress{\state{MA}, \country{USA}}}

\address[2]{\orgdiv{Department of Biostatistics}, \orgname{Yale School of Public Health}, \orgaddress{\state{CT}, \country{USA}}}

\address[3]{\orgdiv{Center on Methods for Implementation and Prevention Science}, \orgname{Yale School of Public Health}, \orgaddress{\state{CT}, \country{USA}}}

% \address[4]{\orgdiv{Department of Population Medicine}, \orgname{Harvard Medical School}, \orgaddress{\state{MA}, \country{USA}}}

\address[4]{\orgdiv{Department of Biostatistics}, \orgname{Harvard School of Public Health}, \orgaddress{\state{MA}, \country{USA}}}

\corres{*Wenze Tang, Department of Epidemiology, Harvard School of Public Health. \email{wtang@g.harvard.edu}; *Molin Wang, Department of Epidemiology, Harvard School of Public Health. \email{stmow@channing.harvard.edu}}

\abstract[Abstract]{In this paper, we investigated the selection of the minimal and most efficient covariate adjustment sets for regression calibration method developed by Carroll, Rupert and Stefanski (CRS, 1992), correcting for bias due to continuous exposure measurement error. We utilized directed acyclic graph to illustrate how subject matter knowledge can aid in the selection of such adjustment sets. Unbiased measurement error correction requires the collection of the following variable sets: (I) common causes of true exposure and outcome and (II) common causes of measurement error and outcome, in both the main study and validation study. For CRS regression calibration method under linear models, researchers need to minimally adjust for covariate set (I) in both measurement error model (MEM) and outcome model and adjust for covariate set (II) at least in MEM. In contrast to the regression calibration method developed by Rosner, Spiegelman and Willet, adjusting for non-risk factors that are correlates of the true exposure or of measurement error in the MEM only is valid and increases efficiency under CRS method. We applied the proposed covariate selection approach to the Health Professional Follow-up Study of the effect of fiber intake on cardiovascular disease. In this study, we also demonstrated the potential issue with data-driven approach to build the MEM that is agnostic to the structural assumptions. We extend the originally proposed estimators to settings where effect modification by a covariate is allowed. Finally, we caution against the use of the regression calibration method to calibrate the true nutrition intake using biomarkers.}


\maketitle

\section{Introduction}\label{sec1}

Regression calibration methods have been used to correct for the bias induced by exposure measurement error 
 when estimating the effect of continuous exposure on the outcome\cite{RSWapp1,RSWapp2,RSWapp3,CRSapp1,CRSapp2,CRSapp3}. In the regression calibration method proposed by Carroll, Ruppert and Stefanski \cite{CRS_book,CRS1}(hereinafter CRS), a measurement error model (hereinafter MEM), also known as a calibration model, is first built linking the mismeasured exposure to the true exposure using the validation study dataset, where both true exposure and mismeasured exposure are available. In a main study/external validation study design where the outcome data is not available in the validation dataset, an outcome model is then built linking the imputed true exposure value based on the previous MEM to the error-free outcome using the main study dataset. 

Typically, some covariates are adjusted in the MEM and the same or a different set of covariates are adjusted in the outcome model. However, it remains unclear what should be the minimal covariate set to be adjusted for in each model to guarantee validity. It is also unknown what would be the optimal covariate set to be adjusted for for optimal efficiency. In this paper, we investigate these questions under a counterfactual framework with the aid of directed acyclic graphs (DAGs). 

Compared with the other form of regression calibration method developed by Rosner, Spiegelman and Willet \cite{RSW1,RSW2} (hereinafter RSW), the CRS method does not require researchers to adjust for the same set of covariates in the two models. In practice, researchers often build MEM entirely independently from the outcome model\cite{lampe2017dietary}, resulting in MEMs of varying complexity \cite{CRSapp2,lampe2017dietary,prentice2019application,Prentice2020}. For example, recent publications \cite{prentice2019application,Prentice2020} used adjusted $R^2$ as model fit statistics to select covariates to include in the MEM model. In this paper, we assess to what extent this data-driven practice is appropriate with respect to the validity of the resulting exposure effect estimate.  

In section 2, we review the causal contrast of interest, the statistical models typically used in the CRS method and relevant DAGs. In section 3, we introduce the identification equations. We then present the results regarding validity and efficiency of alternative CRS estimators under each measurement error structure. In section 4, we give the results from Monte Carlo simulations evaluating finite sample performance of each estimator under various scenarios. In section 5, we follow the proposed covariate selection procedure to study the effect of fiber intake on the risk of CVD in the Health Professional Follow-up Study (HPFS) \cite{gu2022dietary}. We conclude the paper with a summary and further discussion.

\section{Causal Estimands and Regression Calibration estimators} \label{sec2}

\subsection{Causal Estimand of Exposure Effect on the Outcome}
$X$, $Z$, $V$, $Y$ are respectively denoted as true exposure, surrogate exposure, error-free covariate(s) and outcome. For clarity, $X$, $Z$ are assumed to be scalar while $V$ can be either a scalar or a vector. We focus on the main study/external validation study design in this paper\cite{CRS_book} and assume that $(Z,V,Y)$ and $(X,Z,V)$ are available in the main study and the validation study, respectively, and that the MEM parameters estimated within the main study population, had we been able to do so, would be the same as those estimated within the validation study (i.e. transportability condition). We define the potential outcome $Y^x$ as the value of $Y$ that each participant would have experienced had their exposure value been fixed at $X=x$. 

Our primary causal contrast of interest is the conditional average treatment effect $E[Y^x - Y^{x^{\prime}}|V]$, i.e. the mean difference in the potential outcome $Y$ had everyone been assigned to exposure value $x$ compared to $Y$ value had the same population been assigned to exposure value $x'$, within levels of covariate value $V$. CATE is adopted as the primary causal contrast because it motivates the use of the regression framework. But if one is instead interested in the average treatment effect (ATE) $E[Y^x - Y^{x^{\prime}}]$, they can obtain ATE by standardizing CATE over the distribution of covariate $V$. Or if $V$'s are not effect modifiers, CATE equals ATE.

\subsection{CRS Regression Calibration Estimators with Linear Terms}\label{parametricCRSEstimators}
In the typical application of the CRS regression calibration method, one can choose whether to include a given covariate $V$ in the linear MEM (the adjusted model \eqref{eq1} versus the unadjusted model \eqref{eq2}):  
\begin{align}
    E[X|Z,V] & = \alpha_0 + \alpha_1 Z + \alpha_2 V \label{eq1}, \\
    E[X|Z] & = \alpha_0^* + \alpha_1^* Z \label{eq2},
\end{align} and whether to also include $V$ in the outcome model, resulting in four possible outcome models, i.e. \eqref{eq3} through \eqref{eq6}:
\begin{align}
    E[Y|Z,V] & = \beta_0 + \beta_{_{(OM)}} E[X|Z,V] + \beta_2 V  \label{eq3} \\
    E[Y|Z,V] & = \tilde{\beta}_0 + \beta_{_{(-M)}} E[X|Z,V] \label{eq4}\\
    E[Y|Z,V] & = \beta_0^+ + \beta_{_{(O-)}} E[X|Z] + \beta_2^+ V \label{eq5} , \text{ and}\\
    E[Y|Z] & = \beta_0^* + \beta_{_{(--)}} E[X|Z] \label{eq6}.
\end{align}

We use subscripts to distinguish the four different CRS estimators, all estimable through (stacked) estimating equations (e.g. by stacking \eqref{eq1} and \eqref{eq3}). For example, estimator $\hat{\beta}_{_{(OM)}}$ means the estimating procedure where one includes a given $V$ in both outcome model and MEM, i.e. under models \eqref{eq1} and \eqref{eq3}. The subscript in CRS estiamtor $\hat{\beta}_{_{(--)}}$ denotes that the covariate is included in neither model, i.e. under models \eqref{eq2} and \eqref{eq6}. Similarly, we have $\hat{\beta}_{_{(-M)}}$ under models \eqref{eq1} and \eqref{eq4} where $V$ is adjusted in the MEM but not outcome model and $\hat{\beta}_{_{(O-)}}$ under models \eqref{eq2} and \eqref{eq5} where $V$ is only adjusted for in the outcome  model.

\subsection{Directed Acyclic Graphs for Measurement Error Structures}

We considered the inclusion of pre-treatment covariates in the independent, non-differential measurement error process \cite{HernanCole2009}, where the covariate can have one of the several different structural relationships to $Z$, $X$ and $Y$. This results in eight DAGs with $V_1$ through $V_8$ on each DAG. For example, DAG 4 describes the scenario where covariate $V_{4(XZY)}$ is a confounder and directly contributes to measurement error. We write the additional subscript $(XZY)$ to emphasize the fact that the covariate has an arrow to $X, Z$ and $Y$ under DAG 4. In these modified measurement error DAGs, the non-differential measurement error condition means that there is statistical independence between the mismeasured exposure $Z$ and $Y$ conditional on $(X,V)$. This assumption is invalid if, for example, disease status $Y$ affects how people report their fat or sodium intake, the mismeasured exposure ($Z$) of the underlying true intake $X$, such as in a retrosepctive case-control study. 

%figure 1 DAGs%
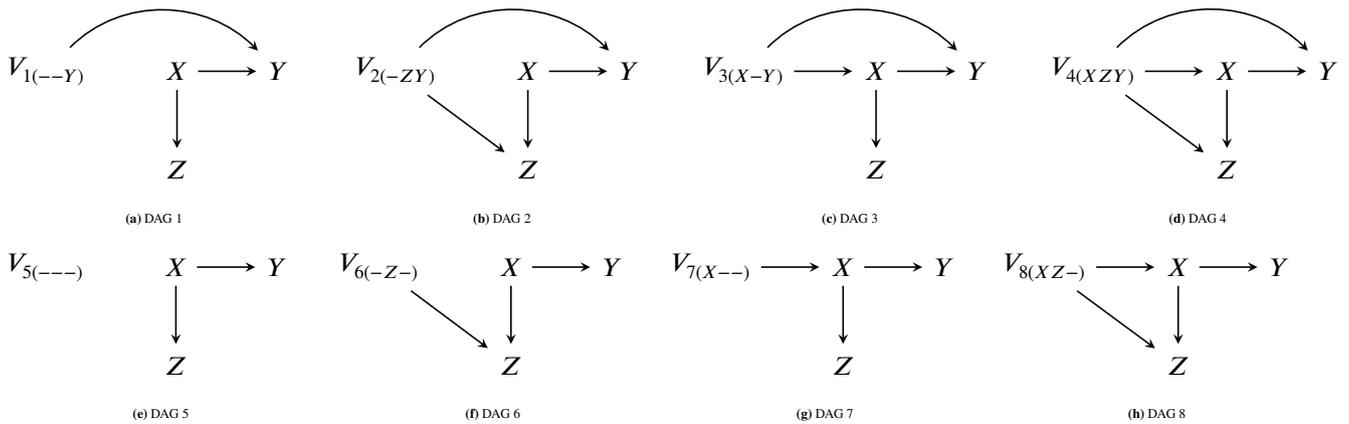
\begin{figure}[!]
     \caption{\label{fig1}: Directed Acyclic Graphs Representing Common Measurement Error Structures}
     \begin{subfigure}[b]{0.23\textwidth}
         \begin{tikzpicture}[>= stealth, shorten >= 1pt, auto, node distance = 0.8 cm, semithick]
            % nodes %
            \node[text centered] (v) {$V_{1(-{}-Y)}$};
            \node[right= of v] (x) {$X$};
            \node[right= of x] (y) {$Y$};
            \node[below= of x] (z) {$Z$};
            % edges %
            \path[->] (x) edge node {} (z);
            \path[->] (v) edge [out=45, in=135] node {} (y);
            \path[->] (x) edge node {} (y);
        \end{tikzpicture}
        \caption{DAG 1}
     \end{subfigure}
     \begin{subfigure}[b]{0.23\textwidth}
         \begin{tikzpicture}[>= stealth, shorten >= 1pt, auto, node distance = 0.8 cm, semithick]
            % nodes %
            \node[text centered] (v) {$V_{2(-ZY)}$};
            \node[right= of v] (x) {$X$};
            \node[right= of x] (y) {$Y$};
            \node[below= of x] (z) {$Z$};
            % edges %
            \path[->] (x) edge node {} (z);
            \path[->] (v) edge [out=45, in=135] node {} (y);
            \path[->] (x) edge node {} (y);
            \path[->] (v) edge node {} (z);
        \end{tikzpicture}
        \caption{DAG 2}
     \end{subfigure}
     \begin{subfigure}[b]{0.23\textwidth}
         \begin{tikzpicture}[>= stealth, shorten >= 1pt, auto, node distance = 0.8 cm, semithick]
            % nodes %
            \node[text centered] (v) {$V_{3(X-Y)}$};
            \node[right= of v] (x) {$X$};
            \node[right= of x] (y) {$Y$};
            \node[below= of x] (z) {$Z$};
            % edges %
            \path[->] (x) edge node {} (z);
            \path[->] (v) edge [out=45, in=135] node {} (y);
            \path[->] (x) edge node {} (y);
            \path[->] (v) edge node {} (x);
        \end{tikzpicture}
        \caption{DAG 3}
     \end{subfigure}
     \begin{subfigure}[b]{0.23\textwidth}
         \begin{tikzpicture}[>= stealth, shorten >= 1pt, auto, node distance = 0.8 cm, semithick]
            % nodes %
            \node[text centered] (v) {$V_{4(XZY)}$};
            \node[right= of v] (x) {$X$};
            \node[right= of x] (y) {$Y$};
            \node[below= of x] (z) {$Z$};
            % edges %
            \path[->] (x) edge node {} (z);
            \path[->] (v) edge [out=45, in=135] node {} (y);
            \path[->] (x) edge node {} (y);
            \path[->] (v) edge node {} (x);
            \path[->] (v) edge node {} (z);
        \end{tikzpicture}
        \caption{DAG 4}
     \end{subfigure}
     \begin{subfigure}[b]{0.24\textwidth}
         \begin{tikzpicture}[>= stealth, shorten >= 1pt, auto, node distance = 0.8 cm, semithick]
            % nodes %
            \node[text centered] (v) {$V_{5(-{}-{}-)}$};
            \node[right= of v] (x) {$X$};
            \node[right= of x] (y) {$Y$};
            \node[below= of x] (z) {$Z$};
            % edges %
            \path[->] (x) edge node {} (z);
            \path[->] (x) edge node {} (y);
        \end{tikzpicture}
        \caption{DAG 5}
     \end{subfigure}
     \begin{subfigure}[b]{0.24\textwidth}
         \begin{tikzpicture}[>= stealth, shorten >= 1pt, auto, node distance = 0.8 cm, semithick]
            % nodes %
            \node[text centered] (v) {$V_{6(-{}Z{}-)}$};
            \node[right= of v] (x) {$X$};
            \node[right= of x] (y) {$Y$};
            \node[below= of x] (z) {$Z$};
            % edges %
            \path[->] (x) edge node {} (z);
            \path[->] (x) edge node {} (y);
            \path[->] (v) edge node {} (z);
        \end{tikzpicture}
        \caption{DAG 6}
     \end{subfigure}
     \begin{subfigure}[b]{0.24\textwidth}
         \begin{tikzpicture}[>= stealth, shorten >= 1pt, auto, node distance = 0.8 cm, semithick]
            % nodes %
            \node[text centered] (v) {$V_{7(X{}-{}-)}$};
            \node[right= of v] (x) {$X$};
            \node[right= of x] (y) {$Y$};
            \node[below= of x] (z) {$Z$};
            % edges %
            \path[->] (x) edge node {} (z);
            \path[->] (x) edge node {} (y);
            \path[->] (v) edge node {} (x);
        \end{tikzpicture}
        \caption{DAG 7}
     \end{subfigure}
     \begin{subfigure}[b]{0.24\textwidth}
         \begin{tikzpicture}[>= stealth, shorten >= 1pt, auto, node distance = 0.8 cm, semithick]
            % nodes %
            \node[text centered] (v) {$V_{8(X{}Z{}-)}$};
            \node[right= of v] (x) {$X$};
            \node[right= of x] (y) {$Y$};
            \node[below= of x] (z) {$Z$};
            % edges %
            \path[->] (x) edge node {} (z);
            \path[->] (x) edge node {} (y);
            \path[->] (v) edge node {} (z);
            \path[->] (v) edge node {} (x);
        \end{tikzpicture}
        \caption{DAG 8}
     \end{subfigure}
\end{figure}

For DAGs 1, 3, 5, 6, 7 and 8, the non-differential measurement error condition (i.e. the surrogacy assumption) $Y \perp Z |X$ is encoded and for all eight DAGs, $Y \perp Z| X, V_j$, where $j=1,\dots,8$\cite{HernanRobinsWhatIf,Richardson2013SWIG}. We can also establish the following exchangeability (i.e. no confounding) assumptions  $Y^x \perp X$ for DAGs 1, 2, 5, 6, 7 and 8 and $Y^x \perp X|V_j$ for all eight DAGs where $j=1,\dots,8$. 

If $V$ does not directly cause changes in $X$ or $Z$ (through causing measurement error) but is associated with $X$ or $Z$ via some (unmeasured) common causes, we can replace the arrows from $V$ into $X$ or $Z$ with a double-headed arrow. However, this modification produces the same conditions listed above and will not change our conclusion on validity and efficiency and therefore will not be discussed separately. 

\section{Validity and Efficiency of Regression Calibration Estimators}

\subsection{Identification of Causal Effect of Interest}
We show in section 1.2 of the online supplemental material (SM) that under the following sufficient conditions:  
\begin{itemize}
    \item[C.1.1] Exchangeability: $Y^x \perp X |V$,
    \item[C.1.2] Non-differential measurement error (surrogacy): $Y \perp Z| (X, V)$,
    \item[C.1.3] Consistency: $Y^x, Z^x$ take observed value $Y$, $Z$ when $X=x$, and
    \item[C.1.4] Linearity: The conditional mean $E[Y|X,V]$ is linear with respect to $X$ and can be rewritten as $E[Y|X,V] = \kappa + \beta(V) X + g(V)$.
\end{itemize} we can identify the CATE defined in 2.1 for all DAGs 1 through 8 using the $\beta (V)$ in the following equation:

\begin{equation} 
\label{identify_formula_V}
E[Y|Z, V] = \kappa + \beta(V) E[X|Z,V] + g(V),
\end{equation}estimable using tools such as estimating equations. 

If we additionally make the following modeling assumption: 
\begin{itemize}
\item[C.1.5] The models for $E[X|Z=z,V=v]$ and $E[Y|Z=z,V=v]$  as in (\ref{eq1}) and (\ref{eq3}) are correct  and $E[V|Z=z]$ is a linear function of $Z$,
\end{itemize} $\beta(V)$ reduces to $\beta$ and is identified as:
\begin{equation} \label{eq10}
\beta (V) = \beta = \beta_{_{(OM)}}. 
\end{equation} 

We also show in the SM that under the following sufficient conditions:  
\begin{itemize}
    \item[C.2.1] Exchangeability: $Y^x \perp X$,
    \item[C.2.2] Non-differential measurement error (surrogacy): $Y \perp Z| X$,
    \item[C.2.3] Consistency: $Y^x, Z^x$ take observed value $Y$, $Z$ when $X=x$, and
    \item[C.2.4] Linearity: the conditional mean $E[Y|X]$ is linear with respect to $X$ and can be rewritten as $E[Y|X] = \kappa + \beta X$,

\end{itemize} we can identify the ATE defined in 2.1 for DAGs 5, 6, 7 and 8 with $\beta$ in the following equation (estimable using estimating equations): 
\begin{equation}  \label{identify_formula_noV}
E[Y|Z] = \kappa + \beta E[X|Z].
\end{equation}

If we make the following modeling assumption: 
\begin{itemize}
    \item[C.2.5] The models for $E[X|Z=z]$ and $E[Y|Z=z]$ as in (\ref{eq2}) and (\ref{eq4}) are correct and $E[V|Z]$ is a linear function of $Z$,
\end{itemize} $\beta$ is identified as: 
\begin{equation} \label{eq9}
\beta=\beta_{_{(--)}}.
\end{equation}

\subsection{Validity and Efficiency of Regression Calibration Estimators} \label{theoreticalResults}
Each of the four CRS regression calibration estimators' validity was evaluated using the definitions and assumptions given in section 2 and 3.1. Whenever more than one estimator is valid for a given DAG, asymptotic relative efficiency was analytically evaluated under the relevant linear models \eqref{eq1} through \eqref{eq6}. The asymptotic variance of CRS estimators is obtained using sandwich formula of estimating equations\cite{CRS_book}. See section 1 and 2 of the SM for details for the proof of validity and relative efficiency. Note that some of the validity and efficiency results rely on a slightly stronger modeling assumption than that required by \eqref{eq1}, i.e. $X$ can be expressed as the sum of conditional mean $E[X|Z,V]$ plus a residual term that has zero correlation with $V$. 

We summarize whether each CRS estimator is valid across the DAGs in Table \ref{tab:validity}, with the most efficient estimators indicated in the parenthesis. The minimal covariate adjustment sets contain $V_{2(-ZY)}, V_{3(X-Y)}, V_{4(XZY)}$, which need to be collected in both the main and validation studies. While $V_{3(X-Y)}$ and $V_{4(XZY)}$ need to be included in both the MEM and the outcome model, $V_{2(-ZY)}$ can be included in either both the MEM and the outcome model or in the MEM alone. Similar to what was found for the RSW method\cite{tang2022causal}, $V_{1(-{}-Y)}$ can be adjusted for in both, neither or either of the outcome model and the MEM, where adjustment in both models and in the outcome model only produce the most efficient estimators. When it comes to $V_{6(-Z-)}$, $V_{7(X-{}-)}$ and $V_{8(XZ-)}$, one can choose to adjust for these covariates in (1) both the outcome model and the MEM, (2) neither the outcome model nor the MEM, or (3) in the MEM only, with option (3) being the most efficient. The covariate adjustment rules for $V_{2(-ZY)}$, $V_{6(-Z-)}$, $V_{7(-{}Z-)}$ and $V_{8(XZ-)}$ differ from that required by the RSW method, where adjusting for these covariates in the MEM only results in bias \cite{tang2022causal}.
\begin{table}[!]
\centering
  \begin{threeparttable}
  \caption{: Validity and Efficiency of CRS Estimators with Linear Models}
  \label{tab:validity}
    \begin{tabular}{|c || c  c  c  c||}
        \hline
        $V_j$ as in \tnote{a} & $\beta_{_{(OM)}}$ & $\beta_{_{(--)}}$ & $\beta_{_{(-M)}}$ & $\beta_{_{(O-)}}$ \\[1ex] 
        \hline
        DAG 1, $V_{1(-{}-{}Y)}$ & valid (efficient) & valid & valid & valid (efficient)\\ 
        DAG 2, $V_{2(-{}Z{}Y)}$ & valid (efficient) & biased & valid\tnote{b} & biased \\
        DAG 3, $V_{3(X{}-{}Y)}$ & valid & biased & biased & biased \\
        DAG 4, $V_{4(X{}Z{}Y)}$ & valid & biased & biased & biased \\
        DAG 5, $V_{5(-{}-{}-)}$ & valid & valid & valid & valid \\ 
        DAG 6, $V_{6(-{}Z{}-)}$ & valid (efficient) & valid & valid (efficient\tnote{b} ) & biased\\ 
        DAG 7, $V_{7(X{}-{}-)}$ & valid & valid & valid  (efficient\tnote{b} ) & biased\\ 
        DAG 8, $V_{8(X{}Z{}-)}$ & valid & valid & valid (efficient\tnote{b} ) & biased\\ 
        \hline
    \end{tabular}
    
    \begin{tablenotes}
        \item[a] The subscript such as $(-{}-{}Y)$ emphasizes how the given covariate relate to $X, Z$ and $Y$. For example, DAG 2 describes a situation where covariate $V_{2(-ZY)}$ systematically affects measurement error and is a risk factor for the outcome.
        \item[b] This validity/efficiency result additionally relies on $X=E[X|Z,V] + \epsilon_{x|v,z}$ where $Cov(\epsilon_{x|v,z}, V)=0$.
    \end{tablenotes}
 \end{threeparttable}
\end{table}

\subsection{Approximation in Generalized Linear Models with Logistic Link} \label{s34}

For binary outcomes modeled by logistic regression, the efficiency results may differ from the results given in section \ref{theoreticalResults} applying to linear models. Specifically, the coefficient estimates in the logistic outcome model for $\hat{X}$ not conditioning on ($V_{1(-{}-{}Y)}$,$V_{2(-{}Z{}Y)}$), i.e. $\hat{\beta}_{_{(--)}}$ and $\hat{\beta}_{_{(-M)}}$, converge to values that are always closer to zero than the value to which the coefficient estimates $\hat{\beta}_{_{(O-)}}$, $\hat{\beta}_{_{(OM)}}$ converge\cite{neuhaus1993geometric}. This is known as the non-collapsibility of odds ratio \cite{sjolander2016note} and has several consequences for the efficiency results, because for binomial and bernoulli outcome data, the variances of the point estimates are functionally dependent on the probability parameter itself. This means, for DAG 1, $Var(\hat{\beta}_{_{(OM)}})$ and $Var(\hat{\beta}_{_{(O-)}})$ are not necessarily smaller than $Var(\hat{\beta}_{_{(-M)}})$ and $Var(\hat{\beta}_{_{(--)}})$ respectively because ${\beta}_{_{(OM)}}^2 \ge {\beta}_{_{(-M)}}^2$, and ${\beta}_{_{(O-)}}^2 \ge {\beta}_{_{(--)}}^2$. Similarly, for DAG 2, $Var(\hat{\beta}_{_{(-M)}})$ is not necessarily greater in value than $Var(\hat{\beta}_{_{(OM)}})$. 

It has been shown that when logistic regression is used to model binary outcomes, under one of the following two condition sets, valid CRS estimators $\hat{\beta}_{(OM)}$ and $\hat{\beta}_{(--)}$ approximately converge to the true effect: 
\begin{enumerate}
    \item Condition set I: $var(X|Z,V)\beta^2$ is small (e.g. less than 0.5), where $var(X|Z,V)$ can be estimated as the mean squared error from the MEM and $\beta$ is the true causal effect \cite{kuha1994corrections,spiegelman2000estimation};
    \item  Condition set II: the disease is rare (e.g. less than 5\%) and the error term in the MEM is homoskedastic\cite{RSW2}. 
\end{enumerate} 

We also show via simulation that under either of these condition sets, the approximation also works well for CRS estimators $\hat{\beta}_{(-M)}$ and $\hat{\beta}_{(O-)}$. 

\section{Simulation Studies}
Monte Carlo simulation experiments were conducted to empirically evaluate the finite sample performance of the theoretical results in section 3, including both validity and efficiency, of all CRS estimators for both continuous and binary outcomes.

\subsection{Simulation Study Design}
Following the data generating process below

\begin{enumerate}[noitemsep,topsep=0pt,parsep=0pt,partopsep=0pt]
    \item $X = \eta_{v} V + \epsilon_x, \epsilon_x \sim N(0,1)$, 
    \item $Z = \theta_{x} X + \theta_{v} V + \epsilon_z, \epsilon_z \sim N(0,0.5)$, 
    \item $Y = \beta_{x} X + \beta_{v}V  + \epsilon_y,\epsilon_y \sim N(0,1)$ for continuous $Y$, and 
    \item $Y \sim Bern(p), log(p/(1-p))= -5 + \beta_{x} X + \beta_{v} V$ for binary outcome $Y$, 
\end{enumerate}as base case scenario, we generated 1,000 samples of size $n=5000$ for continuous outcome $Y$ and $10,000$ for binary $Y$, where a subset of size $n_{VS}=400$ was randomly sampled to form external validation study data. The remaining $n_{MS}=4,600$ or $9,600$ was kept as main study data where only $(V, Z, Y)$ is observed. For binary outcome scenarios, all simulations satisfied at least one of the two condition sets discussed in section \ref{s34}.

Also as base case scenario, the true causal effect of $X$ on $Y$ was set to be $\beta_x = 0.5$, on the additive scale for continuous outcome and on the logit scale for binary outcome (equivalent to an odds ratio of approximately $1.65$). For other parameters, we used the following parameterization: $\eta_{v} = (0.5,0), \theta_{x} = 0.8, \theta_{v} = (0.5, 0), \beta_{v} = (0.5,0)$, which corresponds to each of the scenarios in the $2^3=8$ DAGs in \ref{fig1}. Details for the comprehensive simulation scenarios are given in section 4 of the SM. 

\subsection{Results}
We present the empirical percent biases in Table \ref{tab:basecase} for the point estimates. For the efficiency comparison of the valid estimators, we reported the empirical relative efficiency (ERE) in Table \ref{tab:basecase_efficiency}, calculated as the empirical variance of $\hat{\beta}_{(OM)}$ divided by the empirical variance of the other valid estimators, as well as the empirical variance for the valid point estimates obtained over simulation replicates. Sections 4 and 5 of the SM contain the results for all simulation scenarios, which are briefly summarized below. 

\begin{table}[!]
  \centering
  \begin{threeparttable}
  \caption{: Percent Bias for Point Estimates under Base Case \tnote{a}}
  \label{tab:basecase}
%    \scriptsize
\begin{tabular}{| l |  c c c c | c c c c |}
\hline
 &  \multicolumn{8}{ c |}{\textbf{Percent Bias (\%)}} \\
\textbf{$V_j$ as in} &  \multicolumn{4}{ c |}{\textbf{Continuous Outcome}} &  \multicolumn{4}{ c |}{\textbf{Binary outcome}\tnote{b}} \\
& $\hat{\beta}_{(OM)}$ & $\hat{\beta}_{(--)}$ & $\hat{\beta}_{(-M)}$ & $\hat{\beta}_{(O-)}$ & $\hat{\beta}_{(OM)}$ & $\hat{\beta}_{(--)}$ & $\hat{\beta}_{(-M)}$ & $\hat{\beta}_{(O-)}$  \\
\hline 
                         $V_{1,(-{}-Y)}$                                 & 0 & 0  & 0   & 0  & 0 & -1 & -1  & 0 \\
                         $V_{2,(-{}ZY)}$                                 & 0 & 32 & 0   & 2 & 0 & 31 & -1   & 2 \\
                         $V_{3,(X{}-{}Y)}$                               & 0 & 55 & 97  & -7  & 0 & 52 & 94  & -7 \\
                         $V_{4,(X{}Z{}Y)}$                               & 0 & 78 & 97  & -5 & 0 & 75 & 94  & -5 \\
                         $V_{5,(-{}-{}-)}$                               & 0 & 0  & 0   & 0   & 0 & 1 & 1  & 1\\
                         $V_{6,(-{}Z{}-)}$                               & 0 & 0  & 0 & 2  & 1 & 1 & 1 & 3\\
                         $V_{7,(X{}-{}-)}$                               & 0 & 0  & 0   & -7 & 1 & 2 & 2   & -6  \\
                         $V_{8,(X{}Z{}-)}$                               & 0 & 0  & 0 & -5  & 1 & 2 & 2 & -4 \\
\hline
\end{tabular}
         \begin{tablenotes}
        \item[a] As base case scenario, we generated 1,000 samples of size $n=5000$ for continuous outcome $Y$ and $10,000$ for binary $Y$, where a subset of size $n_{VS}=400$ was randomly sampled to form external validation study data. The remaining $n_{MS}=4,600$ or $9,600$ was kept as main study data, with the following coefficients: $\beta_x = 0.5, \eta_{v} = (0.5,0), \theta_{x} = 0.8, \theta_{v} = (0.5, 0)$ and $\beta_{v} = (0.5,0)$.
        \item[b] Satisfies outcome prevalence less than 5\% or small measurement error condition ${var}(X|Z,V)\beta^2<0.5)$.
    \end{tablenotes}
 \end{threeparttable}
\end{table}

The results for continuous outcome met analytical expectations across all simulation. Adjusting for $V_{1(-{}-Y)}, V_{2(-ZY)}$ in the MEM only, though valid, had coverage probability as low as $0.69$ under the base case. This is likely due to the poor finite sample performance of the variance estimator, reflecting the severity of efficiency loss. The inter-quantile range of the coverage probability for all the other valid estimators was (0.95, 0.95), with the lowest coverage probability of 0.92. For binary outcomes, the approximation worked well for all valid estimators under condition set I or II (see section 3.3) with percent bias $<6\%$ across all scenarios except for one where the percent bias was equal to $12\%$. Excluding the valid estimators where $V_{1(-{}-Y)}, V_{2(-ZY)}$ were adjusted in the MEM only, the interquartile range of the coverage probability for all the other valid estimators was (0.95, 0.95), with the lowest coverage probability of 0.93. We also noticed that adjusting for $V_{1(--Y)}$ in the logistic outcome model alone or in both the outcome and the MEM resulted in similar variance compared with adjusting for such variables in neither model (Table \ref{tab:basecase_efficiency} and section 5 of SM). 

For both continuous and binary outcomes, our simulations confirmed that an efficiency gain is obtained by adjusting for $V_{6(-Z{}-{})}$, $V_{7(X-{}-{})}$ and $V_{8(XZ{}-)}$ in the MEM only ($1.01 \le ERE(\beta_{_{(-M)}}) \le 2.84)$) (section 5 of SM). Also for both outcome types, as long as the correlation between covariates $V_{2(-ZY)}$, $V_{3(X-Y)}$ and $V_{4(-ZY)}$ and measurement error was small (e.g. $\rho_{v,z|x} \le 0.2$), we consistently observed less biased estimates when we adjusted for the covariates in the outcome model only (percent bias $< 10\%$ in most cases) versus if we did not adjust for them in either of the two models (percent bias $> 30\%$ in most cases). 
\begin{table}
  \centering
  \begin{threeparttable}
  \caption{: Empirical Relative Efficiency (ERE) and Empirical Variance under Base Case When More Than One Estimator is Valid}
  \label{tab:basecase_efficiency}
%  \scriptsize
\begin{tabular}{| l l | c c c c |}
\hline
\textbf{Outcome Type} & \textbf{$V_j$ as in} &  \multicolumn{4}{ c |}{\textbf{ERE \tnote{a} (Empirical Variance \tnote{b} )}} \\
& & \textbf{$\hat{\beta}_{(OM)}$}   & \textbf{$\hat{\beta}_{(--)}$} & \textbf{$\hat{\beta}_{(-M)}$} & \textbf{$\hat{\beta}_{(O-)}$} \\
\hline
Continuous & $V_{1,(-{}-Y)}$ &  1 (1.14) & 0.79 (1.44) & 0.24 (4.81) & 1.00 (1.14) \\
           & $V_{2,(-{}Z{}Y)}$ &  1 (1.14) &           & 0.24 (4.81) & \\
           & $V_{5,(-{}-{}-)}$ &  1 (1.14) & 1.00 (1.14)    & 1.01 (1.13)    & 1.00 (1.14) \\
           & $V_{6,(-{}Z{}-)}$ & 1 (1.14) & 0.97 (1.18) &  1.01 (1.13)           &          \\
           & $V_{7,(X{}-{}-)}$ & 1 (1.14) & 1.19 (0.96) &  1.31 (0.87)           &          \\
           & $V_{8,(X{}Z{}-)}$ & 1 (1.14) & 1.28 (0.89) &  1.31 (0.87)           &          \\
\hline
Binary \tnote{c}     & $V_{1,(-{}-Y)}$ & 1 (2.11) & 1.01 (2.09) & 0.92 (2.29) & 1.00 (2.11) \\
           & $V_{2,(-{}Z{}Y)}$ & 1 (2.11) &           & 0.92 (2.29) & \\
           & $V_{5,(-{}-{}-)}$ & 1 (2.90)  & 1.00 (2.91) & 1.00 (2.90) & 1.00 (2.91) \\
           & $V_{6,(-{}Z{}-)}$ & 1 (2.90)  & 0.97 (2.99) & 1.00 (2.89)           &          \\
           & $V_{7,(X{}-{}-)}$ & 1 (2.84) & 1.25 (2.28) &  1.34 (2.13)           &          \\
           & $V_{8,(X{}Z{}-)}$ & 1 (2.84) & 1.31 (2.17) &  1.34 (2.13)           &  \\
\hline
\end{tabular}
         \begin{tablenotes}
        \item[a] Calculated as the empirical variance of $\hat{\beta}_{(OM)}$ over the empirical variance of each estimator under consideration, i.e. $\text{ERE} (\hat{\beta}_{(\cdot)}) = \frac{Var(\hat{\beta}_{_{(OM)}})}{Var(\hat{\beta}_{{(\cdot)}})}$, where $\hat{\beta}_{(\cdot)}$ could be any of the four estimators.
        \item[b] Calculated as the variance of the point estimates $\hat{\beta}_{(\cdot)}$ over simulation replicates for each estimator. $\times 10^{-3}$ for continuous outcome and $\times 10^{-2}$ for binary outcome.  
        \item[c] Satisfies the condition that the outcome prevalence is less than 5\% or small measurement error condition $\widehat{var}(X|Z,V)\beta^2<0.5$.
    \end{tablenotes}
 \end{threeparttable}
\end{table}

\section{Real-data Example}
\subsection{Methods}

We studied the effect of total daily fiber intake on cardiovascular risk, defined as incidence of myocardial infarction or angina, using our proposed covariate selection framework among the participants of the Health Professional Follow-up Study (HPFS). HPFS is an ongoing prospective study of 51,529 US male health professionals 40 to 75 years of age at enrollment in 1986\cite{gu2022dietary,kim2014longitudinal}. We considered self-reported fiber intake assessed by the food frequency questionnaire (FFQ) as the mismeasured exposure and fiber intake obtained through dietary record (DR) in the men's lifestyle validation study as the true exposure \cite{gu2022dietary,kim2014longitudinal,cahill2013prospective,liu2002prospective,pernar2022validity}. A subcohort was created for this example, consisting of participants who completed the sleep duration question in 1987 and sunscreen use question in 1992. Participants who died, reported a CVD event, or any cancer diagnosis other than melanoma in and prior to 1990, or were missing birth year were ineligible for inclusion. 

We encoded subject matter knowledge based on existing empirical evidence into our DAG (Figure  \ref{fig2}). Here, we list the covariates that were considered, with their covariate set indicated in the parenthesis: family history of CVD, marital status ($V_{1(-{}-Y)}$), depression status\cite{mcdermott2009meta,van2007depression}, sleep duration\cite{ma2020association,hoevenaar2011sleep} ($V_{2({}-ZY)}$), energy intake (in calories), baseline hypertension, diabetes and hypercholesterolemia, smoking ($V_{3(X-Y)}$), age, physical exercise/activity (in metabolic equivalent tasks)\cite{mandolesi2018effects}, body mass index (BMI), use of multivitamin supplement\cite{grima2012effects}, alcohol drinking\cite{brennan2020long} ($V_{4(XZY)}$), and frequency of sunscreen use ($V_{7(X-{}-)}$). In contrast to the DAG in figure \ref{fig1}, correlations between covariate sets through an unmeasured common cause $U$ was allowed. We also note that sunscreen use here cannot be assumed to directly cause change in fiber intake therefore the direct arrow in DAG 7 in figure 2 is now replaced by the backdoor path through the $U$ (e.g. embracing a healthy lifestyle) connecting $V_{7(X-{}-)}$ and $X$. Despite this change, $V_{7(X-{}-)}$ still has the defining features that it is not a direct cause of the outcome and that it is independent of measurement error conditional on covariates that cause measurement error. Please refer to section 6 of the SM for more details about the real data example, including the empirical assessment of how each covariate, conditional on all other covariates, was associated with validated fiber intake $X$, measurement error (i.e. mismeasured FFQ fiber intake $Z$ conditional on $X$) and CVD incidence $Y$.

\usetikzlibrary{shapes,decorations,arrows,calc,arrows.meta,fit,positioning}
\begin{figure}
    \begin{center}
     \caption{\label{fig2}: Proposed DAG for the Effect of Fiber Intake on Cardiovascular Disease Incidence}
             \begin{tikzpicture}[>= stealth, shorten >= 1pt, auto, node distance = 0.8 cm, semithick]
                % nodes %
            \draw (0,0) node[above] (x) {fiber, DR (X)};
            \draw (4,0) node[above] (z) {fiber, FFQ (Z)};
            \draw (7,0) node[above] (y) {CVD (Y)};

            \draw (3.5, -1) node[below,text width=6cm] (v2) {{$V_{2(-ZY)}:$ sleep duration, depression}};
            \draw (0, -3) node[below,text width=7cm] (v3) {{$V_{3(X-Y)}$: energy intake, hypertension, diabetes, hypercholesterolemia, smoking status}};
            \draw (-5, 3) node[below,text width=4cm] (v1) {{$V_{1(-{}-Y)}:$ marital status, family history}};
            \draw (-4,-2) node[below,text width=6cm] (v4) {{$V_{4(XZY)}:$ age, physical exercise, BMI, multivitamin use, alcohol use}};
            \draw (-5,1) node[below] (v7) {{$V_{7(X{}-{}-)}:$ sunscreen use}};
            \draw (-8,-1) node[below] (u) {{$U$}};
                % edges %
                \path[->] (x) edge node {} (z);
                \path[->] (x) edge [out=45, in=135] node {} (y);
                \path[->] (v1) edge [out=0, in=135] node {} (y);
                %\path[->] (v2) edge node {} (y);
                %\path[->] (v2) edge node {} (z);
                \path[->] (v3) edge node {} (x);
                \path[->] (v3) edge [out= 0, in=270] node {} (y);
                \path[->] (v4) edge node {} (z);
                \path[->] (v4) edge node {} (x);
                \path[->] (v4) edge [out=0, in=240] node {} (y);
                \path[->] (u) edge [dashed, out=30, in=180] node {} (x);
%                \path[<->] (v4) edge [dashed, out=90, in=180] node {} (x);

               \path[->] (u) edge [dashed, out=270, in=180] node {} (v3);
                \path[->] (u) edge [dashed, out=270, in=180] node {} (v4);
                \path[->] (u) edge [dashed, out=270, in=180] node {} (v7);

                %\path[<->] (v3) edge [dashed, out=135, in=180] node {} (x);
                
%                \path[<->] (v7) edge [ out=45, in=135] node {} (x);
%                \path[<->] (v7) edge [dashed, out= 180, in=180] node {} (v4);
%                \path[<->] (v7) edge [dashed, out= 180, in=180] node {} (v3);

            \path[<->] (v2) edge [dashed, out=180, in=135] node {} (v3);
            \path[<->] (v2) edge [dashed, out=180, in=45] node {} (v4);
%                \path[<->] (v3) edge [dashed, out=180, in=270] node {} (v4);

            \end{tikzpicture}
    \end{center}
\end{figure}
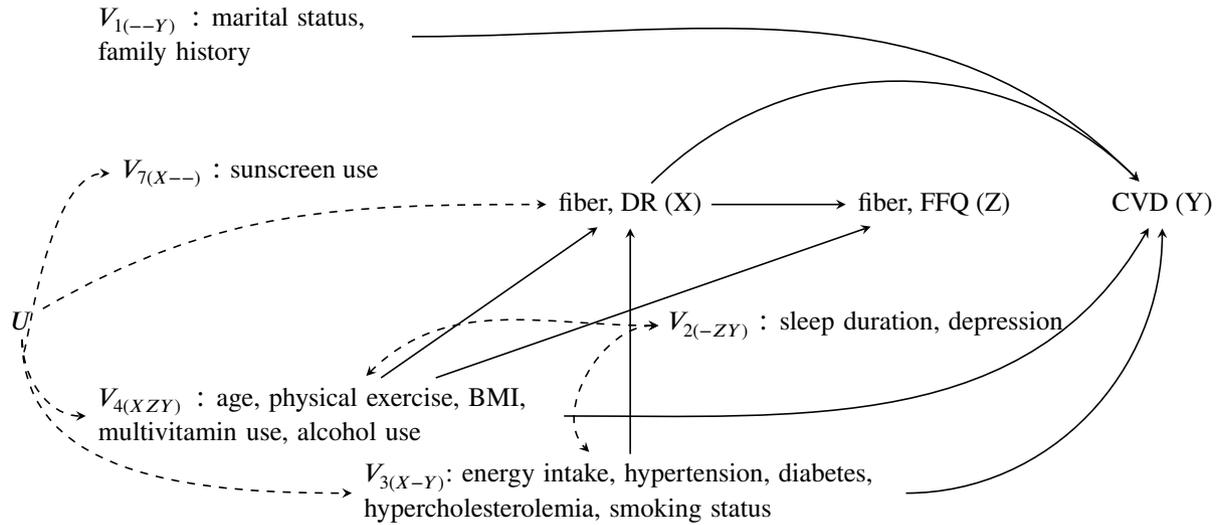

Following our theoretical results for \textit{valid} estimators, we adjusted for covariates sets $V_{2(-ZY)}, V_{3(X-Y)}, V_{4(XZY)}$, i.e., the minimal covariate adjustment set, with $V_{3(X-Y)}, V_{4(XZY)}$ in both the outcome model and the MEM (indicated by OM in Table 4) and $V_{2(-ZY})$ in either both the outcome model and the MEM (OM in Table 4) or in the MEM alone (-M in Table 4). We also adjusted for $V_{1(-{}-Y)}$ in the outcome model alone (O- in Table 4) and $V_{7(X{}-{}-)}$ in both the outcome and the MEM as well as in the MEM only in addition to adjusting for ${V_{2(-ZY)}, V_{3(X-Y)}, V_{4(XZY)}}$ in both the MEM and the outcome model, to evaluate how these additions affect point estimate and efficiency. 

In order to assess the impact of a data-driven variable selection approach in building MEM on the validity and efficiencies of the effect estimate, we built a MEM using LASSO regression, similar in terms of the data-driven principle as using adjusted $R^2$ to perform covariate selection as done in previous published studies\cite{Prentice2017sodiumCVD,prentice2019application,prentice2021biomarker}. Our candidate covariates include all covariates listed above. We used ten-fold cross-validation for tuning parameter selection. We obtained the predicted true exposure value in the main study sample and then adjusted for only $V_{3(X-Y)}, V_{4(X-Y)}$ (i.e. confounding varaibles for the $X-Y$ relationship) in the logistic outcome model. We used bootstrap to obtain the variance and confidence interval estimates. Please see section 7 in the SM for more detailed specification.

\subsection{Results}
A total of 22,379 participants were included in the analysis, among whom 409 participants also had dietary records. A total of 2,090 cardiovascular events were reported over the 16 years of follow-up, giving a cumulative incidence of $9\%$. For empirical assessment of the (partial) correlations between each covariate and $X$, measurement error (and thus $Z$) and $Y$ was presented in Table 5, section 6 of the SM. However, note that such numerical evaluation is only suggestive evidence of the presence or lack of arrow between variables, especially those evaluated within validation study containing limited sample size. 

We summarized the odds ratio representing the effect of increasing fiber intake by 10g daily on the incidence of CVD after correcting for the bias due to measurement error in table \ref{tab:application}. Regressing CVD event on the mismeasured fiber intake with the adjustment for different covariate sets ($V_{3(X-Y)}, V_{4(XZY)}$), ($V_{2(-ZY)},V_{3(X-Y)}, V_{4(XZY)}$), ($V_{1(-{}-Y)}, V_{2(-ZY)},V_{3(X-Y)}, V_{4(XZY)}$) or ($V_{2(-ZY)},V_{3(X-Y)}, V_{4(XZY)},V_{7(X{}-{}-)}$) (as in standard regression analyses) all produce an attenuated odds ratio (OR) estimate of 0.93, with slightly differing (biased) confidence intervals (CIs). Adjusting for ($V_{2(-ZY)},V_{3(X-Y)}, V_{4(XZY)}$) in both outcome model and MEM gives a corrected OR (95\% CI) of $0.83 (0.69,1.00)$. Adjusting for $V_{2(-ZY)}$ in the MEM alone slightly increased the point estimate to 0.85 while decreasing the standard error, compared with adjusting for $V_{2(-ZY)}$ in both models, from $9.480$ to $9.264 \times 10^{-2}$. This could be due to the fact that covariates in $V_{2(-ZY)}$ are only weakly associated with the outcome (ORs range from 0.87 to 0.96 and p values range from 0.12 to 0.66, see Section 6 of the SM). After adjusting for ($V_{2(-ZY)},V_{3(X-Y)}, V_{4(XZY)}$), additionally adjusting for $V_{1(-{}-Y)}$ in the logistic outcome model alone, in the MEM alone or in both models all resulted in similar point estimates of $0.83$ as well as similar standard errors as not adjusting for $V_{1(-{}-Y)}$ in any model. The corrected odds ratio slightly increased to $0.85$ by additionally adjusting for $V_{7(X{}-{}-)}$ in the MEM alone, and a small efficiency gain was observed (4\% reduction in SE compared with adjusting for ($V_{2(-ZY)},V_{3(X-Y)}, V_{4(XZY)}$) in both models).

 The data-driven approach selected the following variables into the MEM: daily total fiber intake measured by FFQ, family history ($V_{1(-{}-Y)}$), energy intake, diabetes, hypercholesteolemia, smoking ($V_{3(X-Y)}$), physical activity, BMI, supplements use, alcohol drinking ($V_{4(XZY)}$) and sun-screen use ($V_{7(X-{}-)}$). The corrected point estimate from this approach (0.90) was substantially different compared with all other (valid) point estimates obtained through subject-knowledge-driven covariate selection. 

\begin{table}
  \begin{threeparttable}
  \caption{: Valid Effect Estimates of 10g/d Increase in Fiber Intake on CVD Risk}
  \label{tab:application}
  %\scriptsize
    \begin{tabular}{| r | c c | c c c |}
        \hline
        CRS Adjusted Sets (Models) \tnote{a} & \multicolumn{2}{c|}{Uncorrected Analysis}  & \multicolumn{3}{c|}{Corrected Analysis} \\
        {} & Uncorrected OR & 95\% CI & Corrected OR & 95\% CI & SE ($10^{-2}$) \\
        \hline
        $V_{2(-ZY)}$ (-M) , $V_{3(X-Y)}, V_{4(XZY)}$ (OM) & 0.93 & (0.86, 1.00) & 0.85 & (0.71, 1.02) &  9.264\\
        $V_{2(-ZY)}, V_{3(X-Y)}, V_{4(XZY)}$ (OM) & 0.93 & (0.87, 1.00) & 0.83 & (0.69, 1.00) &  9.480  \\
        $V_{1(-{}-{}Y)}, V_{2(-ZY)}, V_{3(X-Y)}, V_{4(XZY)}$ (OM) & 0.93 & (0.86, 1.00) & 0.83 & (0.69, 1.00) &  9.488\\
        $V_{1(-{}-{}Y)}$ (-M) , $V_{2(-ZY)}, V_{3(X-Y)}, V_{4(XZY)}$ (OM) & 0.93 & (0.87, 1.00) & 0.83 & (0.69, 1.00) &  9.418\\
        $V_{1(-{}-Y)}$ (O-), $V_{2(-ZY)}, V_{3(X-Y)}, V_{4(XZY)}$ (OM) & 0.93 & (0.86, 1.00) & 0.83 & (0.69, 1.00) & 9.492\\
        $V_{2(-ZY)}, V_{3(X-Y)}, V_{4(XZY)}, V_{7(X{}-{}-)}$ (OM) & 0.93 & (0.87, 1.00) & 0.83 & (0.69, 1.01) & 9.560\\
        $V_{7(X{}-{}-)}$ (-M), $V_{2(-ZY)}, V_{3(X-Y)}, V_{4(XZY)}$ (OM) &  0.93 & (0.87, 1.00) & 0.85 & (0.71, 1.02) & 9.119\\
        LASSO (-M) + $V_{3(X-Y)}, V_{4(XZY)}$ (O-) & 0.93 & (0.86, 1.00) & 0.90 & (0.71, 1.11) & 11.340\\
        \hline
    \end{tabular}

    \begin{tablenotes}
        \item[a] (OM), (-M) and (O-) indicate whether the covariate(s) are adjusted in both outcome model and MEM, MEM only or outcome model only. For example, $V_{2(-ZY)}$ (-M) , $V_{3(X-Y)}, V_{4(XZY)}$ (OM) indicates that $V_{2(-ZY)}$ is adjusted in MEM only and $V_{3(X-Y)}, V_{4(XZY)}$ are adjusted in both MEM and outcome model. 
    \end{tablenotes}
    \end{threeparttable}
\end{table}

\section{Discussion}
\subsection{Summary of Main Results}
First, our findings suggest that in order to obtain a valid causal effect, it is necessary to collect data on the minimal set of pre-treatment covariates $V=(V_{2(-ZY)}, V_{3(X-Y)}, V_{4(XZY)})$ that account for the common causes of true exposure and outcome, as well as the common causes of measurement error and outcome, within both the main and validation studies. However, the covariate adjustment rule differs slightly between CRS and RSW methods \cite{RSW1,RSW2,tang2022causal}, both of which fall under the regression calibration umbrella. For both methods, it is important to include $V_{3(X-Y)}$ and $V_{4(XZY)}$, i.e. conventional confounding variables, in both the MEM and the outcome model. In contrast to RSW method, it is not necessary to include $V_{2(-ZY)}$ in the outcome model for the validity of CRS method under linear models. However, our simulation suggests that the finite sample performance is poor when $V_{2(-ZY)}$ (or $V_{1(-{}-Y)}$) is adjusted in the MEM only. Therefore, in practice, we recommend including the minimal covariate set in both the MEM and the outcome model.

Having to include confounders such as $V_{3(X-Y)}$ in the MEM may seem counterintuitive, as they do not contribute to measurement error. However, they are necessary for imputing the correct true exposure value. In practice, we see researchers (incorrectly) including variables in the nature of $V_{3(X-Y)}$ either in the MEM only (e.g. total energy intake \cite{prentice2022four}) or in the outcome model only (e.g. age \cite{gonzalez2006fruit}). 

Second, for both RSW and CRS methods under linear outcome model, we may gain statistical efficiency by including $V_{1(-{}-Y)}$ in the outcome model alone relative to omitting $V_{1(-{}-Y)}$ in either model. For CRS method, including $V_{6(Z{}-{}-)},V_{7(-{}X{}-)},V_{8(Z{}X{}-)}$ in the MEM alone (i.e., $\beta_{(-M)}$) are also more efficient compared with $\beta_{(--)}$ and $\beta_{(OM)}$, for both continuous and binary outcomes. In contrast, for RSW, including these covariates in the MEM alone would result in bias. Note, however, that $V_{6(Z{}-{}-)},V_{7(-{}X{}-)},V_{8(Z{}X{}-)}$ need to be collected in both main study and validation study samples such that the true exposure can be appropriately imputed in the main study, while $V_{1(-{}-Y)}$ is only needed in main study sample. The efficiency finding for $V_{6(Z{}-{}-)},V_{7(-{}X{}-)},V_{8(Z{}X{}-)}$ is consistent with that indicated in the missing data literature, where the covariates $V_{6(Z{}-{}-)},V_{7(-{}X{}-)},V_{8(Z{}X{}-)}$ can be thought of as auxiliary variables for the purpose of imputing missing true exposure data \cite{white2010bias}. 

Lastly, our results suggest that the data-driven covariate selection approach for building MEM model may result in bias when covariates of the nature $V=(V_{2(-ZY)}, V_{3(X-Y)}, V_{4(XZY)})$ are omitted from the MEM model. In our real-data example, we saw that the LASSO-based MEM forced the coefficient of the covariate age to zero due to its weak predictive power for the DR-based fiber intake conditional on FFQ-based fiber intake and other covariates in the validation study (section 7 of the SM). Per our simulations, we know that omitting such a covariate from the MEM will lead to obvious bias when the variable is strongly associated with both the measurement error and the outcome. In our real-data example, age is the strongest risk factor for CVD incidence (adjusted OR of $1.054$ with p-value $<0.001$) and has a strong correlation with the measurement error (partial correlation of $0.131$ with p-value $=0.008$). 

\subsection{Effect Modification by Covariate $V$} \label{s35}
In section 3.1, we obtained the identification equation $E[Y|Z,V] = \kappa + \beta (V) E[X|Z,V] + g(V)$ under assumptions C.1.1 through C.1.4, where both $\beta (V)$ and $g(V)$ can be flexible functions of covariate $V$. For example, if we believe there is a simple $X-V$ interaction and $g(V)$ is a linear function of $V$, we can modify model \eqref{eq3} as $E[Y|Z,V] = \gamma_0 + \gamma_1 E[X|Z,V] + \gamma_2 E[X|Z,V] V + \gamma_3 V$, where $\kappa = \gamma_0, \beta(V) = \gamma_1 + \gamma_2 V, g(V) =  \gamma_3 V$. 

When we are not willing to assume any parametric form for $\beta(V)$, $E[X|Z,V]$ or $g(V)$, we can then extend the CRS estimators (i.e. estimating equations) in section \ref{parametricCRSEstimators} to less restricted semi-parametric estimators, motivated by the following equations:

\begin{equation} \label{eq11}
E[Y|Z,V] = \kappa_1 + \beta_{_{(OM)}} (V) E[X|Z,V] + g(V)
\end{equation}
\begin{equation} \label{eq12}
E[Y|Z] = \kappa_2 + \beta_{_{(--)}} (V) E[X|Z]
\end{equation}
\begin{equation} \label{eq13}
E[Y|Z,V] = \kappa_3 + \beta_{_{(-M)}} (V) E[X|Z,V]
\end{equation}
\begin{equation} \label{eq14}
E[Y|Z,V] = \kappa_4 + \beta_{_{(O-)}} (V) E[X|Z] + g(V)
\end{equation}

We show in section 8 of the SM that the validity result in \ref{theoreticalResults} still apply in all cases with the following exception: in order for for $\beta_{_{(--)}} (V)$ in equation \eqref{eq12} to equal $\beta (V)$ under DAG 1 and for $\beta_{_{(-M)}} (V)$ in equation \eqref{eq13} to equal $\beta (V)$ under both DAG 1 and DAG 2, additional modeling assumptions are needed, including that the true exposure $X$ can be expressed as its mean conditional on $Z,V$ plus some error term that is uncorrelated with $V$. See details of the proof and a summary table in section 8 of the SM. The analytical relative efficiency results obtained under linear models do not necessarily hold for the above estimators and the proof of non-parametric (relative) efficiency is beyond the scope of this paper.

\subsection{Other Measurement Error Structures}
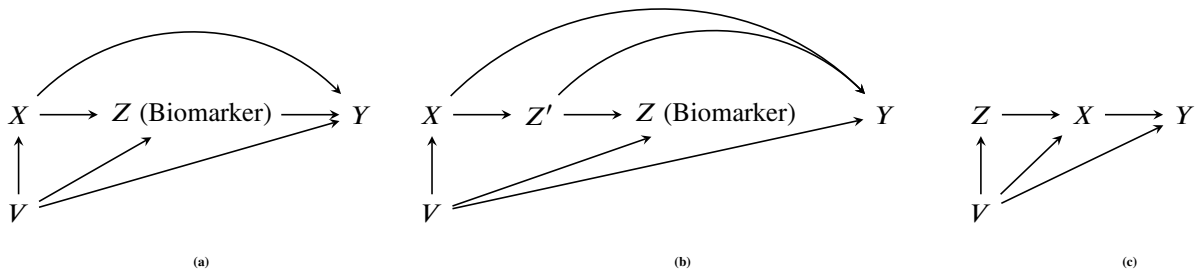
\begin{figure}
     \caption{: Other Measurement Error Structures\label{fig3}} 
     \begin{subfigure}[b]{0.3\textwidth} 
         \begin{tikzpicture}[>= stealth, shorten >= 1pt, auto, node distance = 0.8 cm, semithick]
            % nodes %
            \node[text centered] (x) {$X$};
            \node[right= of x] (z) {$Z$ (Biomarker)};
            \node[right= of z] (y) {$Y$};
            \node[below= of x] (v) {$V$};
            % edges %
            \path[->] (x) edge node {} (z);
            \path[->] (x) edge [out=45, in=135] node {} (y);
            \path[->] (z) edge node {} (y);
            \path[->] (v) edge node {} (x);
            \path[->] (v) edge node {} (z);
            \path[->] (v) edge node {} (y);
        \end{tikzpicture}
        \caption{\label{fig3a}}
     \end{subfigure}
     \begin{subfigure}[b]{0.4\textwidth}
         \begin{tikzpicture}[>= stealth, shorten >= 1pt, auto, node distance = 0.8 cm, semithick]
            % nodes %
            \node[text centered] (x) {$X$};
            \node[right= of x] (zprime) {$Z'$};
            \node[right= of zprime] (z) {$Z$ (Biomarker)};
            \node[right= of z] (y) {$Y$};
            \node[below= of x] (v) {$V$};
            % edges %
            \path[->] (x) edge node {} (zprime);
            \path[->] (x) edge [out=45, in=135] node {} (y);
            \path[->] (zprime) edge node {} (z);

            \path[->] (zprime) edge [out=45, in=135] node {} (y);
            \path[->] (v) edge node {} (x);
            \path[->] (v) edge node {} (z);
            \path[->] (v) edge node {} (y);
        \end{tikzpicture}
        \caption{\label{fig3b}}
     \end{subfigure}
         \begin{subfigure}[b]{0.25\textwidth}
         \begin{tikzpicture}[>= stealth, shorten >= 1pt, auto, node distance = 0.8 cm, semithick]
            % nodes %
            \node[text centered] (z) {$Z$};
            \node[right= of z] (x) {$X$};
            \node[right= of x] (y) {$Y$};
            \node[below= of z] (v) {$V$};
            % edges %
            \path[->] (z) edge node {} (x);
            \path[->] (x) edge node {} (y);
            \path[->] (v) edge node {} (y);
            \path[->] (v) edge node {} (x);
            \path[->] (v) edge node {} (z);
        \end{tikzpicture}
        \caption{\label{fig3c}}
     \end{subfigure}
\end{figure}

Finally, we want to caution against the use of biomarkers as a surrogate exposure. Particularly, recent nutrition studies frequently saw the use of feeding technique to obtain the true nutrient intake ($X$) and its biomarkers as surrogate exposure ($Z$)\cite{prentice2019application}. We note, however, this type of design does not necessarily share the same measurement error process as where one uses FFQ-based nutrient intake as surrogate exposure (as in Figure \ref{fig1}); biomarkers could act as both a surrogate exposure as well as a mediator of the nutrient's effect on the outcome, thus strongly indicating an arrow from surrogate $Z$ to outcome $Y$, as depicted in DAG \ref{fig3a}. Alternatively, the surrogate $Z$ could be a proxy or byproduct of a true but unmeasured biomarker $Z'$ that affects outcome, as in DAG \ref{fig3b}. Both of these situations violate surrogacy condition, but not as a result of the outcome affecting measurements of the surrogate exposure as in the typical case of differentiality \cite{HernanCole2009}. For example, serum beta carotene has been used as a surrogate to calibrate true beta carotene intake determined from feeding study\cite{Prentice2020}, although serum beta carotene can be converted into vitamin A, which improves the health of multiple organs and thus prevent chronic diseases\cite{grune2010beta}. In another study, (biomarkers-calibrated) carbohydrate density (i.e. percentage of energy intake from carbonhydrate) was found to be protective of a range of chronic disease outcomes including coronary heart disease and breast cancer among post-menopausal women\cite{prentice2021biomarker}. Here, on top of the issue that there might be no good biomarkers for carbohydrates, the bioarmkers could also serve as a mediator between true carbohydrate intake and chronic diseases\cite{BiochemicalIndicators2012}. Given that the validity of any of the CRS estimators considered in this paper (including the originally proposed estimator) rely on the surrogacy assumption, it is an open question whether these CRS estimators are still valid for identifying CATE or ATE in the presence of such differentiality. 

Another common measurement error structure not considered here is the one represented by air pollution exposure measurement where ambient measures, $Z$, collected through regional monitors are the surrogate exposure for air pollutants $X$ at the individual level\cite{weisskopf2017trade,weisskopf2015air} (depicted in DAG \ref{fig3c}). We will discuss the measurement error structures in Figure \ref{fig3} and their implication for covariate adjustment using regression calibration methods in separate papers. 

%\section{Funding} 
\noindent{\large{\textbf{Funding}}}

Supported in part through the National Institute of Health grants 1R01CA279175-01A1,5R01ES026246, R01 DC017717, and NCI U01 167552.. All statements in this article, including its findings and conclusions, are solely those of the authors and do not necessarily represent the views of National Institute of Health.
\\

% conflict of interest
%\section{Conflict of Interest} 
\noindent{\large{\textbf{Conflict of Interest}}}

The authors report no conflicts of interest.
\\

%acknowledgement
%\section{Acknowledgment} 
\noindent{\large{\textbf{Acknowledgement}}}

We thank Drs Jessica Young, James Robins, and Abrania Marrero (Harvard University) for helpful comments on different versions of this article and its supplemental material.
\\

%data availability
%\section{Data Sharing} 
\noindent{\large{\textbf{Data Sharing}}}

Data for the real-data example can be obtained by submitting data access request to the coordinating center of the Health Professional Follow-up Study. Codes for simulation are available upon request to the authors.

%\section{Bibliography}
%\nocite{*}% Show all bib entries - both cited and uncited; comment this line to view only cited bib entries;
\bibliography{WileyNJD-AMA}
%\begin{enumerate}[1]
%\item Use \verb"\bibliography{wileyNJD-AMA}" BST file for AMA reference style
%\item Use \verb"\bibliography{wileyNJD-APA}" BST file for APA reference style
%\item Use \verb"\bibliography{wileyNJD-AMS}" BST file for AMS reference style
%\item Use \verb"\bibliography{wileyNJD-VANCOUVER}" BST file for Vancouver reference style
%\item Use \verb"\bibliography{wileyNJD-ACS}" BST file for Chemistry reference style
%\end{enumerate}

%The normal commands for producing the reference list are:

%\begin{verbatim}
%\begin{thebibliography}{99}
%\bibitem{<x-ref label>}
%         <Reference details>

%\end{thebibliography}
%\end{verbatim}

\end{document}

\title{Causally Select Covariates in Regression Calibration for Mismeasured Continuous Exposure: Web Supplemental Material}

\authormark{Tang et al.}

\section{Validity of CRS Estimators}

\subsection{Data Set-up}\label{A1}
As specified in the manuscript, we assume that the underlying data generating law, $P(\mathbf{O})$, as consistent with each DAG, gives data $\mathbf{O}=(X,Y,Z,V)$, where $X$ is true exposure, $Z$ is mismeasured exposure, $V$ is a set of covariates, $Y$ outcome and $(X,Y,V)$ and $(X,Z,V)$ are observed in main study and validation study respectively, following a MS/EVS design. Because the main and validation study data are both generated by $\mathbf{O}=(X,Y,Z,V)$, transportability is guaranteed for all RSW estimators considered in the paper.  This automatically gives us transportability between the main and validation studies for all RSW estimators. All statistical models in section \ref{A3} are defined with respect to law $P(\mathbf{O})$. 

For simplicity but without loss of generality, we treat $V$ as univariate but it can be easily extended to multvariate case and we use $V$ to generically indicate the covariate set $V_j$ in each DAG where $j=1,2,\dots, 8$.

\subsection{Identification} \label{A2}
\subsubsection{Identification under DAG 1 through 8} \label{A21}
Same as the manuscript, we are interested in either the conditional average treatment effect (CATE) $E[Y^x - Y^{x^{'}}|V]$ or the average treatment effect (ATE) $E[Y^x - Y^{x^{'}}]$ or the  $\beta(V) = E[Y^x - Y^{x^{'}}|V]$. 

We note that if we can identify the marginal counterfactual mean given V, i.e. $E[Y^x|V]$ for any value $X=x$, then CATE is identified. To do so, we consider the following transformation: 

\begin{align*}
        E[Y^x|V] & = E[Y^x|X=x, V] & \text{(C.1.1: exchangeability condition $Y^x \perp X | V$)} \\
        & = E[Y|X=x, V] & \text{(C.1.2: consistency)}\\
        & = E[Y|X=x, Z, V] & \text{(C.1.3: surrogacy assumption: $Y \perp Z| (X,V)$)}\\
        & = \kappa + \beta(V) X + g(V) & \text{(C.1.4: $E[Y|X,V]$ is linear in $X$ modeling assumption)}
\end{align*}

 Before we proceed, we note the following from above results: 
 \begin{align*} \tag{A1} \label{target_x}
 E[Y^x|V] = \kappa + \beta (V)x + g(V),
\end{align*} and this implies:
\begin{align*} 
 \beta(V) (x-x') & = E[Y^x - Y^{x'}|V],
\end{align*} where $\beta(V)$ is the effect of increasing $X$ by one unit and if $\beta(V)$ is identified, then CATE is identified. If we are willing to further assume that $V$ does not modify effect of $X$ on $Y$ then we can replace condition C.1.4. with: 
\begin{align*} \tag{A2} \label{target_x_alternate}
 E[Y^x|V] = E[Y|X=x, V] = \kappa + \beta x + g(V), 
\end{align*} which implies
\begin{align*}  
 \beta (x-x') & = E[Y^x - Y^{x'}|V] = E[Y^x - Y^{x'}]. 
\end{align*}
 
From our previous results under sufficient conditions C.1.1 through C.1.4., we know that $E[Y|X=x,Z,V]= \kappa + \beta (V) x + g(V)$ holds for all realized values $x$ of observed random variable $X$. This allows us to treat $X$ as a random variable and rewrite the equation as $E[Y|X,Z,V] = {{\kappa}}+\beta(V) X + g(V)$. Taking expectation with respect to $X|Z,V$ on both sides gives us the following equation: 

\begin{align*} \tag{A3}  \label{identify_formula_V}
    E[Y|Z,V] = \kappa + \beta (V) E[X|Z,V] + g(V),
\end{align*} where we can estimate $\beta (V)$ through estimating equations and $\beta(V)$ for example can be specified as $\beta (V)= \beta_1 + \beta_2 V$. See section 7 of supplemental material. If we are willing to assume that the (unit) conditional treatment effect $\beta(V)$ is in fact homogeneous across levels of $V$, then the equation \eqref{identify_formula_V} reduces to:
\begin{align*} \tag{A4}  \label{identify_formula_V_noEMM}
    E[Y|Z,V] = \kappa + \beta E[X|Z,V] + g(V),
\end{align*} where unit treatment effect no longer depends on $V$ therefore is also average treatment effect. 

Next we briefly show how equation \eqref{identify_formula_V} is related to the identification formula under RSW regression calibration method. Particularly, the equation \eqref{identify_formula_V} holds for any two realization of $Z$, $z \ne z'$. This gives: $$\kappa + g(V) = E[Y|Z=z,V] - \beta(V) E[X|Z=z,V] = E[Y|Z=z',V] - \beta(V) E[X|Z=z',V].$$ Rearranging this equation we have the identification formula for the conditional average treatment effect parameter $\beta(V)$ under RSW method:
$$\beta (V) = \frac{E[Y|Z=z, V] - E[Y|Z=z^{'},V]}{E[X|Z=z, V] - E[X|Z=z^{'},V]}$$

\subsubsection{Identification under DAG 1, 5, 6, 7 and 8} \label{A22}
Alternatively, we can attempt to identify ATE in some of the DAGs by noticing that under DAGs 1, 5, 6, 7 and 8, the marginal counterfactual outcome $E[Y^x]$ can be re-expressed as: 

\begin{align*}
      E[Y^x]  & = E[Y^x|X=x] & \text{(C.2.1: exchangeability condition $Y^x \perp X$)} \\
        & = E[Y|X=x] & \text{(C.2.2: consistency)}\\
        & = E[Y|X=x, Z] & \text{(C.2.3: surrogacy assumption: $Y \perp Z|X$)}\\
        & = \kappa + \beta x& \text{(C.2.4: $E[Y|X]$ is linear in $X$ modeling assumption)}\\
\end{align*}

We note the following from the above transformation:  
\begin{align*} \tag{A5} \label{target_x}
 E[Y^x] & = \kappa + \beta x,
\end{align*} where $\beta$ is the unit average treatment effect and if $\beta$ is identified then all ATE is identified using the relationship: 
\begin{align*} 
 E[Y^x - Y^{x'}] & = \beta (x-x').
\end{align*}

The resulting equation $E[Y|X=x,Z]= \beta x$ under sufficient conditions C.2.1 through C.2.4 holds for all realized values $x$ of observed random variable $X$. This allows us to treat $X$ as a random variable and rewrite the equation as $E[Y|X,Z] = {{\kappa}}+ \beta X$. Taking expectation with respect to $X|Z$ on both sides gives us the following equation: 

\begin{align*} \tag{A6}  \label{identify_formula_noV}
    E[Y|Z] = \kappa + \beta E[X|Z],
\end{align*} where we can estimate $\beta$ through estimating equations. 

Next we briefly show how equation \eqref{identify_formula_noV} is related to the identification formula under RSW regression calibration method. Particularly, the equation \eqref{identify_formula_noV} holds for any two realization of $Z$, $z \ne z'$. This gives: $$\kappa = E[Y|Z=z] - \beta E[X|Z=z] = E[Y|Z=z'] - \beta E[X|Z=z'].$$ Rearranging this equation we have the identification formula for the (marginal) unit treatment effect parameter $\beta$ under RSW method:
$$\beta = \frac{E[Y|Z=z] - E[Y|Z=z^{'}]}{E[X|Z=z] - E[X|Z=z^{'}]}$$

To avoid confusion, we denote $\beta$ under DAG 1, 5, 6, 7 and 8 as $\beta'$.

\subsection{Statistical Models and CRS Estimators} \label{A3}
Now suppose the quantities such as $E[Y|Z,V]$ and $E[X|Z,V]$ specified in identification formula \eqref{identify_formula_V} and \eqref{identify_formula_noV} can be modeled with several linear models. For example, suppose conditional mean of $X$ given $(Z,V)$ in \eqref{identify_formula_V} can be modeled with linear terms as: 
\begin{align*}  \tag{A7}  \label{EX_ZV}
    X & = \alpha_0 + \alpha_1 Z + \alpha_2 V + \epsilon_X, \text{ which gives}\\
    E[X|Z,V; \boldsymbol{\alpha}] & = \alpha_0 + \alpha_1 Z + \alpha_2 V,
\end{align*} where $\boldsymbol{\alpha} =  (\alpha_0, \alpha_1, \alpha_2)$, $E[\epsilon_X|Z,V]=0$ and  $Cov(\epsilon_X,V)=Cov(\epsilon_X,Z)=0$.

Suppose $E[V|Z]$ is a linear function of $Z$, i.e.:
$$E[V|Z] = \lambda_0 + \lambda_1 Z,$$
then we have the conditional mean of $X$ model given $Z$ in \eqref{identify_formula_noV} as:
\begin{equation} \tag{A8}  \label{EX_Z}
    E[X|Z; \boldsymbol{\alpha^*}] = \alpha_0^* + \alpha_1^* Z,
\end{equation} where $\boldsymbol{\alpha^*} =  (\alpha_0^*, \alpha_1^*)$.

Similarly, the conditional mean of $Y$ given $(Z,V)$ in \eqref{identify_formula_V} can be modeled with linear terms as: 
\begin{align*}  \tag{A9}  \label{Y_EX_ZV_V}
    E[Y|Z,V; \boldsymbol{\alpha, \beta}] & = \beta_0 + \beta_1 E[X|Z,V; \boldsymbol{\alpha}] + \beta_2 V, \text{ or more restrictively}\\
    Y & = \beta_0 + \beta_1 E[X|Z,V; \boldsymbol{\alpha}] + \beta_2 V + \epsilon_Y
\end{align*} where $\boldsymbol{\beta}=(\beta_0, \beta_1, \beta_2)$, $E[\epsilon_Y|E[X|Z,V; \boldsymbol{\alpha}],V]=0$, $\beta (V) = \beta_1$, $g(V) = \beta_2 V$, $Cov(\epsilon_Y,E[X|Z,V])=Cov(\epsilon_Y,Z) = Cov(\epsilon_Y,V)=0$. If $E[V|E[X|Z,V]]$ is a linear function of $E[X|Z,V]$, then \eqref{Y_EX_ZV_V} can be re-expressed as:
\begin{align*}  \tag{A10}  \label{Y_EX_ZV}
    E[Y|Z,V; \boldsymbol{\alpha, \tilde{\beta}}] & = E[Y|E[X|Z,V]; \boldsymbol{\alpha, \tilde{\beta}}] \\
    & = \tilde{\beta}_0 + \tilde{\beta}_1 E[X|Z,V; \boldsymbol{\alpha}], \text{ or more restrictively}\\
    Y & = \tilde{\beta}_0 + \tilde{\beta}_1 E[X|Z,V; \boldsymbol{\alpha}] + \tilde{\epsilon}_Y,
\end{align*} where $\boldsymbol{\tilde{\beta}}=(\tilde{\beta}_0, \tilde{\beta}_1)$, $E[\tilde{\epsilon}_Y|E[X|Z,V]]=0$ and $Cov(\tilde{\epsilon}_Y,E[X|Z,V])=0$.

Last, because of model specification under \eqref{EX_ZV}, \eqref{EX_Z} and \eqref{Y_EX_ZV_V}, we can also model $E[Y|Z,V]$ as:  
\begin{align*}  \tag{A11}  \label{Y_EX_Z_V}
    Y & =  \beta_0^+ + \beta_1^+ E[X|Z; \boldsymbol{\alpha^*}] + \beta_2^+ V + \epsilon_Y^+ , \text{ which gives} \\
    E[Y|Z,V; \boldsymbol{\alpha^*, \beta^+}] & = \beta_0^+ + \beta_1^+ E[X|Z; \boldsymbol{\alpha^*}] + \beta_2^+ V,
\end{align*} where $\boldsymbol{\beta^+}=(\beta_0^+, \beta_1^+, \beta_2^+)$, $E[\epsilon_Y^+|E[X|Z],V]=0$ and $Cov(\epsilon_Y^+, E[X|Z])=Cov(\epsilon_Y^+, V)=0$. 

Finally, because of model specification under \eqref{EX_Z} and \eqref{Y_EX_Z_V}, the conditional mean of $Y$ given $Z$ in \eqref{identify_formula_noV} can be modeled as: 
\begin{align*}  \tag{A12}  \label{Y_EX_Z}
    Y & = \beta_0^* + \beta_1^* E[X|Z; \boldsymbol{\alpha^*}] + \epsilon_Y^*, \text{ which gives}\\
    E[Y|Z; \boldsymbol{\alpha^*, \beta^*}] & = \beta_0^* + \beta_1^* E[X|Z; \boldsymbol{\alpha^*}],
\end{align*} where $\boldsymbol{\beta^*}=(\beta_0^*,\beta_1^*)$, $E[\epsilon_Y^*|E[X|Z]]=0$, and $Cov(\epsilon_Y^*, E[X|Z])=0$.

From \eqref{EX_ZV} to \eqref{Y_EX_Z} we have the following four CRS estimators based on estimating equations and convergence: 

\begin{enumerate}
\item Based on models \eqref{EX_ZV} and \eqref{Y_EX_ZV_V}, we construct the following estimating equations: 
\begin{align*}
        S_1 (\boldsymbol{\alpha}) & = \sum_{i\in V} W_{1i}[X_i - E[X_i|Z_i,V_i;\boldsymbol{\alpha}]] & = & \sum_{i\in V} W_{1i}[X_i - ({\alpha_0} + {\alpha_1} Z_i + {\alpha_2} V_i)], \text{ and} \\
        S_2 (\boldsymbol{\alpha},\boldsymbol{\beta}) & = \sum_{i\in M} W_{2i}[Y_i - E[Y_i|Z_i,V_i; \boldsymbol{\alpha, \beta}] ] & = & \sum_{i\in M} W_{2i}[Y_i - (\beta_0 + \beta_1 E[X_i|Z_i,V_i;\boldsymbol{\alpha}] + \beta_2 V_i)],
\end{align*} where $i$ indexes participants, $V$ and $M$ are the sets of participants for the validation and main  study, respectively. $W_{1i}$ and $W_{2i}$ are weights for the estimating functions following appropriate forms for standard linear regression. By standard estimating equations theory, $\boldsymbol{\hat{\alpha}} \overset{p}{\to} \boldsymbol{{\alpha}},\boldsymbol{\hat{\beta}} \overset{p}{\to} \boldsymbol{\beta}$. In particular, $\hat{\beta}_1 = \hat{\beta}_{_{(OM)}} \overset{p}{\to} \beta_1 = \beta_{_{(OM)}}$, where $\beta_{_{(OM)}}$ is the CRS estimator where covariate $V$ is included in both MEM and outcome model, thus the subscript (OM). 

\item Based on models \eqref{EX_Z} and \eqref{Y_EX_Z}, we construct the following estimating equations: 
\begin{align*}
        S_1 (\boldsymbol{\alpha^*}) & = \sum_{i\in V} W_{1i}[X_i - E[X_i|Z_i; \boldsymbol{\alpha^*}]] & = & \sum_{i\in V} W_{1i}[X_i - ({\alpha_0}^* + {\alpha_1}^* Z_i)], \text{ and}  \\
        S_2 (\boldsymbol{\alpha^*},\boldsymbol{\beta^*}) & = \sum_{i\in M} W_{2i}[Y_i - E[Y_i|Z_i; \boldsymbol{\alpha^*, \beta^*}] ] & = & \sum_{i\in M} W_{2i}[Y_i - (\beta_0^* + \beta_1^* E[X_i|Z_i;\boldsymbol{\alpha^*}])].
\end{align*} 
Similar to (a), by standard estimating equations theory, $\boldsymbol{\hat{\alpha}^*} \overset{p}{\to} \boldsymbol{{\alpha}^*},\boldsymbol{\hat{\beta}^*} \overset{p}{\to} \boldsymbol{\beta^*}$. In particular, $\hat{\beta}_1^* = \hat{\beta}_{_{(--)}} \overset{p}{\to} \beta_1^* = \beta_{_{(--)}}$, where $\beta_{_{(--)}}$ is the CRS estimator where covariate $V$ is included in neither MEM nor outcome model, thus the subscript (-{ }-). 

\item Based on models \eqref{EX_ZV} and \eqref{Y_EX_ZV}, we construct the following estimating equations: 
\begin{align*}
        S_1 (\boldsymbol{\alpha}) & = \sum_{i\in V} W_{1i}[X_i - E[X_i|Z_i,V_i;\boldsymbol{\alpha}]] & = & \sum_{i\in V} W_{1i}[X_i - ({\alpha_0} + {\alpha_1} Z_i + {\alpha_2} V_i)], \text{ and} \\
        S_2 (\boldsymbol{\alpha},\boldsymbol{\tilde{\beta}}) & = \sum_{i\in M} W_{2i}[Y_i - E[Y_i|Z_i, V_i; \boldsymbol{\alpha, \tilde{\beta}}] ] & = & \sum_{i\in M} W_{2i}[Y_i - (\tilde{\beta}_0 + \tilde{\beta}_1 E[X_i|Z_i,V_i;\boldsymbol{\alpha}])].
\end{align*} 
Similar to (a), by standard estimating equations theory, $\boldsymbol{\hat{\alpha}} \overset{p}{\to} \boldsymbol{{\alpha}},\boldsymbol{\hat{\tilde{\beta}}} \overset{p}{\to} \boldsymbol{\tilde{\beta}}$. In particular, $\hat{\tilde{\beta}}_1 = \hat{\beta}_{_{(-M)}} \overset{p}{\to} \tilde{\beta}_1 = \beta_{_{(-M)}}$, where $\beta_{_{(-M)}}$ is the CRS estimator where covariate $V$ is included in the MEM but not the outcome model, thus the subscript (-M). 

\item Based on models \eqref{EX_Z} and \eqref{Y_EX_Z_V}, we construct the following estimating equations: 
\begin{align*}
        S_1 (\boldsymbol{\alpha^*}) & = \sum_{i\in V} W_{1i}[X_i - E[X_i|Z_i; \boldsymbol{\alpha^*}]] & = & \sum_{i\in V} W_{1i}[X_i - ({\alpha_0}^* + {\alpha_1}^* Z_i)], \text{ and}  \\
        S_2 (\boldsymbol{\alpha^*},\boldsymbol{{\beta}^+}) & = \sum_{i\in M} W_{2i}[Y_i - E[Y_i|Z_i, V_i; \boldsymbol{\alpha^*, \beta^+}] ] & = & \sum_{i\in M} W_{2i}[Y_i - (\beta_0^+ + \beta_1^+ E[X_i|Z_i;\boldsymbol{\alpha^*}] + \beta_2^+ V_i)].
\end{align*} 
Similar to (a), by standard estimating equations theory, $\boldsymbol{\hat{\alpha}^*} \overset{p}{\to} \boldsymbol{{\alpha}^*},\boldsymbol{\hat{{\beta}^+}} \overset{p}{\to} \boldsymbol{\beta^+}$. In particular, $\hat{\beta}_1^+ = \hat{\beta}_{_{(O-)}} \overset{p}{\to} \beta_1^+ = \beta_{_{(O-)}}$, where $\beta_{_{(O-)}}$ is the CRS estimator where covariate $V$ is included in the outcome model but not the MEM, thus the subscript (O-).
\end{enumerate} 

\subsection{Validity of CRS estimators}
\subsubsection{Estimator 1: $\hat{\beta}_{_{(OM)}}$}

$\hat{\beta}_1=\hat{\beta}_{_{(OM)}}$ is obtained by jointly solving unbiased estimating functions:

\begin{align}
        S_1 (\boldsymbol{\alpha}) & = \sum_{i\in V} W_{1i}[X_i - ({\alpha_0} + {\alpha_1} Z_i + {\alpha_2} V_i)], \text{ and} \\
        S_2 (\boldsymbol{\alpha},\boldsymbol{\beta}) & = \sum_{i\in M} W_{2i}[Y_i - (\beta_0 + \beta_1 E[X_i|Z_i,V_i;\boldsymbol{\alpha}] + \beta_2 V_i)].
\end{align}

If the models \eqref{EX_ZV} and \eqref{Y_EX_ZV_V} are in needed the true statistical models for $E[X|Z,V]$ and $E[Y|Z,V]$ in equation \eqref{identify_formula_V}, then under DAG 1 through 8, $\hat{\beta}_1 = \hat{\beta}_{_{(OM)}} \overset{p}{\to} \beta_1 = \beta_{_{(OM)}} = \beta (V) = \beta$, where {{we assume in the true causal model}} the unit treatment effect in \eqref{identify_formula_V} no longer depends on $V$.   

From model specification of \eqref{EX_ZV}, {{taking the covariance between $V$ on both sides of the equations}} we have: 

\begin{align*} \tag{A13} \label{A13}
Cov(X,V)& =\alpha_1 Cov(V,Z) + \alpha_2 Var(V),
\end{align*} and from \eqref{Y_EX_ZV_V}, {{taking covariance with $E(X|Z,V)$ on both sides of the equation, and if we further replace $E[X|Z,V]$ with model \eqref{EX_ZV}}}, $\beta_1$ can be expressed as the following : 
\begin{align*} \tag{A14} \label{A14}
\beta_1 & = \frac{Cov(Y, E(X|Z,V))}{Var(E(X|Z,V))} - \beta_2 \frac{Cov(V, E(X|Z,V))}{Var(E(X|Z,V))} \\
 & = \frac{[\alpha_1 Cov(Y,Z) + \alpha_2 Cov(Y,V)] - \beta_2 [\alpha_1 Cov(V,Z) + \alpha_2 Var(V)]}{\alpha_1^2 Var(Z) + \alpha_2^2 Var(V) + 2 \alpha_1 \alpha_2 Cov(Z,V)}.
\end{align*}

These relationships will be useful later.

\subsubsection{Estimator 2: $\hat{\beta}_{_{(- -)}}$}

$\hat{\beta}_1^*=\hat{\beta}_{_{(-{}-)}}$ is obtained by jointly solving the estimating functions:

\begin{align}
        S_1 (\boldsymbol{\alpha^*}) & = \sum_{i\in V} W_{1i}[X_i - ({\alpha_0}^* + {\alpha_1}^* Z_i)], \text{ and}  \\
        S_2 (\boldsymbol{\alpha^*},\boldsymbol{\beta^*}) & = \sum_{i\in M} W_{2i}[Y_i - (\beta_0^* + \beta_1^* E[X_i|Z_i;\boldsymbol{\alpha^*}])].
\end{align}

If the models \eqref{EX_Z} and \eqref{Y_EX_Z} are indeed the true statistical models for $E[X|Z]$ and $E[Y|Z]$ in equation \eqref{identify_formula_noV}, then under DAG 1, 5, 6, 7 and 8, $\hat{\beta}_1^* = \hat{\beta}_{_{(-{}-)}} \overset{p}{\to} \beta_1^* = \beta_{_{(-{}-)}} = \beta'$, the unit treatment effect defined in \eqref{identify_formula_noV}.  

From model \eqref{Y_EX_Z}, we can also express $\beta_1^*$ as the following where we replace $E[X|Z]$ with model \eqref{EX_Z}:
\begin{align*} \tag{A15} \label{A15}
\beta_1^* & = \frac{Cov(Y, E[X|Z])}{Var(E[X|Z])}\\
 & = \frac{Cov(Y, Z)}{\alpha_1^*Var(Z)}. 
\end{align*}

For DAG 2, 3 and 4, let us now write out a more restrictive version of the model of \eqref{EX_Z}, i.e. $X= \alpha_0^* + \alpha_1^* Z + \epsilon_X^*$, where in contrast to \eqref{EX_Z} now we require $E(\epsilon_X|Z)=0$ and $Cov(\epsilon_X, Z)=0$. From this we have: 

\begin{align*} \tag{A16} \label{A16}
\alpha_1^*=\frac{Cov(X,Z)}{Var(Z)}. 
\end{align*}

To see if $\beta_1^* = \beta_1 = \beta$, we investigate whether $\textstyle{\frac{Cov(Y, Z)}{\alpha_1^*Var(Z)}=\frac{[\alpha_1 Cov(Y,Z) + \alpha_2 Cov(Y,V)] - \beta_2 [\alpha_1 Cov(V,Z) + \alpha_2 Var(V)]}{\alpha_1^2 Var(Z) + \alpha_2^2 Var(V) + 2 \alpha_1 \alpha_2 Cov(Z,V)}}$ by result \eqref{A14} and \eqref{A15}. Using the results \eqref{A13} and \eqref{A16}, this equation can be reformulated as:  

\begin{align*} \tag{A17} \label{A17}
\frac{Cov(Y, Z)}{Cov(X,Z)}=\frac{\alpha_1 Cov(Y,Z) + \alpha_2 Cov(Y,V) - \beta_2 Cov(X,V)}{\alpha_1^2 Var(Z) + \alpha_2^2 Var(V) + 2 \alpha_1 \alpha_2 Cov(Z,V)}.
\end{align*}

Now we prove that the above equation is generally not true under DAG 2,3 and 4 using counterexample. We assume the following data generating process:
    \begin{align*}
        V & \sim N(0,1) \\
        X & = bV + e_x, e_x \sim N(0,1), e_x \perp V \\
        Z & = X + a V + e_z, e_z \sim N(0,1), e_z \perp (X,V) \\
        Y & = X + V + e_y, e_y \sim N(0,1), e_y \perp (X,V,Z)
    \end{align*}
    
    Note in DAG 2, $a \ne 0, b = 0$ and in DAG 3, $a =0, b \ne 0$ We can immediately find the joint multivariate distribution of $(X,Z,V)$ and $(Y,X,V)$ given the above data generating process as:
    
    \[\begin{pmatrix}
    X \\
    Z \\
    V
    \end{pmatrix}\sim N\left(\begin{pmatrix}
    0 \\
    0 \\
    0 
    \end{pmatrix},\begin{pmatrix}
    b^2 + 1 & b^2 + ab + 1 & b \\
    b^2 + ab + 1 & a^2 + b^2 + 2ab + 2 & a + b \\
    b & a + b & 1
    \end{pmatrix}\right).
    \]
    
      \[\begin{pmatrix}
    Y \\
    X \\
    V
    \end{pmatrix}\sim N\left(\begin{pmatrix}
    0 \\
    0 \\
    0 
    \end{pmatrix},\begin{pmatrix}
    Var(Y) &  b^2 + b + 1 & b + 1 \\
    b^2 + b + 1 & b^2 + 1 & b \\
    b + 1 & b & 1
    \end{pmatrix}\right).
    \]
    
    We also have: 
    $$Cov(Y,Z) = b^2 + ab + a + b + 1.$$

    And we can obtain: 
    
    \begin{align*}
        E[{{X}}|Z,V] & =  \frac{1}{2} Z + \frac{b-a}{2}  V , \\
        E[Y|X,V] & = X + V.
    \end{align*}

    {{Therefore, we know $\alpha_1=\frac{1}{2}$, $\alpha_2=\frac{b-a}{2}$, $\beta_2=1$, and therefore   }}Equation \eqref{A17} becomes:

    \begin{align*} \tag{A18} \label{A18}
        \frac{b^2 + ab + a + b + 1}{b^2 + ab + 1} & = & \frac{\frac{1}{2} (b^2 + ab + a + b + 1) + \frac{1}{2}(b-a)(b+ 1) -b}{ \frac{1}{4}(a^2 + b^2 + 2ab + 2) + \frac{1}{4} (b-a)^2 + {{2}}*\frac{1}{2}*\frac{1}{2}(b-a)({{a+b}})},         
    \end{align*}
    
%or equivalently: 
%\begin{align*} \tag{A18} \label{A18}
%b^4 + ab^3 + 2ab^2 + 2b^2 + a + b = 0.
%\end{align*}

\begin{itemize}
    \item For DAG 2, $a\ne 0, b=0$, \eqref{A18} reduces to $a=0$, which contradicts DAG 2's $a \ne 0$. Thus \eqref{A17} is not true under DAG 2. 
    \item For DAG 3, $a = 0, b \ne 0$, \eqref{A18} reduces to {{$b=0$, which contradicts DAG 3's $b \ne 0$. Thus \eqref{A17} is not true under DAG 3. }} %$b(b^3 + b^2 + 2b + 1) = 0$. Since $b\ne 0$, \eqref{A20} is not generally true under DAG 3. 
    \item For DAG 4, $a \ne 0, b \ne 0$, \eqref{A18} is not generally true under DAG 4. {{For example, when $a=b=1$, LHS of \eqref{A18} is 5/3, while RHS of \eqref{A18} is 1.}} 
\end{itemize}
Thus we proved that under the submodel of \eqref{EX_Z} (that gives result \eqref{A15}), the equation \eqref{A18} does not hold, therefore the equation would not hold under model \eqref{EX_Z} and \eqref{Y_EX_Z} either. In summary, $\beta_{(--)}$ is only valid for DAG 1, 5, 6, 7 and 8.

\subsubsection{Estimator 3: $\hat{\beta}_{_{(- M)}}$}
$\hat{\tilde{\beta}}_1=\hat{\beta}_{_{(-{}M)}}$ is obtained by jointly solving estimating functions:
\begin{align}
        S_1 (\boldsymbol{\alpha}) & = \sum_{i\in V} W_{1i}[X_i - ({\alpha_0} + {\alpha_1} Z_i + {\alpha_2} V_i)], \text{ and} \\
        S_2 (\boldsymbol{\alpha},\boldsymbol{\tilde{\beta}}) & =  \sum_{i\in M} W_{2i}[Y_i - (\tilde{\beta}_0 + \tilde{\beta}_1 E[X_i|Z_i,V_i;\boldsymbol{\alpha}])].
\end{align}
From model \eqref{Y_EX_ZV}, we can express $\tilde{\beta}_1$ as the following where we replace $E[X|Z,V]$ with  model \eqref{EX_ZV}: 
\begin{align*} \tag{A19} \label{A19}
\tilde{\beta}_1 & = \frac{Cov(Y, E(X|Z,V))}{Var(E(X|Z,V))} \\
& = \frac{\alpha_1 Cov(Y,Z) + \alpha_2 Cov(Y,V)}{Var(E(X|Z,V))}. 
\end{align*}

Therefore, to prove $\hat{\beta}_{_{(-M)}}$ is consistent for $\beta$, we would need to prove $\tilde{\beta}_1 = \beta$ or equivalently by results \eqref{A14} and \eqref{A19}: $\textstyle{\frac{\alpha_1 Cov(Y,Z) + \alpha_2 Cov(Y,V)}{Var(E(X|Z,V))}=\frac{[\alpha_1 Cov(Y,Z) + \alpha_2 Cov(Y,V)] - \beta_2 [\alpha_1 Cov(V,Z) + \alpha_2 Var(V)]}{Var(E(X|Z,V))}}$. Rearranging the terms and using result \eqref{A13}, we have: 

\begin{align*}  \tag{A20} \label{A20}
    \beta_2 [\alpha_1 Cov(V,Z) + \alpha_2 Var(V)] = \beta_2 Cov(X,V) = 0.
\end{align*}

Equation \eqref{A20} is true if and only if (i) $Cov(X,V)=0$ or (ii) $\beta_2 = 0$ (i.e. $Cov(Y,V|X)=0)$. For each DAG under consideration: 

\begin{itemize}
    \item DAG 1 and 2 satisfy condition (i). 
    \item DAG 5, 6, 7 and 8 satisfy condition (ii). 
    \item DAG 3 and 4 do not satisfy condition (i) or (ii).  
\end{itemize}

In summary, $\beta_{(M-)}$ is only valid for DAG 1, 2, 5, 6, 7 and 8.

\subsubsection{Estimator 4: $\hat{\beta}_{_{(O -)}}$}

$\hat{\beta}_1^+ = \hat{\beta}_{_{(O-)}}$ is obtained by jointly solving estimating functions:

\begin{align}
    S_1 (\boldsymbol{\alpha^*}) & = \sum_{i\in V} W_{1i}[X_i - ({\alpha_0^*} + {\alpha_1^*} Z_i)] \\
    S_2 (\boldsymbol{\alpha^*}, \boldsymbol{\beta^+}) & = \sum_{i\in M} W_{2i}[Y_i - ({\beta_0^+} + {\beta_1^+}E[X_i|Z_i;\boldsymbol{\alpha^*}] + \beta_2^+ V_i)]
\end{align}

From model \eqref{Y_EX_Z_V}, we can express $\beta_1^+$ as the following where we replace $E[X|Z]$ with model \eqref{EX_Z}: 
\begin{align*} \tag{A21} \label{A21}
{\beta_1^+} & = \frac{Cov(Y, E[X|Z]) - \beta_2^+ Cov({{V}},E[X|Z])}{Var(E[X|Z])} \\
& = \frac{Cov(Y, E[X|Z])Var(V) - Cov(Y,V) Cov(E[X|Z],V)}{Var(E[X|Z])Var(V) - Cov(E[X|Z],V)^2} \\
& = \frac{Cov(Y,Z)Var(V) - Cov(Y,V)Cov(Z,V)}{\alpha_1^* Var(Z)Var(V) - \alpha_1^* Cov(Z,V)^2},
\end{align*} where $\beta_2^+ = \frac{Cov(Y,V) - \beta_1^+ Cov(E[X|Z],V)}{Var(V)}$.

To prove $\hat{\beta}_{_{(O-)}}$ is consistent for $\beta$, we would need to prove ${\beta_1^+} = \beta$ or equivalently by results \eqref{A14} and \eqref{A21}: $\textstyle{\frac{Cov(Y,Z)Var(V) - Cov(Y,V)Cov(Z,V)}{\alpha_1^* Var(Z)Var(V) - \alpha_1^* Cov(Z,V)^2} =\frac{[\alpha_1 Cov(Y,Z) + \alpha_2 Cov(Y,V)] - \beta_2 [\alpha_1 Cov(V,Z) + \alpha_2 Var(V)]}{\alpha_1^2 Var(Z) + \alpha_2^2 Var(V) + 2 \alpha_1 \alpha_2 Cov(Z,V)}}$. Rearranging the terms and by result \eqref{A13} we have: 
\begin{align*}  \tag{A22} \label{A22}
    & \frac{Cov(Y,Z)Var(V) - Cov(Y,V)Cov(Z,V)}{Cov(X,Z) Var(V) - Cov(X,Z)Cov(Z,V)^2/Var(Z)} = & \frac{[\alpha_1 Cov(Y,Z) + \alpha_2 Cov(Y,V)] - \beta_2 Cov(X,V)}{\alpha_1^2 Var(Z) + \alpha_2^2 Var(V) + 2 \alpha_1 \alpha_2 Cov(Z,V)}
\end{align*}

For DAG 1 and 5, $Cov(Z,V)=Cov(X,V)=0$, \eqref{A22} can be simplified as $\frac{Cov(Y,Z)}{\alpha_1^* Var(Z)} = \frac{\alpha_1 Cov(Y,Z)+ \alpha_2 Cov(Y,V)}{\alpha_1^2 Var(Z) + \alpha_2^2 Var(V)}$. Also under DAG 1 and 5, $X \perp V|Z$ therefore  $E[X|Z] = E[X|Z,V]$ and thus under model \eqref{EX_ZV}  and \eqref{EX_Z}, we have $\alpha_2=0$ and $\alpha_1^* = \alpha_1$. Thus equation \eqref{A22} holds under DAG 1 and 5. 
Next we use counterexample to show that for DAG 2, 3, 4, 6, 7 and 8, $\beta_1^+ = \beta$ is generally not true. We assume the following data generating process: 

    \begin{align*}
        V & \sim N(0,1) \\
        X & = bV + e_x, e_x \sim N(0,1), e_x \perp V \\
        Z & = X + a V + e_z, e_z \sim N(0,1), e_z \perp (X,V) \\
        Y & = X + c V + e_y, e_y \sim N(0,1), e_y \perp (X,V,Z)
    \end{align*}
    
We can immediately find the joint multivariate distribution of $(X,Z,V)$ and $(Y,X,V)$ given the above data generating process as:
    
    \[\begin{pmatrix}
    X \\
    Z \\
    V
    \end{pmatrix}\sim N\left(\begin{pmatrix}
    0 \\
    0 \\
    0 
    \end{pmatrix},\begin{pmatrix}
    b^2 + 1 & b^2 + ab + 1 & b \\
    b^2 + ab + 1 & a^2 + b^2 + 2ab + 2 & a + b \\
    b & a + b & 1
    \end{pmatrix}\right).
    \]

      \[\begin{pmatrix}
    Y \\
    X \\
    V
    \end{pmatrix}\sim N\left(\begin{pmatrix}
    0 \\
    0 \\
    0 
    \end{pmatrix},\begin{pmatrix}
    Var(Y) & b^2 + bc + 1 & b + {{c}} \\
    b^2 + bc + 1 & b^2 + 1 & b \\
    b + {{c}} & b & 1
    \end{pmatrix}\right).
    \]
    We can also obtain $Cov(Y,Z)= b^2 + ab + ac + bc + 1$. It follows that:
    
    \begin{align*}
        E[X|Z,V] & = \frac{1}{2}Z + \frac{b-a}{2} V \\
        E[Y|X,V] & = {{X+cV}}\\
    \end{align*}
  
Thus under the specified data generating process, $\alpha_1 = \frac{1}{2}, \alpha_2 = \frac{b-a}{2}, Var(Z) = a^2 + b^2 + 2ab + 2, Var(V)=1$ and $Cov(X,V)=b, Cov(Y,V)=b+{{c}}, Cov(X,Z)=b^2 + ab + 1, Cov(V,Z)=a+b$. 

Equation \eqref{A22} can be rewritten as:
\begin{align*} \tag{A23} \label{A23}
 & \frac{(b^2 + ab + ac + bc + 1) - (b+{{c}})(a+b)}{(b^2 + ab + 1)[1 - (a+b)^2/(a^2 + 2ab + b^2 + 2)]} \\ = & \frac{\frac{1}{2}(b^2 + ab + ac + bc + 1) + \frac{1}{2}(b-a)(b+{{c}}) - {{c}}b}{\frac{1}{4}(a^2 + b^2 + 2ab + 2) + \frac{1}{4}(b-a)^2 + 2*\frac{1}{2}*\frac{1}{2} (b-a) (b+a)}
\end{align*}
%Let $f(a,b,c)$ be a function of $a, b, c$ then the equation \eqref{A22} can be rearranged to: \begin{align*} \tag{A23} \label{A23}
%f(a,b,c) & = (c-1)(2b^5 + b^3 - 2a^3 b^2 - 6 a^2 b^3 - a^3 - a^2b - 3ab^2) + a^2 - b^2 + 2ab^4c + ab^2 + 2a^2b^2 \\
%& =0
%\end{align*}

\begin{itemize}
    \item Under DAG 2, $a\ne 0, b=0, c\ne0$. If we let $c=1$ then {{for \eqref{A23} to hold, we have $a=0$, which contradicts the condition of $a\ne 0$ under DAG2. }}%$f(a,b,c)=a^2 =0$.
    Thus \eqref{A22} do not hold in general. 
    \item Under DAG 3, $a=0,b \ne 0, c \ne 0$. If we let $c=1$ {{for \eqref{A23} to hold, we have $b=0$, which contradicts the condition of $b\ne 0$ under DAG3. }} Thus \eqref{A22} does not hold in general. 
    \item Under DAG 4, $a \ne 0,b \ne 0, c \ne 0$. %If we let $c=1$ then $f(a,b,c)=a^2 -b^2 + 2ab^4 + ab^2 + 2a^2b^2 =0$ for any value of $a\ne 0, b\ne 0$. Thus \eqref{A22} and \eqref{A23} do not hold in general. 
     {{For instance, when we take $a=1, b=2, c=1$, the LHS of \eqref{A23} is not equal to RHS of \eqref{A23}.}}

    \item Under DAG 6, $a\ne 0, b=0, c=0$. {{For \eqref{A23} to hold, $a=0$, which contradicts the condition of $a\ne 0$ under DAG6.}} Thus \eqref{A22} does not hold in general. 
    \item Under DAG 7, $a=0,b \ne 0, c= 0$. {{For \eqref{A23} to hold, $b=0$, which contradicts the condition of $b\ne 0$ under DAG7.}} Thus \eqref{A22} does not hold in general. 
    \item Under DAG 8, $a \ne 0,b \ne 0, c = 0$, then \eqref{A22} and \eqref{A23} do not hold in general. {{For instance, when we take $a=1, b=2$, the LHS of \eqref{A23} is not equal to RHS of \eqref{A23}.}}
\end{itemize}

In summary, $\beta_{_{(O-)}}$ is only valid for DAG 1 and 5.

\newpage

\section{Comparison of Efficiency for CRS Estimators Under Linear Model}

\subsection{Set-up}

We use the same set up as in section 1, including the modeling assumptions specified in section 1.3 and a main study/external validation study design. In this section, we compare the efficiency of the valid CRS estimators under each DAG analytically, with different conditional relationships implied by each DAG. Without loss of generality, we assume covariate V is a scalar but the result extends to p-dimensional $V$ where $p>1$. We denote the following:

\begin{itemize}
    \item $n, m$ denotes the sample size of validation study and main study respectively. 
    \item {{$\mathcal{M}, \mathcal{V}$ denote the collection of participants in the main and validation study, respectively. }}
    \item $\eta_{_M}$ is short for $\eta_i (\boldsymbol{\alpha}) = \alpha_0 + \alpha_1 Z_i + \alpha_2 V_i$ and $\eta_{_-}$ is short for $\eta_i (\boldsymbol{\alpha^*}) = \alpha_0^* + \alpha_1^* Z_i$. 
    \item $P_n, P_m$ denotes the empirical mean within validation study and main study sample, i.e. $\frac{1}{n} \sum_{i \in \mathcal{V}}, \frac{1}{m} \sum_{i \in \mathcal{M}}$, respectively. For example, $P_m \eta_{_M} V = \frac{1}{m} \sum_{i \in \mathcal{M}} \eta_i(\boldsymbol{\alpha}) V_i$. 
\end{itemize}

%We can then derive quantities below following definitions in section 1 of the supplemental material, plus some additional mild assumptions: 
We give the formula for the sandwich variance of the estimated regression parameters.

\begin{enumerate}
\item For $\beta_{_{(OM)}}$, let 

    \begin{align*}
        g_{_{(OM),i}}(\boldsymbol{\alpha,\beta}) & = \begin{pmatrix}  S_{1,i} (\boldsymbol{\alpha}) \\ S_{2,i} (\boldsymbol{\alpha}, \boldsymbol{{\beta}}) \end{pmatrix}  
        & = \begin{pmatrix} 
            \begin{pmatrix} 1 \\ Z_i \\ V_i \end{pmatrix} & (X_i - \eta_M){{ I(i\in \mathcal{V} )}} \\
            \begin{pmatrix} 1 \\ \eta_M \\ V_i \end{pmatrix} & (Y_i - {\beta_0} - {\beta_1} \eta_M - {\beta_2} V_i){{ I(i\in\mathcal{M} )}}
        \end{pmatrix} 
        & = 0, 
    \end{align*}
{{Let $A_{_{OM}}(\boldsymbol{\alpha}, \boldsymbol{\beta})= E \left[ \frac{\partial{g_{_{(OM), i}}(\boldsymbol{\alpha,{\beta}})}}{\partial{ (\boldsymbol{\alpha,{\beta}})}}\right]$, and $B_{_{OM}}(\boldsymbol{\alpha}, \boldsymbol{\beta})=Cov( g_{_{OM, i}}(\boldsymbol{\alpha,{\beta}}))$ }}, and the two matrices can be estimated as follows:

    \begin{align*} \tag{B1} \label{B1}
        {{\widehat{A}_{_{OM}}(\boldsymbol\alpha,\boldsymbol\beta)}} %& = P_{(\cdot)}\Big\{ E \left[ \frac{\partial{g_{_{(OM), i}}(\boldsymbol{\alpha,{\beta}})}}{\partial{ (\boldsymbol{\alpha,{\beta}})}}\right] \Big\}\\ 
        & = - \begin{pmatrix}  
            P_n \begin{pmatrix}
            1 & Z & V \\
            Z & Z^2 & ZV \\
            V & ZV & V^2
            \end{pmatrix} 
            & 
            \begin{pmatrix}
            0 & 0 & 0 \\
            0 & 0 & 0 \\
            0 & 0 & 0
            \end{pmatrix} \\
            {\beta_1} P_m \begin{pmatrix}
            1 & Z & V \\
            \eta_{_M} & \eta_{_M} Z & \eta_{_M} V \\
            V & ZV & V^2
            \end{pmatrix} 
            & 
            P_m \begin{pmatrix}
            1 & \eta_{_M} & V \\
            \eta_{_M} & \eta_{_M}^2 & \eta_{_M} V \\
            V & \eta_{_M}V & V^2
            \end{pmatrix} 
        \end{pmatrix} \\
        & = \begin{pmatrix}
            \widehat{A}_{_{OM,11}} & \widehat{A}_{_{OM,12}} \\
            \widehat{A}_{_{OM,21}} & \widehat{A}_{_{OM,22}} 
        \end{pmatrix} \\
    \end{align*}
    
    where $\widehat{A}_{_{OM,11}}, \widehat{A}_{_{OM,12}}, \widehat{A}_{_{OM,21}}, \widehat{A}_{_{OM,22}}$ are the four block matrices, with for example:
    $$A_{_{OM,11}}= - P_n \begin{pmatrix}
            1 & Z & V \\
            Z & Z^2 & ZV \\
            V & ZV & V^2
            \end{pmatrix},$$
    
    and 
    
    \begin{align*} \tag{B2} \label{B2}
        {{\widehat{B}_{_{OM}}(\boldsymbol\alpha,\boldsymbol\beta)}} %& = P_{(\cdot)} Cov( g_{_{OM, i}}(\boldsymbol{\alpha,{\beta}})) \\
        %& = \mathbf{E} \begin{pmatrix}  
        %\begin{pmatrix}
        %S_{1,i}(\boldsymbol{\alpha})S^T_{1,i}(\boldsymbol{\alpha}) & S_{1,i}(\boldsymbol{\alpha}) S^T_{2,i}(\boldsymbol{\alpha}, \boldsymbol{{\beta}}) \\
        %S_{1,i}(\boldsymbol{\alpha}) S_{2,i}(\boldsymbol{\alpha}, \boldsymbol{{\beta}}) & S_{2,i}(\boldsymbol{\alpha}, \boldsymbol{{\beta}})^2\\       
        %\end{pmatrix} 
        %\end{pmatrix} \\
        & = \begin{pmatrix}
                n^{-1} \widehat\sigma_{_{X,M}}^2 P_n \begin{pmatrix}
                    1 & Z & V \\
                    Z & Z^2 & ZV \\
                    V & ZV & V^2
                \end{pmatrix}   
                & 
                \begin{pmatrix}
                    0 & 0 & 0 \\
                    0 & 0 & 0 \\
                    0 & 0 & 0
                \end{pmatrix}  \\
                \begin{pmatrix}
                    0 & 0 & 0 \\
                    0 & 0 & 0 \\
                    0 & 0 & 0
                \end{pmatrix} &
                m^{-1} \widehat\sigma_{_{Y,(OM)}}^2 P_m \begin{pmatrix}
                    1 & \eta_{_M} & V \\
                    \eta_{_M} & \eta_{_M}^2 & \eta_{_M}V \\
                    V & \eta_{_M}V & V^2
                \end{pmatrix}   
            \end{pmatrix} \\
        & = \begin{pmatrix}
            \widehat{B}_{_{OM,11}} & \widehat{B}_{_{OM,12}} \\
            \widehat{B}_{_{OM,21}} & \widehat{B}_{_{OM,22}} 
        \end{pmatrix}, \\
    \end{align*}
    
where we assume as in section 1 that (1) samples within validation and main study are independently sampled such that $E[S_1(\boldsymbol{\alpha}) S_2(\boldsymbol{\alpha}, \boldsymbol{{\beta}})] = 0$ and additionally that (2) variance are constant thus $\sigma_{_{X,M}}^2 = Var(X_i - \eta_{_M})$ and $\sigma_{_{Y,(OM)}}^2 = Var(Y_i - {\beta_0} - {\beta_1} \eta_{_M} - {\beta_2} V_i)$ are the true variance of the regression residuals that can be estimated from validation study and main study samples, respectively. 
    
Following the sandwich variance formula based on estimating equations (Carroll et al., 2004) where for us $B_{_{OM,21}}=\mathbf{0}$: $$Var(\boldsymbol{\hat{{\beta}}}) = \widehat{A}_{_{OM,22}}^{-1}[\widehat{B}_{_{OM,22}} - \widehat{A}_{_{OM,21}}\widehat{A}_{_{OM,11}}^{-1}\widehat{B}_{_{OM,11}}(\widehat{A}_{_{OM,11}}^{-1})^t \widehat{A}_{_{OM,21}}^{t}](\widehat{A}_{_{OM,22}}^{-1})^t $$.

{{To obtain the asymptotic variance of $\widehat{\beta}_1$, we replace the }}
block matrices with those defined in \eqref{B1} and \eqref{B2} plus further simplification we have under DAG 1 thorugh 8: 
\begin{align*}
& {{m}}Var(\hat{\beta}_1)\\
= &  \sigma_{_{Y,(OM)}}^2 \hat{C}_{A_{(OM,22)}} [P_m V^2 - (P_m V)^2]  +  {{\frac{m}{n}}}\sigma_{_{X,M}}^2 \beta_1^2 \hat{C}_{A_{(OM,11)}} \hat{C}_{A_{(OM,22)}}^2  \hat{K}_{_{OM}}^2 [P_n V^2 - (P_n V)^2]\\
= &  \sigma_{_{Y,(OM)}}^2 \hat{C}_{A_{(OM,22)}}  \widehat{Var}_m (V)  + {{\frac{m}{n}}}\sigma_{_{X,M}}^2 \beta_1^2  \hat{C}_{A_{(OM,11)}} \hat{C}_{A_{(OM,22)}}^2  \hat{K}_{_{OM}}^2 {\widehat{Var}_n}(V)\\
\overset{p}\to & \sigma_{_{Y,(OM)}}^2 C_{A_{(OM,22)}}  {Var}_m (V)  +{{\frac{1}{\rho}}} \sigma_{_{X,M}}^2 \beta^2 C_{A_{(OM,11)}} C_{A_{(OM,22)}}^2  K_{_{OM}}^2 {{Var}_n}(V)   ,
\end{align*} {{where $\rho=\lim_{n,m\to\infty}\frac{n}{m}$,}} and by Slutsky Theorem and Continuous Mapping Theorem:

\begin{itemize}
    \item $\hat{C}_{A_{(OM,11)}} = [P_n Z^2 P_n V^2 - (P_n ZV)^2 - P_n Z(P_n Z P_n V^2 - P_n V P_n ZV) + P_n V (P_n Z P_n ZV - P_n Z^2 P_n V)]^{-1} = (\widehat{Var}_n (Z) \widehat{Var}_n (V) - \widehat{Cov}_n (Z,V)^2)^{-1} \overset{p}\to C_{A_{(OM,11)}} = ({Var}_n (Z) {Var}_n (V) - {Cov}_n (Z,V)^2)^{-1}$,
    \item $\hat{C}_{A_{(OM,22)}} = [P_m \eta_{_M}^2 P_m V^2 - (P_m \eta_{_M} V)^2 - P_m \eta_{_M}(P_m \eta_{_M} P_m V^2 - P_m \eta_{_M}V P_m V) + P_m V(P_m \eta_{_M} P_m \eta_{_M}V - P_m \eta_{_M}^2 P_m V)]^{-1} = (\widehat{Var}_m (\eta_{_M}) \widehat{Var}_m (V) - \widehat{Cov}_m (\eta_{_M},V)^2)^{-1} \overset{p}\to C_{A_{(OM,22)}} = ({Var}_m (\eta_{_M}) {Var}_m (V) - {Cov}_m (\eta_{_M},V)^2)^{-1}$,  
    \item $\hat{K}_{_{OM}} = P_m Z (P_m \eta_{_M}V P_m V - P_m \eta_{_M} P_m V^2) + P_m \eta_{_M} Z (P_m V^2 - (P_m V)^2) + P_m VZ (P_m V P_m \eta_{_M} - P_m \eta_{_M} V) = \widehat{Cov}_m (\eta_{_M},Z) \widehat{Var}_m (V) - \widehat{Cov}_m (V,Z) \widehat{Cov}_m (\eta_{_M},V) \overset{p}\to K_{_{OM}} = {Cov}_m (\eta_{_M},Z) {Var}_m (V) - {Cov}_m (V,Z) {Cov}_m (\eta_{_M},V)$, and  
    \item ${\beta_1} = \beta$ under DAG 1 through 8. 
\end{itemize}

{{Note that, with an abuse of notation, we use $Var_n(), Cov_n()$ to denote the variance and covariance in the population from which the validation study is sampled, and $Var_m(), Cov_m()$ to denote the variance and covariance in the population from which the main study is sampled. }}
 
%Note that as an example, $\widehat{Var}_n (V), \widehat{Var}_m (V)$ converge in probability to the true $Var(V)$ regardless of whether it is estimated from main study or validation study under independent sample assumption. Therefore later we do not distinguish the sources of data for brevity. 
We derive the formula for other CRS estimators under the same assumptions (1) through (3). 

\item For $\beta_{_{(--)}}$, let

    \begin{align*}
        g_{_{(--)}, {{ i}} }(\boldsymbol{\alpha^*,{\beta^*}}) & = \begin{pmatrix}  S_1(\boldsymbol{\alpha^*}) \\ S_2(\boldsymbol{\alpha^*}, \boldsymbol{{\beta}^*}) \end{pmatrix}  
        & = \begin{pmatrix} 
            \begin{pmatrix} 1 \\ Z_i \end{pmatrix} & (X_i - \eta_{-}){{ I(i\in\mathcal{V} )}} \\
             \begin{pmatrix} 1 \\ \eta_{-} \end{pmatrix} & (Y_i - {\beta_0}^* - {\beta_1}^* \eta_{-}){{ I(i\in\mathcal{M} )}}
        \end{pmatrix} 
        & = 0,
    \end{align*}

{{Let $A_{_{--}}(\boldsymbol{\alpha^*}, \boldsymbol{\beta^*})= E \left[ \frac{\partial{g_{_{(--), i}}(\boldsymbol{\alpha^*,{\beta^*}})}}{\partial{ (\boldsymbol{\alpha^*,{\beta^*}})}}\right]$, and $B_{_{--}}(\boldsymbol{\alpha^*}, \boldsymbol{\beta^*})=Cov( g_{_{--, i}}(\boldsymbol{\alpha^*,{\beta^*}}))$ }}, and the two matrices can be estimated as follows:

    \begin{align*} \tag{B3} \label{B3}
       {{\widehat{A}_{_{--}}(\boldsymbol{\alpha}^\ast, \boldsymbol{\beta}^\ast)}} %& = P_{(\cdot)}\Big\{ E \left[ \frac{\partial{g_{_{(--), i}}(\boldsymbol{\alpha^*,{\beta^*}})}}{\partial{ (\boldsymbol{\alpha^*,{\beta^*}})}} \right] \Big\}  \\ 
        & = - \begin{pmatrix}  
            P_n \begin{pmatrix}
            1 & Z \\
            Z & Z^2
            \end{pmatrix} 
            & 
            \begin{pmatrix}
            0 & 0 \\
            0 & 0 
            \end{pmatrix} \\
            {\beta_1^*} P_m \begin{pmatrix}
            1 & Z \\
            \eta_{_-} & \eta_{_-} Z
            \end{pmatrix} 
            & 
            P_m \begin{pmatrix}
            1 & \eta_{_-} \\
            \eta_{_-} & \eta_{_-}^2
            \end{pmatrix} 
        \end{pmatrix} \\
        & = \begin{pmatrix}
            \widehat{A}_{_{--,11}} & \widehat{A}_{_{--,12}} \\
            \widehat{A}_{_{--,21}} & \widehat{A}_{_{--,22}} 
        \end{pmatrix}, \\
    \end{align*}

    \begin{align*} \tag{B4} \label{B4}
        {{\widehat{B}_{_{--}}(\boldsymbol{\alpha}^\ast,\boldsymbol{\beta}^\ast)}} %& = P_{(\cdot)} Cov(g_{_{(--), i}}(\boldsymbol{\alpha^*,{\beta^*}})) \\
        & = \begin{pmatrix}
                n^{-1} \widehat\sigma_{_{X,-}}^2 P_n \begin{pmatrix}
                    1 & Z\\
                    Z & Z^2
                \end{pmatrix}   
                & 
                \begin{pmatrix}
                    0 & 0 \\
                    0 & 0
                \end{pmatrix}  \\
                \begin{pmatrix}
                    0 & 0 \\
                    0 & 0
                \end{pmatrix} &
                m^{-1} \widehat\sigma_{_{Y,(--)}}^2 P_m \begin{pmatrix}
                    1 & \eta_{_-} \\
                    \eta_{_-} & \eta_{_-}^2
                \end{pmatrix}   
            \end{pmatrix} \\
        & = \begin{pmatrix}
            \widehat{B}_{_{--,11}} & \widehat{B}_{_{--,12}} \\
            \widehat{B}_{_{--,21}} & \widehat{B}_{_{--,22}} 
        \end{pmatrix}, \\
    \end{align*}

and $$Var(\boldsymbol{\hat{\beta}^*}) = \widehat{A}_{_{--,22}}^{-1}[\widehat{B}_{_{--,22}} - \widehat{A}_{_{--,21}}\widehat{A}_{_{--,11}}^{-1}\widehat{B}_{_{--,11}}\widehat{A}_{_{--,11}}^{-t}\widehat{A}_{_{--,21}}^{t}]\widehat{A}_{_{--,22}}^{-t} $$

Replacing block matrices with those defined in \eqref{B3} and \eqref{B4} plus further simplification we have: 

\begin{align*}
{{m}}Var(\hat{{\beta}_1^*}) & = &  \sigma_{_{Y,(--)}}^2 \widehat{Var}_m (\eta_{_-})^{-1}  +{{\frac{m}{n}}}\sigma_{_{X,-}}^2 {\beta_1^*}^2 \widehat{Var}_m (\eta_{_-})^{-2} \widehat{Var}_n (Z)^{-1} \widehat{Cov}_m (\eta_{_-},Z)^2 \\
& \overset{p} \to &  \sigma_{_{Y,(--)}}^2 {Var}_m (\eta_{_-})^{-1}  + {{\frac{1}{\rho}}}\sigma_{_{X,-}}^2 {\beta}^2 {Var}_m (\eta_{_-})^{-2} {Var}_n (Z)^{-1} {Cov}_m (\eta_{_-},Z)^2 ,
\end{align*} where:

\begin{itemize}
    \item $\sigma_{_{X,-}}^2 = Var(X_i - \eta_{_-})$ and $\sigma_{_{Y,(--)}}^2 = Var(Y_i - {\beta_0^*} - {\beta_1^*} \eta_{_-})$, and
    \item ${\beta_1^*} = \beta$ under DAG 1, 5, 6, 7 and 8. 
\end{itemize}
    
\item For $\beta_{_{(-M)}}$, let
    \begin{align*}
        g_{_{(-M), i}}(\boldsymbol{\alpha,\tilde{\beta}}) & = \begin{pmatrix}  S_{1,i}(\boldsymbol{\alpha}) \\ S_{2,i}(\boldsymbol{\alpha}, \boldsymbol{\tilde{\beta}}) \end{pmatrix}  
        & = \begin{pmatrix} 
             \begin{pmatrix} 1 \\ Z_i \\ V_i \end{pmatrix} & (X_i - \eta_M){{ I(i\in\mathcal{V} )}} \\
            \begin{pmatrix} 1 \\ \eta_M  \end{pmatrix} & (Y_i - \tilde{\beta}_0 - \tilde{\beta}_1 \eta_M){{ I(i\in\mathcal{M} )}}
        \end{pmatrix} 
        & = 0,
    \end{align*}
{{Let $A_{_{-M}}(\boldsymbol{\alpha}, \boldsymbol{\tilde\beta})= E \left[ \frac{\partial{g_{_{(-M), i}}(\boldsymbol{\alpha,\tilde{\beta}})}}{\partial{ (\boldsymbol{\alpha,\tilde{\beta}})}}\right]$, and $B_{_{-M}}(\boldsymbol{\alpha}, \boldsymbol{\tilde\beta})=Cov( g_{_{-M, i}}(\boldsymbol{\alpha,\tilde{\beta}}))$ }}, and the two matrices can be estimated as follows:
    \begin{align*} \tag{B5} \label{B5}
        \widehat{A}_{_{-M}}(\boldsymbol{\alpha,\tilde{\beta}}) %& = P_{(\cdot)} \Big\{ E \left[ \frac{\partial{g_{_{(-M), i}}(\boldsymbol{\alpha,\tilde{\beta}})}}{\partial{ (\boldsymbol{\alpha,\tilde{\beta}})}} \right] \Big\}\\ 
        & = - \begin{pmatrix}  
            P_n \begin{pmatrix}
            1 & Z & V \\
            Z & Z^2 & ZV \\
            V & ZV & V^2
            \end{pmatrix} 
            & 
            \begin{pmatrix}
            0 & 0 & 0 \\
            0 & 0 & 0 \\
            0 & 0 & 0
            \end{pmatrix} \\
            \tilde{\beta}_1 P_m \begin{pmatrix}
            1 & Z & V \\
            \eta_{_M} & \eta_{_M} Z & \eta_{_M} V
            \end{pmatrix} 
            & 
            P_m \begin{pmatrix}
            1 & \eta_{_M} \\
            \eta_{_M} & \eta_{_M}^2
            \end{pmatrix} 
        \end{pmatrix} \\
        & = \begin{pmatrix}
            \widehat{A}_{_{-M,11}} & \widehat{A}_{_{-M,12}} \\
            \widehat{A}_{_{-M,21}} & \widehat{A}_{_{-M,22}} 
        \end{pmatrix}, \\
    \end{align*}
 
    \begin{align*} \tag{B6} \label{B6}
        \widehat{B}_{_{-M}}(\boldsymbol{\alpha,\tilde{\beta}}) %& = P_{(\cdot)}  Cov( g_{_{(-M, i)}}(\boldsymbol{\alpha,\tilde{\beta}})) \\
        & = \begin{pmatrix}
                n^{-1} \widehat\sigma_{_{X,M}}^2 P_n \begin{pmatrix}
                    1 & Z & V \\
                    Z & Z^2 & ZV \\
                    V & ZV & V^2
                \end{pmatrix}   
                & 
                \begin{pmatrix}
                    0 & 0 \\
                    0 & 0 \\
                    0 & 0
                \end{pmatrix}  \\
                \begin{pmatrix}
                    0 & 0 & 0 \\
                    0 & 0 & 0
                \end{pmatrix} &
                m^{-1} \widehat\sigma_{_{Y,(-M)}}^2 P_m \begin{pmatrix}
                    1 & \eta_{_M} \\
                    \eta_{_M} & \eta_{_M}^2 
                \end{pmatrix}   
            \end{pmatrix} \\
        & = \begin{pmatrix}
            \widehat{B}_{_{-M,11}} & \widehat{B}_{_{-M,12}} \\
            \widehat{B}_{_{-M,21}} & \widehat{B}_{_{-M,22}} 
        \end{pmatrix}, \\
    \end{align*}
and
     $$Var(\boldsymbol{\hat{\tilde{\beta}}}) = \widehat{A}_{_{-M,22}}^{-1}[\widehat{B}_{_{-M,22}} - \widehat{A}_{_{-M,21}}\widehat{A}_{_{-M,11}}^{-1}\widehat{B}_{_{-M,11}}\widehat{A}_{_{-M,11}}^{-t}\widehat{A}_{_{-M,21}}^{t}]\widehat{A}_{_{-M,22}}^{-t} $$

Replacing block matrices with those defined in \eqref{B5} and \eqref{B6} plus further simplification we have: 

\begin{align*}
{{m}}Var(\hat{\tilde{\beta}}_1)
& = & \sigma_{_{Y,(-M)}}^2 \widehat{Var}_m ({\eta}_{_M})^{-1} 
+ {{\frac{m}{n}}} \sigma_{_{X,M}}^2 \tilde{\beta_1}^2 \hat{C}_{A_{(-M,11)}} \hat{K}_{_{-M}} \widehat{Var}_m (\eta_{_M})^{-2} \\
& \overset{p} \to &  \sigma_{_{Y,(-M)}}^2 {Var}_m ({\eta}_{_M})^{-1} 
+ {{\frac{1}{\rho}}} \sigma_{_{X,M}}^2 \beta^2 {C}_{A_{(-M,11)}} {K}_{_{-M}} {Var}_m (\eta_{_M})^{-2} 
\end{align*}

where:
\begin{itemize}
    \item $\sigma_{_{Y,(-M)}}^2 = Var(Y_i - \tilde{\beta}_0 - \tilde{\beta}_1 \eta_{_M})$,
    \item $\hat{C}_{A_{(-M,11)}}  = \hat{C}_{A_{(OM,11)}} =  [P_n Z^2 P_n V^2 - (P_n ZV)^2 - P_n Z(P_n Z P_n V^2 - P_n V P_n ZV) + P_n V (P_n Z P_n ZV - P_n Z^2 P_n V)]^{-1} = (\widehat{Var}_n(Z) \widehat{Var}_n (V) - \widehat{Cov}_n (Z,V)^2)^{-1} \overset{p} \to ({Var}_n(Z) {Var}_n (V) - {Cov}_n (Z,V)^2)^{-1}$, and
    \item $\hat{K}_{_{-M}} = (\widehat{Cov}_m (\eta_{_M}, Z)^2 \widehat{Var}_n (V) - 2 \widehat{Cov}_m (\eta_{_M}, Z) \widehat{Cov}_m (\eta_{_M}, V) \widehat{Cov}_n (V,Z) + \widehat{Cov}_m (\eta_{_M}, V)^2 \widehat{Var}_n (Z)) \overset{p} \to ({Cov}_m (\eta_{_M}, Z)^2 {Var}_n (V) - \\2 {Cov}_m (\eta_{_M}, Z) {Cov}_m (\eta_{_M}, V) {Cov}_n (V,Z) + {Cov}_m (\eta_{_M}, V)^2 {Var}_n (Z))$, and
    \item $\tilde{\beta}_1 = \beta$ under DAG 1, 2, 5, 6, 7 and 8. 
\end{itemize}

\item For $\beta_{_{(O-)}}$, let
    \begin{align*}
        g_{_{(O-),i}}(\boldsymbol{\alpha^*,{\beta}^+}) & = \begin{pmatrix}  S_{1,i}(\boldsymbol{\alpha}^*) \\ S_{2,i}(\boldsymbol{\alpha}^*, \boldsymbol{{\beta}^+}) \end{pmatrix}  
        & = \begin{pmatrix} 
             \begin{pmatrix} 1 \\ Z_i \end{pmatrix} & (X_i - \eta_{-}){{ I(i\in\mathcal{V} )}} \\
             \begin{pmatrix} 1 \\ \eta_{-} \\ V_i \end{pmatrix} & (Y_i - {\beta_0}^+ - {\beta_1}^+ \eta_{-} - {\beta_2}^+ V_i){{ I(i\in\mathcal{M} )}}
        \end{pmatrix} 
        & = 0,
    \end{align*}
{{Let $A_{_{O-}}(\boldsymbol{\alpha^*}, \boldsymbol{\tilde\beta^+})= E \left[ \frac{\partial{g_{_{(O-), i}}(\boldsymbol{\alpha^*,{\beta}^+})}}{\partial{ (\boldsymbol{\alpha^*,{\beta}^+})}}\right]$, and $B_{_{O-}}(\boldsymbol{\alpha^*}, \boldsymbol{\beta}^+)=Cov( g_{_{O-, i}}(\boldsymbol{\alpha^*,{\beta}^+}))$ }}, and the two matrices can be estimated as follows:
    \begin{align*} \tag{B7} \label{B7}
        \widehat{A}_{_{O-}}(\boldsymbol{\alpha^*,\beta^+})% & = P_{(\cdot)} \Big\{ E \left[ \frac{\partial{g_{_{(O-),i}}(\boldsymbol{\alpha^*,\beta^+})}}{\partial{ (\boldsymbol{\alpha^*,\beta^+})}} \right] \Big\} \\ 
        & = - \begin{pmatrix}  
            P_n \begin{pmatrix}
            1 & Z \\
            Z & Z^2
            \end{pmatrix} 
            & 
            \begin{pmatrix}
            0 & 0 & 0\\
            0 & 0 & 0
            \end{pmatrix} \\
            \beta_1^+ P_m \begin{pmatrix}
            1 & Z \\
            \eta_{_-} & \eta_{_-} Z \\
            V & ZV
            \end{pmatrix} 
            & 
            P_m \begin{pmatrix}
            1 & \eta_{_-} & V \\
            \eta_{_-} & \eta_{_-}^2 & \eta_{_-} V \\
            V & \eta_{_-}V & V^2
            \end{pmatrix} 
        \end{pmatrix} \\
        & = \begin{pmatrix}
            \widehat{A}_{_{O-,11}} & \widehat{A}_{_{O-,12}} \\
            \widehat{A}_{_{O-,21}} & \widehat{A}_{_{O-,22}} 
        \end{pmatrix}, \\
    \end{align*}

    \begin{align*} \tag{B8} \label{B8}
        \widehat{B}_{_{O-}}(\boldsymbol{\alpha^*,\beta^+})  %& = P_{(\cdot)} Cov( g_{_{(O-), i}}(\boldsymbol{\alpha^*,\beta^+})) \\
        & = \begin{pmatrix}
                n^{-1} \widehat{\sigma}_{_{X,M}}^2 P_n \begin{pmatrix}
                    1 & Z \\
                    Z & Z^2
                \end{pmatrix}   
                & 
                \begin{pmatrix}
                    0 & 0 & 0\\
                    0 & 0 & 0
                \end{pmatrix}  \\
                \begin{pmatrix}
                    0 & 0\\
                    0 & 0\\
                    0 & 0
                \end{pmatrix} &
                m^{-1} \widehat{\sigma}_{_{Y,(O-)}}^2 P_m \begin{pmatrix}
                    1 & \eta_{_-} & V \\
                    \eta_{_-} & \eta_{_-}^2 & \eta_{_-}V \\
                    V & \eta_{_-}V & V^2
                \end{pmatrix}   
            \end{pmatrix} \\
        & = \begin{pmatrix}
            \widehat{B}_{_{O-,11}} & \widehat{B}_{_{O-,12}} \\
            \widehat{B}_{_{O-,21}} & \widehat{B}_{_{O-,22}} 
        \end{pmatrix}, \\
    \end{align*}
and
$$Var(\boldsymbol{\hat{\beta}^+}) = \widehat{A}_{_{O-,22}}^{-1}[\widehat{B}_{_{O-,22}} - \widehat{A}_{_{O-,21}}\widehat{A}_{_{O-,11}}^{-1}\widehat{B}_{_{O-,11}}\widehat{A}_{_{O-,11}}^{-t}\widehat{A}_{_{O-,21}}^{t}]\widehat{A}_{_{O-,22}}^{-t} $$

Replacing block matrices with those defined in \eqref{B7} and \eqref{B8} plus further simplification we have: 

\begin{align*}
&{{m}} Var(\hat{\beta_1}^+)\\
= &  \sigma_{_{Y,(O-)}}^2 \hat{C}_{A_{(O-,22)}}  \widehat{Var}_m (V)  + {{\frac{m}{n}}}\sigma_{_{X,-}}^2 {\beta_1^+}^2 \widehat{Var}_n (Z)^{-1} \hat{C}_{A_{(O-,22)}}^2  \hat{K}_{_{O-}}^2 \\
\overset{p}\to &  \sigma_{_{Y,(O-)}}^2 {C}_{A_{(O-,22)}}  {Var}_m (V)  + {{\frac{1}{\rho}}}\sigma_{_{X,-}}^2 {\beta}^2 {Var}_n (Z)^{-1} {C}_{A_{(O-,22)}}^2  {K}_{_{O-}}^2  \\
\end{align*} where under DAG 1 and DAG 5:
\begin{itemize}
    \item $\sigma_{_{Y,(O-)}}^2 = Var(Y_i - \beta_0^+ - \beta_1^+ \eta_{_-} - \beta_2^+ V_i)$,
    \item $\hat{C}_{A_{(O-,22)}} = [P_m \eta_{_-}^2 P_m V^2 - (P_m \eta_{_-} V)^2 - P_m \eta_{_-}(P_m \eta_{_-} P_m V^2 - P_m \eta_{_-}V P_m V) + P_m V(P_m \eta_{_-} P_m \eta_{_-}V - P_m \eta_{_-}^2 P_m V)]^{-1} = (\widehat{Var}_m (\eta_{_-}) \widehat{Var}_m (V) - \widehat{Cov}_m (\eta_{_-},V)^2)^{-1} \overset{p}\to {C}_{A_{(O-,22)}} = ({Var}_m (\eta_{_-}) {Var}_m (V) - {Cov}_m (\eta_{_-},V)^2)^{-1} = ({Var}_m (\eta_{_-}) {Var}_m (V))^{-1}$ as ${Cov}_m (\eta_{_-},V)=0$ under DAG 1 and 5, and
    \item $\hat{K}_{_{O-}}= P_m Z (P_m \eta_{_-}V P_m V - P_m \eta_{_-} P_m V^2) + P_m \eta_{_-} Z (P_m V^2 - (P_m V)^2) + P_m VZ (P_m V P_m \eta_{_-} - P_m \eta_{_-} V) =  \widehat{Cov}_m (\eta_{_-},Z) \widehat{Var}_m (V) - \widehat{Cov}_m (V,Z) \widehat{Cov}_m (\eta_{_-},V) \overset{p}\to K_{_{O-}} = {Cov}_m (\eta_{_-},Z) {Var}_m (V)$ as ${Cov}_m (\eta_{_-},V)=0, Cov_m(V,Z)=0$ under DAG 1 and 5, and 
    \item ${\beta_1^+} = \beta$ under DAG 1 and 5.
\end{itemize}

\end{enumerate} 

Last, we note that the CRS estimators ${\beta}_{_{(OM)}}$ and $\beta_{_{(--)}}$ have algebraic equivalent using methods developed independently by Rosner et al \cite{thurston2003equivalence} under the modeling assumptions made in section 1. And it has been proved that for Rosner estimators $Var(\hat{\beta}_{_{(OM)}}) \le Var(\hat{\beta}_{_{(--)}})$ under DAG 1, 6 and $Var(\hat{\beta}_{_{(OM)}}) \ge Var(\hat{\beta}_{_{(--)}})$ under DAG 7 \cite{Tang2022RSW}. Therefore, the inequalities hold for CRS estimators $\beta_{_{(OM)}}$ and $\beta_{_{(--)}}$ as well under DAG 1, 6 and 7. 

\subsection{Efficiency Comparison}
We only compare variance for DAGs where more than one estimator are valid. 

\subsubsection{Under DAG 1}
Under DAG 1:

\begin{itemize}
    \item $\alpha_2=0$ and thus $\eta_{_M}=\eta_{_-}=\eta$, $\sigma_{_{X,-}}=\sigma_{_{X,M}}=\sigma_{_{X}}$, $\sigma_{_{Y,(OM)}}=\sigma_{_{Y,(O-)}}=\sigma_{_{Y,O}}$ and $\sigma_{_{Y,(-M)}}=\sigma_{_{Y,(--)}}=\sigma_{_{Y,-}}$.
    \item ${Cov}(\eta_{_M},V) = {Cov}(\eta_{_-},V)=0$ and ${Cov}(Z,V)=0$. 
    \item $C_{A_{(OM,11)}}=C_{A_{(-M,11)}} = ({Var_n}(Z) {Var_n}(V))^{-1}$.
    \item $C_{A_{(OM,22)}} = C_{A_{(O-,22)}} = ({Var_m}(\eta) {Var_m}(V))^{-1}$.
    \item $K_{_{OM}}=K_{_{O-}}= {Cov_m}(\eta,Z){Var_m}(V)$.
    \item $K_{_{-M}} = {Cov_m}(\eta, Z)^2 {Var_{{n}}}(V)$.
\end{itemize}

We now compare $Var(\hat{\beta}_1) = Var(\hat{\beta}_{_{OM}})$, $Var(\hat{\beta_1}^+) = Var(\hat{\beta}_{_{O-}})$ and $Var(\hat{\tilde{\beta_1}}) = Var(\hat{\beta}_{_{-M}})$. Per results from section 2.1, we have:

\begin{itemize}
   \item ${{m}}Var(\hat{\beta}_1) = {{m}}Var(\hat{\beta}_{_{OM}}) \overset{p}\to \sigma_{_{Y,O}}^2 Var_m(\eta)^{-1}  + {{\frac{1}{\rho}}} \sigma_{_{X}}^2 \beta^2 Var_n (Z)^{-1} Var_m (\eta)^{-2} Cov_m (\eta, Z)^2 $,
    \item ${{m}}Var(\hat{\beta}_1^+) = {{m}}Var(\hat{\beta}_{_{O-}}) \overset{p} \to n^{-1}  \sigma_{_{Y,O}}^2 Var_m(\eta)^{-1} + {{\frac{1}{\rho}}} \sigma_{_{X}}^2 \beta^2 Var_n (Z)^{-1} Var_m (\eta)^{-2} Cov_m (\eta, Z)^2 $ and,
    \item ${{m}}Var(\hat{\tilde{\beta}}_1) = {{m}}Var(\hat{\beta}_{_{-M}})  \overset{p} \to  \sigma_{_{Y,-}}^2 {Var_m}(\eta)^{-1} + {{\frac{1}{\rho}}} \sigma_{_{X}}^2 \beta^2 Var_n (Z)^{-1} Var_m (\eta)^{-2} Cov_m (\eta, Z)^2 $.
\end{itemize}

Because $V$ is only a risk factor of outcome (thus a precision variable) and independent of $X$ and $Z$, under linear regression framework, $\sigma_{_{Y,O}}^2 \le \sigma_{_{Y,-}}^2$. Thus asymptotically ${{m}}Var(\hat{\beta}_1) = {{m}}Var(\hat{\beta_1}^+) \le {{m}}Var(\hat{\tilde{\beta}}_1)$. 

Recall that under DAG 1, it has been proved that $Var(\hat{\beta}_{_{(OM)}})\le Var(\hat{\beta}_{_{(--)}})$. Thus $\hat{\beta}_{_{(O-)}}$ and $\hat{\beta}_{_{(OM)}}$ are the most efficient estimators under DAG 1.

\subsubsection{Under DAG 2 and 6}
Under DAG 2, $\beta_{_{(OM)}}=\beta_{_{(-M)}}=\beta$ and under DAG 6 $\beta_{_{(--)}}=\beta_{_{(OM)}}=\beta$ and for both DAG 2 and 6:
\begin{itemize}
    \item ${Cov}(\eta_{_M},V)=0$ because $Cov(X,V)=0$ and we assumed in our model \eqref{EX_ZV} that $X = \eta_{_M} + \epsilon_{_{X|Z,V}}$ where  $\epsilon_{_{X|Z,V}}$ is the error in $X$ after conditioning on $(Z,V)$ and $\epsilon_{_{X|Z,V}} \perp V$ (this is true for example when (V,X,Z) are generated jointly normal as in the simulation), 
    \item $C_{A_{(OM,11)}}=C_{A_{(-M,11)}} = ({Var_n}(Z) {Var_n}(V) - Cov_n (Z,V)^2)^{-1}$,
    \item $C_{A_{(OM,22)}} = ({Var_m}(\eta_{_M}) {Var_m}(V))^{-1}$,
    \item $K_{_{OM}} = {Cov_m}(\eta_{_M},Z){Var_m}(V)$, and 
    \item $K_{_{-M}} = {Cov_m}(\eta_{_M},Z)^2 {Var_{{n}}}(V)$.
\end{itemize}

Per results from section 2.1: 
\begin{itemize}
    \item ${{m}}Var(\hat{\beta}_1) \overset{p} \to \sigma_{_{Y,(OM)}}^2 {Var_m}(\eta_{_M})^{-1} + \\
    {{\frac{1}{\rho}}}\sigma_{_{X,M}}^2 \beta^2  ({Var_n}(Z) {Var_n}(V) - Cov_n (Z,V)^2)^{-1} ({Var_m}({{\eta_M}}))^{-2} {Cov_m}(\eta_{_M},Z)^2 {Var_n} (V) $.
    \item ${{m}}Var(\hat{\tilde{\beta}}_1) \overset{p} \to  \sigma_{_{Y,(-M)}}^2 {Var_m}(\eta_{_M})^{-1}  + \\ {{\frac{1}{\rho}}}\sigma_{_{X,M}}^2 \beta^2 ({Var_n}(Z) {Var_n}(V) - Cov_n (Z,V)^2)^{-1} ({Var_m}({{\eta_M}}))^{-2} {Cov_m}(\eta_{_M},Z)^2 {Var_n} (V) $.
\end{itemize}

For DAG 2, $V$ is a risk factor of outcome and independent of $X$ thus a precision covariate. Under regression framework, $\sigma_{_{Y,(OM)}}^2 \le \sigma_{_{Y,(-M)}}^2$. Thus asymptotically ${{m}}Var(\hat{\beta}_{_{(OM)}})={{m}}Var(\hat{\beta}_1) \le {{m}}Var(\hat{\tilde{\beta}}_1)={{m}}Var(\hat{\beta}_{_{(-M)}})$ and $\hat{\beta}_{_{(OM)}}$ is the most efficient estimator under DAG 2.

For DAG 6, $V$ is NOT a risk factor of outcome and independent of $X$. Under regression framework, $\sigma_{_{Y,(OM)}}^2 \approx \sigma_{_{Y,(-M)}}^2$ ($V$ has no additional information on Y). Thus asymptotically ${{m}}Var(\hat{\tilde{\beta}}_1)={{m}}Var(\hat{\beta}_{_{(-M)}}) \approx {{m}}Var(\hat{\beta}_{_{(OM)}}) = {{m}}Var(\hat{\beta}_1) \le {{m}}Var(\hat{\beta}_{_{(--)}})$, where ${{m}}Var(\hat{\beta}_{_{(OM)}}) \le {{m}}Var (\hat{\beta}_{_{(--)}})$ under DAG 6 has been proven from previous study, and thus $\hat{\beta}_{_{(-M)}}$ and $\hat{\beta}_{_{(OM)}}$ are the most efficient estimators under DAG 6.

\subsubsection{Under DAG 7 and 8}
Under DAG 7 and 8, $\hat{\beta}_{_{(OM)}}$, $\hat{\beta}_{_{(--)}}$ and $\beta_{_{(-M)}}$ are consistent for the true effect $\beta$ and from section 2.1:

\begin{itemize}
    \item ${{m}}Var(\hat{\beta}_1) \overset{p}\to  \sigma_{_{Y,(OM)}}^2 ({Var_m}(\eta_{_M}) {Var_m}(V) - {Cov_m}(\eta_{_M},V)^2)^{-1}Var_m (V) + \\{{\frac{1}{\rho}}} \sigma_{_{X,M}}^2 \beta^2 ({Var_n}(Z) {Var_n}(V) - Cov_n(Z,V)^2)^{-1} ({Cov_m}(\eta_{_M},Z){Var_m}(V) - {Cov_m}(V,Z){Cov_m}(\eta_{_M},V))^2 \\ ({Var_m}(\eta_{_M}) {Var_m}(V) - Cov_m (\eta_{_M},V)^2)^{-2} Var_n (V) $.
    \item ${{m}}Var(\hat{\beta}_1^*) \overset{p} \to \sigma_{_{Y,(--)}}^2 {Var_m}(\eta_{_-})^{-1}  +\\ {{\frac{1}{\rho}}}\sigma_{_{X,-}}^2 \beta^2 {Var_m}(\eta_{_-})^{-2} {Var_n}(Z)^{-1} {Cov_m}(\eta_{_-},Z)^2 $.
    \item ${{m}}Var(\hat{\tilde{\beta}}_1) \overset{p}\to  \sigma_{_{Y,(-M)}}^2 {Var_m}(\eta_{_M})^{-1} + \\{{\frac{1}{\rho}}} \sigma_{_{X,M}}^2 \beta^2 ({Var_n}(Z) {Var_n}(V) - {Cov_n}(Z,V)^2)^{-1} [{Cov_m}(\eta_{_M}, Z)^2 {Var_n}(V) - \\ 2{Cov_m}(\eta_{_M}, Z) {Cov_m}(\eta_{_M}, V) {Cov_n}(V,Z) + {Cov_m} (\eta_{_M}, V)^2 {Var_n}(Z)] {Var_m}(\eta_{_M})^{-2}  \big\}$.
\end{itemize}

Recall that main study sample and validation study sample are independently sampled from the same source of population, therefore variance and covariance estimate from the two samples should converge to the same population variance and covariance, e.g. $Cov_m (V,Z)= Cov_n (V,Z)$. In the following proof, we will therefore omit the subscript $n$ and $m$. First we compare $Var(\hat{\beta}_1^*)$ and $Var(\hat{\tilde{\beta}}_1)$. 

\begin{itemize}
    \item We can prove that ${Var}(\eta_{_M}) \ge {Var}(\eta_{_-})$ by noticing that ${Var}(\eta_{_M}) = {Var}(E[X|Z,V]) = Var(X) - E(Var[X|Z,V])$, \\ where  $E(Var[X|Z,V])=Var(X|Z) - Var(E[X|Z,V]) \le Var(X|Z)$ by total of total variance and \\ $E(Var[X|Z,V]) \le E(Var(X|Z))$ and thus ${Var}(\eta_{_M}) = Var(X) - E(Var[X|Z,V]) \ge Var(X) - E(Var[X|Z]) = {Var}(\eta_{_-})$.
    \item We can thus prove that $\textstyle{\sigma_{_{Y,(--)}}^2 {Var}(\eta_{_-})^{-1} \ge \sigma_{_{Y,(-M)}}^2 {Var}(\eta_{_M})^{-1}}$ as \\ $\textstyle{\sigma_{_{Y,(--)}}^2 {Var}(\eta_{_-})^{-1}= \frac{Var(Y-\beta \eta_{_-})}{{Var}(\eta_{_-})} = \frac{Var(Y) + \beta^2 {Var}(\eta_{_-}) - 2\beta Cov(\eta_{_-},Y)}{{Var}(\eta_{_-})} = \frac{Var(Y)}{{Var}(\eta_{_-})} - \beta^2}$ and similarly $\textstyle{\sigma_{_{Y,(-M)}}^2{Var}(\eta_{_M})^{-1}=\frac{Var(Y)}{{Var}(\eta_{_M})} - \beta^2}$ (recall that $Cov(\eta_{-},Y) = \beta Var(\eta_{-}) + Cov (\eta_{-}, \epsilon_Y^*) = \beta Var(\eta_{-})$ and $Cov(\eta_{M},Y) = \beta Var(\eta_{M}) + Cov (\eta_{M}, \epsilon_Y^*) = \beta Var(\eta_{M})$).
    \item Under regression framework, $\sigma_{_{X,(M)}}^2 \le \sigma_{_{X,(-)}}^2$ as $V$ is predictive of $X$.
    \item We can also prove that $$\frac{[{Cov}(\eta_{_M}, Z)^2 {Var}(V) - 2{Cov}(\eta_{_M}, Z) {Cov}(\eta_{_M}, V) {Cov}(V,Z) + {Cov} (\eta_{_M}, V)^2 {Var}(Z)]}{({Var}(Z) {Var}(V) - {Cov}(V,Z)^2)({Var}(\eta_{_M}))} \le \Big\{\frac{{Cov}(\eta_{_-},Z)^2}{{Var}(\eta_{_-}){Var_n}(Z)} \Big\}^{-1}$$ because $$0 \le \frac{{Cov}(\eta_{_-},Z)^2}{{Var}(\eta_{_-}){Var}(Z)} \le 1$$ and $$\frac{[{Cov}(\eta_{_M}, Z)^2 {Var}(V) - 2{Cov}(\eta_{_M}, Z) {Cov}(\eta_{_M}, V) {Cov}(V,Z) + {Cov} (\eta_{_M}, V)^2 {Var}(Z)]}{({Var}(Z) {Var}(V) - {Cov}(Z,V)^2)({Var}(\eta_{_M}))} =1 \text{ under model \eqref{EX_ZV}}.\footnote{ This equation holds because under model \eqref{EX_ZV} $Var (\eta_{_M})= \alpha_1^2 Var (Z) + \alpha_2^2 Var (V) + 2 \alpha_1 \alpha_2 Cov (V,Z), Cov (\eta_{_M},V) = \alpha_1 Cov (V,Z) + \alpha_2 Var (V)$ and \\ $Cov (\eta_{_M},Z) = \alpha_1 Var (Z) + \alpha_2 Cov (V,Z)$ and terms cancel out to one. We note that the second equation is not true if samples from main and validation study sample are not from the same source population.}$$
    
    Therefore ${{m}}Var(\hat{\beta}_1^*) \ge {{m}}Var(\hat{\tilde{\beta}}_1)$. Now we compare ${{m}}Var(\hat{\beta}_1)$ and ${{m}}Var(\hat{\tilde{\beta}}_1)$: 

    \item Under linear regression framework, $\sigma_{_{Y,(OM)}}^2 \ge \sigma_{_{Y,(-M)}}^2$ as $V$ is not a predictor of outcome $Y$ but associated with exposure $X$. 
    \item $({Var}(\eta_{_M}) {Var}(V) - {Cov}(\eta_{_M},V)^2)^{-1}Var (V) \ge ({Var}(\eta_{_M}) {Var}(V))^{-1}Var (V) =  {Var}(\eta_{_M})^{-1}$.
    \item We have proved that the second component in $Var(\hat{\tilde{\beta}}_1)$, $\textstyle{\frac{[{Cov}(\eta_{_M}, Z)^2 {Var}(V) - 2{Cov}(\eta_{_M}, Z) {Cov}(\eta_{_M}, V) {Cov}(V,Z) + {Cov} (\eta_{_M}, V)^2 {Var}(Z)]}{({Var}(Z) {Var}(V) - {Cov}(Z,V)^2)({Var}(\eta_{_M}))}=1}$ under model \eqref{EX_ZV} and provided that main study and validation study is sampled from the same source population.
    \item We can also prove that the second component in $Var(\hat{\beta}_1)$:
    $$\frac{({Cov}(\eta_{_M},Z){Var}(V) - {Cov}(V,Z){Cov}(\eta_{_M},V))^2 Var (V) Var (\eta_{_M})}{({Var}(Z) {Var}(V) - Cov(Z,V)^2) ({Var}(\eta_{_M}) {Var}(V) - Cov (\eta_{_M},V)^2)^{2}} \ge 1 \text{ under model \eqref{EX_ZV}}. \footnote{This inequality holds because under model \eqref{EX_ZV} $Var (\eta_{_M})= \alpha_1^2 Var (Z) + \alpha_2^2 Var (V) + 2 \alpha_1 \alpha_2 Cov (V,Z), Cov (\eta_{_M},V) = \alpha_1 Cov (V,Z) + \alpha_2 Var (V)$ and $Cov (\eta_{_M},Z) = \alpha_1 Var (Z) + \alpha_2 Cov (V,Z)$ and $\frac{({Cov}(\eta_{_M},Z){Var}(V) - {Cov}(V,Z){Cov}(\eta_{_M},V))^2 Var (V) Var (\eta_{_M})}{({Var}(Z) {Var}(V) - Cov(Z,V)^2) ({Var}(\eta_{_M}) {Var}(V) - Cov (\eta_{_M},V)^2)^{2}} = \frac{\alpha_1^2 Var(Z) Var(V) + \alpha_2^2 Var(V)^2 + 2 \alpha_1 \alpha_2 Cov(V,Z) Var(V)}{\alpha_1^2 Var(Z) Var(V)  - \alpha_1^2 Cov(V,Z)} \ge 1$, where the last inequality holds because:
    \begin{align*}
       & \alpha_1^2 Var(Z) Var(V) + \alpha_2^2 Var(V)^2 + 2 \alpha_1 \alpha_2 Cov(V,Z) Var(V) \ge \alpha_1^2 Var(Z) Var(V)  - \alpha_1^2 Cov(V,Z)^2\\%\text{, where $\rho_{Z,V}^2 = \frac{Cov(V,Z)^2}{Var(V)Var(Z)} \ne 1$}\\
    \Longleftrightarrow & \alpha_1^2 Cov(V,Z) + \alpha_2^2 Var(V)^2 + 2 \alpha_1 \alpha_2 Cov(V,Z) Var(V) \ge 0\\
    \Longleftrightarrow & (\alpha_1 Cov(V,Z) + \alpha_2 Var(V))^2 \ge 0 \\
    \end{align*}}$$
    \item Combining the above two results gives: $$\frac{({Cov_m}(\eta_{_M},Z){Var_m}(V) - {Cov_m}(V,Z){Cov_m}(\eta_{_M},V))^2 Var_n (V) Var (\eta_{_M})}{({Var_n}(Z) {Var_n}(V) - Cov_n(Z,V)^2) ({Var}(\eta_{_M}) {Var_m}(V) - Cov_m (\eta_{_M},V)^2)^{2}} \frac{1}{Var(\eta_{_M})}\ge \frac{1}{Var(\eta_{_M})}$$  $$= \frac{[{Cov}(\eta_{_M}, Z)^2 {Var}(V) - 2{Cov}(\eta_{_M}, Z) {Cov}(\eta_{_M}, V) {Cov}(V,Z) + {Cov} (\eta_{_M}, V)^2 {Var}(Z)]}{({Var}(Z) {Var}(V) - {Cov}(Z,V)^2)({Var}(\eta_{_M}))} \frac{1}{Var(\eta_{_M})}.$$ 
\end{itemize}
    Therefore $Var(\hat{\beta}_1) \ge Var(\hat{\tilde{\beta}}_1)$. Thus we conclude that $\beta_{_{(-M)}}$ is the most efficient estimator under DAG 7 and 8. 

%\end{landscape}

\newpage
\section{Design of Simulation Studies} 
As extension for the base case, we allowed $V$ to distributed as $Bern(0.4)$ instead of $N(0,1)$ and varied values of $\eta_{v},\theta_{x},\theta_{v},\beta_{x},\beta_{v}$ to represent varying strength and direction of $\rho_{_{V,X}}, \rho_{_{X,Y|V}}, \rho_{_{X,Z|V}}, \rho_{_{V,Z|X}}$ and $\rho_{_{V,Y|X}}$. Note that when one or more of these coefficients are set to zero, the data generating process becomes compatible with a single DAG. For example, $\eta_{v}=\theta_{v}=0$ corresponds to DAG 1 ($V_{1(-{}-{}Y)}$), and $\beta_{v}=0$ in addition to $\eta_{v}=\theta_{v}=0$  corresponds to DAG 5. We summarize the possible coefficient parameterization and thus combinations of different data generating process under each DAG in the table below, where we also give their corresponding approximated (conditional) correlation coefficient $\rho_{(\cdot)}$. For the base case, the parameterization gives us $\rho_{_{V,X}} \approx (0.37, 0) , \rho_{_{X,Y|V}} \approx 0.45, \rho_{_{X,Z|V}} \approx 0.71, \rho_{_{V,Z|X}} \approx (0.2, 0.18, 0)$ and $\rho_{_{V,Y|X}} \approx (0.62, 0.6, 0)$. We also varied $n_{MS}$ and $n_{VS}$ as the variance calculation depends on them. 

We also note that for the binary outcomes generated by the logistic link function, due to non-collapsibility, the marginal effect of $X$ on $Y$ differs from the conditional effect of $X$ on $Y$ given $V$, the true data generating process under DAG 1 through 4. With a logit link and a continuous exposure, one cannot analytically obtain the marginal effect by integrating the conditional effect over covariate $V$, as the marginal effect would vary by values of $X$ and thus is not comparable to the constant conditional effect of $\beta_{x}$. However, since the simulation guarantees a rare disease with an outcome prevalence less than 5 \%, the marginal effect will be approximately $\beta_{x}$. Thus, even if for DAG 1 where $\beta_{(O-)}$ and $\beta_{(OM)}$ converge to slightly different parameter values than values that $\beta_{(--)}$ and $\beta_{(-M)}$ would converge to, we use $\beta_{x}$ as the truth for bias assessment.

\newpage
  \begin{threeparttable}
  \caption{Parameters Used in Monte Carlo Simulation}
  \label{tab:scenario}
    \scriptsize	
 \begin{tabular}{||c|c|ccccc|ccccc||}
 \hline
$V_j$ as in & Scenario \tnote{\textdagger} & $\eta_v$ & $\theta_x$ & $\theta_v$ & $\beta_x$ & $\beta_v$ & $\rho_{x,v}$ & $\rho_{x,z|v}$ & $\rho_{v,z|x}$ & $\rho_{v,y|x}$ \tnote{\textdaggerdbl} & $\rho_{x,y|v}$ \tnote{\textdaggerdbl}\\
\hline
DAG 1  & base case  & 0    & 0.5     & 0     & 0.5    & 0.8    & 0  & 0.71 & 0  & 0.62 & 0.45 \\
  &  small $\rho_{x,z|v}$     & 0    & 0.2     & 0     & 0.5    & 0.8    & 0  & 0.37 & 0  & 0.62 & 0.45 \\
  &  small effect $\beta_x$     & 0    & 0.5     & 0     & 0.1    & 0.8    & 0  & 0.71 & 0  & 0.62 & 0.1  \\
  & V is weak risk factor of Y& 0    & 0.5     & 0     & 0.5    & 0.2    & 0  & 0.71 & 0  & 0.2  & 0.45 \\
  & binary covariate V     & 0    & 0.5     & 0     & 0.5    & 0.8    & 0  & 0.71 & 0  & 0.36 & 0.45 \\
\hline
DAG 2  & base case  & 0    & 0.5     & 0.1     & 0.5    & 0.8    & 0  & 0.71 & 0.2  & 0.62 & 0.45 \\
 &  small $\rho_{x,z|v}$     & 0    & 0.2     & 0.1     & 0.5    & 0.8    & 0  & 0.37 & 0.2  & 0.62 & 0.45 \\
  &  small effect $\beta_x$     & 0    & 0.5     & 0.1     & 0.1    & 0.8    & 0  & 0.71 & 0.2  & 0.62 & 0.1  \\
  & large measurement error   $\rho_{v,z|x}$     & 0    & 0.5     & 2     & 0.5    & 0.8    & 0  & 0.71 & 0.97 & 0.62 & 0.45 \\
  & V is weak risk factor of Y    & 0    & 0.5     & 0.1     & 0.5    & 0.2    & 0  & 0.71 & 0.2  & 0.2  & 0.45 \\
  & binary covariate V     & 0    & 0.5     & 0.1     & 0.5    & 0.8    & 0  & 0.71 & 0.1  & 0.36 & 0.45 \\
\hline
DAG 3  & base case  & 0.4    & 0.5     & 0     & 0.5    & 0.8    & 0.37    & 0.71 & 0  & 0.6  & 0.45 \\
   &  small $\rho_{x,z|v}$     & 0.4    & 0.2     & 0     & 0.5    & 0.8    & 0.37    & 0.37 & 0  & 0.6  & 0.45 \\
   &  small effect $\beta_x$     & 0.4    & 0.5     & 0     & 0.1    & 0.8    & 0.37    & 0.71 & 0  & 0.6  & 0.1  \\
   & negative $\rho_{v,x}$     & -0.4   & 0.5     & 0     & 0.5    & 0.8    & -0.37     & 0.71 & 0  & 0.6  & 0.45 \\
   & small $\rho_{v,x}$     & 0.2    & 0.5     & 0     & 0.5    & 0.8    & 0.20    & 0.71 & 0  & 0.62 & 0.45 \\
   & V is weak risk factor of Y    & 0.4    & 0.5     & 0     & 0.5    & 0.2    & 0.37    & 0.71 & 0  & 0.18 & 0.45 \\
   & binary covariate V     & 0.4    & 0.5     & 0     & 0.5    & 0.8    & 0.19    & 0.71 & 0  & 0.36 & 0.45 \\
\hline
DAG 4  & base case  & 0.4    & 0.5     & 0.1     & 0.5    & 0.8    & 0.37    & 0.71 & 0.18 & 0.6  & 0.45 \\
   &  small $\rho_{x,z|v}$     & 0.4    & 0.2     & 0.1     & 0.5    & 0.8    & 0.37    & 0.37 & 0.18 & 0.6  & 0.45 \\
   &  small effect $\beta_x$     & 0.4    & 0.5     & 0.1     & 0.1    & 0.8    & 0.37    & 0.71 & 0.18 & 0.6  & 0.1  \\
   & negative $\rho_{v,x}$     & -0.4   & 0.5     & 0.1     & 0.5    & 0.8    & -0.37     & 0.71 & 0.18 & 0.6  & 0.45 \\
   & small $\rho_{v,x}$     & 0.2    & 0.5     & 0.1     & 0.5    & 0.8    & 0.20    & 0.71 & 0.19 & 0.62 & 0.45 \\
   & large measurement error   $\rho_{v,z|x}$     & 0.4    & 0.5     & 2     & 0.5    & 0.8    & 0.37    & 0.71 & 0.97 & 0.6  & 0.45 \\
   & V is weak risk factor of Y    & 0.4    & 0.5     & 0.1     & 0.5    & 0.2    & 0.37    & 0.71 & 0.18 & 0.18 & 0.45 \\
   & binary covariate V     & 0.4    & 0.5     & 0.1     & 0.5    & 0.8    & 0.19    & 0.71 & 0.1  & 0.36 & 0.45 \\
\hline
DAG 5  & base case  & 0    & 0.5     & 0     & 0.5    & 0    & 0  & 0.71 & 0  & 0  & 0.45 \\
   &  small $\rho_{x,z|v}$     & 0    & 0.2     & 0     & 0.5    & 0    & 0  & 0.37 & 0  & 0  & 0.45 \\
   &  small effect $\beta_x$     & 0    & 0.5     & 0     & 0.1    & 0    & 0  & 0.71 & 0  & 0  & 0.1  \\
   & binary covariate V     & 0    & 0.5     & 0     & 0.5    & 0    & 0  & 0.71 & 0  & 0  & 0.45 \\
\hline
DAG 6  & base case  & 0    & 0.5     & 0.1     & 0.5    & 0    & 0  & 0.71 & 0.2  & 0  & 0.45 \\
   &  small $\rho_{x,z|v}$     & 0    & 0.2     & 0.1     & 0.5    & 0    & 0  & 0.37 & 0.2  & 0  & 0.45 \\
   &  small effect $\beta_x$     & 0    & 0.5     & 0.1     & 0.1    & 0    & 0  & 0.71 & 0.2  & 0  & 0.1  \\
   & large measurement error   $\rho_{v,z|x}$     & 0    & 0.5     & 2     & 0.5    & 0    & 0  & 0.71 & 0.97 & 0  & 0.45 \\
   & binary covariate V     & 0    & 0.5     & 0.1     & 0.5    & 0    & 0  & 0.71 & 0.1  & 0  & 0.45 \\
\hline
DAG 7  & base case  & 0.4    & 0.5     & 0     & 0.5    & 0    & 0.37    & 0.71 & 0  & 0  & 0.45 \\
   &  small $\rho_{x,z|v}$     & 0.4    & 0.2     & 0     & 0.5    & 0    & 0.37    & 0.37 & 0  & 0  & 0.45 \\
   &  small effect $\beta_x$     & 0.4    & 0.5     & 0     & 0.1    & 0    & 0.37    & 0.71 & 0  & 0  & 0.1  \\
   & negative $\rho_{v,x}$     & -0.4   & 0.5     & 0     & 0.5    & 0    & -0.37     & 0.71 & 0  & 0  & 0.45 \\
   & small $\rho_{v,x}$     & 0.2    & 0.5     & 0     & 0.5    & 0    & 0.20    & 0.71 & 0  & 0  & 0.45 \\
   & binary covariate V     & 0.4    & 0.5     & 0     & 0.5    & 0    & 0.19    & 0.71 & 0  & 0  & 0.45 \\
\hline
DAG 8  & base case  & 0.4    & 0.5     & 0.1     & 0.5    & 0    & 0.37    & 0.71 & 0.18 & 0  & 0.45 \\
   &  small $\rho_{x,z|v}$     & 0.4    & 0.2     & 0.1     & 0.5    & 0    & 0.37    & 0.37 & 0.18 & 0  & 0.45 \\
   &  small effect $\beta_x$     & 0.4    & 0.5     & 0.1     & 0.1    & 0    & 0.37    & 0.71 & 0.18 & 0  & 0.1  \\
   & negative $\rho_{v,x}$     & -0.4   & 0.5     & 0.1     & 0.5    & 0    & -0.37     & 0.71 & 0.18 & 0  & 0.45 \\
   & small $\rho_{v,x}$     & 0.2    & 0.5     & 0.1     & 0.5    & 0    & 0.20    & 0.71 & 0.19 & 0  & 0.45 \\
   & large measurement error   $\rho_{v,z|x}$     & 0.4    & 0.5     & 2     & 0.5    & 0    & 0.37    & 0.71 & 0.97 & 0  & 0.45 \\
   & binary covariate V     & 0.4    & 0.5     & 0.1     & 0.5    & 0    & 0.19    & 0.71 & 0.1  & 0  & 0.45 \\
\hline
    \end{tabular}
        \begin{tablenotes}
        \item[\textdagger] We also varied the sample size: For main study, we reduced the sample size from $n_{MS} = 5,000$ to $n_{MS} = 2,000$ for continuous outcome and from $n_{MS} = 10,000$ to $n_{MS} = 5,000$ for binary outcome. For validation study, we reduced the sample size from $n_{VS}=400$ to $n_{VS} = 150$ for both continuous and binary outcome. 
        \item[\textdaggerdbl] These correlations only apply to continuous outcome generated under linear model. 
    \end{tablenotes}
 \end{threeparttable}

\newpage
\section{All Simulations Results: Percent Bias for Point Estimates}
\scriptsize	
\begin{longtable}{|| ll| lrrrrr | c c c c ||}
\hline
\textbf{Y Type} & \textbf{DAG} & \textbf{Scenario} & $\rho_{v,x}$ & $\rho_{x,z|v}$ & $\rho_{v,z|x}$ & $\beta_x$ & $\beta_v$ & \textbf{$\beta_{(OM)}$} & \textbf{$\beta_{(--)}$} & \textbf{$\beta_{(-M)}$} & \textbf{$\beta_{(O-)}$} \\
\hline
\endhead
Continuous & DAG 1& base case & 0 & 0.71& 0 & 0.5& 0.8& 0 & 0 & 0 & 0 \\
 && small $\rho_{x,z|v}$& 0 & 0.37& 0 & 0.5& 0.8& 2 & 1 & 2 & 2 \\
 && small effect $\beta_x$& 0 & 0.71& 0 & 0.1& 0.8& 1 & 0 & 1 & 1 \\
 && weak risk factor V, small $\beta_v$ & 0 & 0.71& 0 & 0.5& 0.2& 0 & 0 & 0 & 0 \\
 && binary covariate V& 0 & 0.71& 0 & 0.5& 0.8& 0 & 0 & 0 & 0 \\
 && $n_{MS}$ = 2,000 vs $n_{MS}$ = 5,000& 0 & 0.71& 0 & 0.5& 0.8& 0 & 0 & 0 & 0 \\
 && $n_{VS}$ = 150 vs $n_{VS}$ = 400& 0 & 0.71& 0 & 0.5& 0.8& 1 & 1 & 1 & 1 \\
 & DAG 2& base case & 0 & 0.71& 0.2 & 0.5& 0.8& 0 & 32& 0 & 2 \\
 && small $\rho_{x,z|v}$& 0 & 0.37& 0.2 & 0.5& 0.8& 2 & 83& 2 & 5 \\
 && small effect $\beta_x$& 0 & 0.71& 0.2 & 0.1& 0.8& 1 & 161 & 1 & 3 \\
 && large $\rho_{v,z|x}$ & 0 & 0.71& 0.97& 0.5& 0.8& 0 & 670 & 0 & 838 \\
 && weak risk factor V, small $\beta_v$ & 0 & 0.71& 0.2 & 0.5& 0.2& 0 & 8 & 0 & 2 \\
 && binary covariate V& 0 & 0.71& 0.1 & 0.5& 0.8& 0 & 7 & 0 & 1 \\
 && $n_{MS}$ = 2,000 vs $n_{MS}$ = 5,000& 0 & 0.71& 0.2 & 0.5& 0.8& 0 & 32& 0 & 2 \\
 && $n_{VS}$ = 150 vs $n_{VS}$ = 400& 0 & 0.71& 0.2 & 0.5& 0.8& 1 & 33& 1 & 3 \\
 & DAG 3& base case & 0.37& 0.71& 0 & 0.5& 0.8& 0 & 55& 97& -7\\
 && small $\rho_{x,z|v}$& 0.37& 0.37& 0 & 0.5& 0.8& 2 & 57& 214 & -11 \\
 && small effect $\beta_x$& 0.37& 0.71& 0 & 0.1& 0.8& 1 & 276 & 485 & -6\\
 && negative $\rho_{v,x}$ & -0.37 & 0.71& 0 & 0.5& 0.8& 0 & -55 & -97 & -6\\
 && small $\rho_{v,x}$& 0.20& 0.71& 0 & 0.5& 0.8& 0 & 31& 60& -2\\
 && weak risk factor V, small $\beta_v$ & 0.37& 0.71& 0 & 0.5& 0.2& 0 & 14& 24& -7\\
 && binary covariate V& 0.19& 0.71& 0 & 0.5& 0.8& 0 & 13& 25& -1\\
 && $n_{MS}$ = 2,000 vs $n_{MS}$ = 5,000& 0.37& 0.71& 0 & 0.5& 0.8& 0 & 56& 97& -7\\
 && $n_{VS}$ = 150 vs $n_{VS}$ = 400& 0.37& 0.71& 0 & 0.5& 0.8& 1 & 56& 97& -6\\
 & DAG 4& base case & 0.37& 0.71& 0.18& 0.5& 0.8& 0 & 78& 97& -5\\
 && small $\rho_{x,z|v}$& 0.37& 0.37& 0.18& 0.5& 0.8& 2 & 108 & 214 & -17 \\
 && small effect $\beta_x$& 0.37& 0.71& 0.18& 0.1& 0.8& 1 & 388 & 485 & -4\\
 && negative $\rho_{v,x}$ & -0.37 & 0.71& 0.18& 0.5& 0.8& 0 & -29 & -97 & -5\\
 && small $\rho_{v,x}$& 0.20& 0.71& 0.19& 0.5& 0.8& 0 & 60& 60& 0 \\
 && large $\rho_{v,z|x}$ & 0.37& 0.71& 0.97& 0.5& 0.8& 0 & 256 & 97& 288 \\
 && weak risk factor V, small $\beta_v$ & 0.37& 0.71& 0.18& 0.5& 0.2& 0 & 20& 24& -5\\
 && binary covariate V& 0.19& 0.71& 0.1 & 0.5& 0.8& 0 & 20& 25& -1\\
 && $n_{MS}$ = 2,000 vs $n_{MS}$ = 5,000& 0.37& 0.71& 0.18& 0.5& 0.8& 0 & 78& 97& -5\\
 && $n_{VS}$ = 150 vs $n_{VS}$ = 400& 0.37& 0.71& 0.18& 0.5& 0.8& 1 & 78& 97& -4\\
 & DAG 5& base case & 0 & 0.71& 0 & 0.5& 0& 0 & 0 & 0 & 0 \\
 && small $\rho_{x,z|v}$& 0 & 0.37& 0 & 0.5& 0& 2 & 2 & 0 & 2 \\
 && small effect $\beta_x$& 0 & 0.71& 0 & 0.1& 0& 1 & 1 & 1 & 1 \\
 && binary covariate V& 0 & 0.71& 0 & 0.5& 0& 0 & 0 & 0 & 0 \\
 && $n_{MS}$ = 2,000 vs $n_{MS}$ = 5,000& 0 & 0.71& 0 & 0.5& 0& 0 & 0 & 0 & 0 \\
 && $n_{VS}$ = 150 vs $n_{VS}$ = 400& 0 & 0.71& 0 & 0.5& 0& 1 & 1 & 0 & 1 \\
 & DAG 6& base case & 0 & 0.71& 0.2 & 0.5& 0& 0 & 0 & 0 & 2 \\
 && small $\rho_{x,z|v}$& 0 & 0.37& 0.2 & 0.5& 0& 2 & 2 & 0 & 5 \\
 && small effect $\beta_x$& 0 & 0.71& 0.2 & 0.1& 0& 1 & 1 & 1 & 3 \\
 && large $\rho_{v,z|x}$ & 0 & 0.71& 0.97& 0.5& 0& 0 & 4 & 0 & 838 \\
 && binary covariate V& 0 & 0.71& 0.1 & 0.5& 0& 0 & 0 & 0 & 1 \\
 && $n_{MS}$ = 2,000 vs $n_{MS}$ = 5,000& 0 & 0.71& 0.2 & 0.5& 0& 0 & 0 & 0 & 2 \\
 && $n_{VS}$ = 150 vs $n_{VS}$ = 400& 0 & 0.71& 0.2 & 0.5& 0& 1 & 1 & 0 & 3 \\
 & DAG 7& base case & 0.37& 0.71& 0 & 0.5& 0& 0 & 0 & 0 & -7\\
 && small $\rho_{x,z|v}$& 0.37& 0.37& 0 & 0.5& 0& 2 & 1 & 0 & -11 \\
 && small effect $\beta_x$& 0.37& 0.71& 0 & 0.1& 0& 1 & 1 & 0 & -6\\
 && negative $\rho_{v,x}$ & -0.37 & 0.71& 0 & 0.5& 0& 0 & 1 & 0 & -6\\
 && small $\rho_{v,x}$& 0.20& 0.71& 0 & 0.5& 0& 0 & 0 & 0 & -2\\
 && binary covariate V& 0.19& 0.71& 0 & 0.5& 0& 0 & 0 & 0 & -1\\
 && $n_{MS}$ = 2,000 vs $n_{MS}$ = 5,000& 0.37& 0.71& 0 & 0.5& 0& 0 & 0 & 0 & -7\\
 && $n_{VS}$ = 150 vs $n_{VS}$ = 400& 0.37& 0.71& 0 & 0.5& 0& 1 & 0 & 0 & -6\\
 & DAG 8& base case & 0.37& 0.71& 0.18& 0.5& 0& 0 & 0 & 0 & -5\\
 && small $\rho_{x,z|v}$& 0.37& 0.37& 0.18& 0.5& 0& 2 & 1 & 0 & -17 \\
 && small effect $\beta_x$& 0.37& 0.71& 0.18& 0.1& 0& 1 & 0 & 0 & -4\\
 && negative $\rho_{v,x}$ & -0.37 & 0.71& 0.18& 0.5& 0& 0 & 1 & 0 & -5\\
 && small $\rho_{v,x}$& 0.20& 0.71& 0.19& 0.5& 0& 0 & 0 & 0 & 0 \\
 && large $\rho_{v,z|x}$ & 0.37& 0.71& 0.97& 0.5& 0& 0 & 0 & 0 & 288 \\
 && binary covariate V& 0.19& 0.71& 0.1 & 0.5& 0& 0 & 0 & 0 & -1\\
 && $n_{MS}$ = 2,000 vs $n_{MS}$ = 5,000& 0.37& 0.71& 0.18& 0.5& 0& 0 & 0 & 0 & -5\\
 && $n_{VS}$ = 150 vs $n_{VS}$ = 400& 0.37& 0.71& 0.18& 0.5& 0& 1 & 0 & 0 & -4\\
Binary & DAG 1& base case & 0 & 0.71& 0 & 0.5& 0.8& 0 & -1& -1& 0 \\
 && small $\rho_{x,z|v}$& 0 & 0.37& 0 & 0.5& 0.8& 1 & 0 & 0 & 1 \\
 && small effect $\beta_x$& 0 & 0.71& 0 & 0.1& 0.8& -5& -6& -5& -5\\
 && weak risk factor V, small $\beta_v$ & 0 & 0.71& 0 & 0.5& 0.2& 1 & 1 & 1 & 1 \\
 && binary covariate V& 0 & 0.71& 0 & 0.5& 0.8& -1& -1& -2& -1\\
 && $n_{MS}$ = 5,000 vs $n_{MS}$ = 10,000 & 0 & 0.71& 0 & 0.5& 0.8& 0 & -1& -1& 0 \\
 && $n_{VS}$ = 150 vs $n_{VS}$ = 400& 0 & 0.71& 0 & 0.5& 0.8& 0 & -1& -2& 0 \\
 & DAG 2& base case & 0 & 0.71& 0.2 & 0.5& 0.8& 0 & 31& -1& 2 \\
 && small $\rho_{x,z|v}$& 0 & 0.37& 0.2 & 0.5& 0.8& 1 & 81& 0 & 5 \\
 && small effect $\beta_x$& 0 & 0.71& 0.2 & 0.1& 0.8& -5& 154 & -5& -3\\
 && large $\rho_{v,z|x}$ & 0 & 0.71& 0.97& 0.5& 0.8& 0 & 675 & -1& 839 \\
 && weak risk factor V, small $\beta_v$ & 0 & 0.71& 0.2 & 0.5& 0.2& 1 & 9 & 1 & 3 \\
 && binary covariate V& 0 & 0.71& 0.1 & 0.5& 0.8& -1& 4 & -2& -1\\
 && $n_{MS}$ = 5,000 vs $n_{MS}$ = 10,000 & 0 & 0.71& 0.2 & 0.5& 0.8& 0 & 30& -1& 2 \\
 && $n_{VS}$ = 150 vs $n_{VS}$ = 400& 0 & 0.71& 0.2 & 0.5& 0.8& 0 & 31& -2& 2 \\
 & DAG 3& base case & 0.37& 0.71& 0 & 0.5& 0.8& 0 & 52& 94& -7\\
 && small $\rho_{x,z|v}$& 0.37& 0.37& 0 & 0.5& 0.8& 1 & 54& 210 & -11 \\
 && small effect $\beta_x$& 0.37& 0.71& 0 & 0.1& 0.8& -5& 270 & 479 & -11 \\
 && negative $\rho_{v,x}$ & -0.37& 0.71& 0 & 0.5& 0.8& -1& -56 & -98 & -7\\
 && small $\rho_{v,x}$& 0.20& 0.71& 0 & 0.5& 0.8& 0 & 29& 57& -2\\
 && weak risk factor V, small $\beta_v$ & 0.37& 0.71& 0 & 0.5& 0.2& 0 & 14& 25& -7\\
 && binary covariate V& 0.19& 0.71& 0 & 0.5& 0.8& -1& 9 & 18& -3\\
 && $n_{MS}$ = 5,000 vs $n_{MS}$ = 10,000 & 0.37& 0.71& 0 & 0.5& 0.8& -1& 52& 93& -8\\
 && $n_{VS}$ = 150 vs $n_{VS}$ = 400& 0.37& 0.71& 0 & 0.5& 0.8& 0 & 53& 94& -7\\
 & DAG 4& base case & 0.37& 0.71& 0.18& 0.5& 0.8& 0 & 75& 94& -5\\
 && small $\rho_{x,z|v}$& 0.37& 0.37& 0.18& 0.5& 0.8& 1 & 103 & 210 & -17 \\
 && small effect $\beta_x$& 0.37& 0.71& 0.18& 0.1& 0.8& -5& 382 & 479 & -9\\
 && negative $\rho_{v,x}$ & -0.37& 0.71& 0.18& 0.5& 0.8& -1& -31 & -98 & -6\\
 && small $\rho_{v,x}$& 0.20& 0.71& 0.19& 0.5& 0.8& 0 & 57& 57& 0 \\
 && large $\rho_{v,z|x}$ & 0.37& 0.71& 0.97& 0.5& 0.8& 0 & 256 & 94& 287 \\
 && weak risk factor V, small $\beta_v$ & 0.37& 0.71& 0.18& 0.5& 0.2& 0 & 20& 25& -5\\
 && binary covariate V& 0.19& 0.71& 0.1 & 0.5& 0.8& -1& 14& 18& -2\\
 && $n_{MS}$ = 5,000 vs $n_{MS}$ = 10,000 & 0.37& 0.71& 0.18& 0.5& 0.8& -1& 74& 93& -6\\
 && $n_{VS}$ = 150 vs $n_{VS}$ = 400& 0.37& 0.71& 0.18& 0.5& 0.8& 0 & 75& 94& -5\\
 & DAG 5& base case & 0 & 0.71& 0 & 0.5& 0& 1 & 1 & 1 & 1 \\
 && small $\rho_{x,z|v}$& 0 & 0.37& 0 & 0.5& 0& 4 & 4 & 2 & 4 \\
 && small effect $\beta_x$& 0 & 0.71& 0 & 0.1& 0& -3& -3& -3& -3\\
 && binary covariate V& 0 & 0.71& 0 & 0.5& 0& 1 & 1 & 1 & 1 \\
 && $n_{MS}$ = 5,000 vs $n_{MS}$ = 10,000 & 0 & 0.71& 0 & 0.5& 0& -1& -1& -2& -1\\
 && $n_{VS}$ = 150 vs $n_{VS}$ = 400& 0 & 0.71& 0 & 0.5& 0& 1 & 1 & 1 & 1 \\
 & DAG 6& base case & 0 & 0.71& 0.2 & 0.5& 0& 1 & 1 & 1 & 3 \\
 && small $\rho_{x,z|v}$& 0 & 0.37& 0.2 & 0.5& 0& 4 & 4 & 2 & 7 \\
 && small effect $\beta_x$& 0 & 0.71& 0.2 & 0.1& 0& -3& -1& -3& -1\\
 && large $\rho_{v,z|x}$ & 0 & 0.71& 0.97& 0.5& 0& 1 & 12& 1 & 850 \\
 && binary covariate V& 0 & 0.71& 0.1 & 0.5& 0& 1 & 1 & 1 & 1 \\
 && $n_{MS}$ = 5,000 vs $n_{MS}$ = 10,000 & 0 & 0.71& 0.2 & 0.5& 0& -1& -1& -2& 1 \\
 && $n_{VS}$ = 150 vs $n_{VS}$ = 400& 0 & 0.71& 0.2 & 0.5& 0& 1 & 2 & 1 & 4 \\
 & DAG 7& base case & 0.37& 0.71& 0 & 0.5& 0& 1 & 2 & 2 & -6\\
 && small $\rho_{x,z|v}$& 0.37& 0.37& 0 & 0.5& 0& 4 & 4 & 3 & -9\\
 && small effect $\beta_x$& 0.37& 0.71& 0 & 0.1& 0& -3& 0 & 3 & -10 \\
 && negative $\rho_{v,x}$ & -0.37& 0.71& 0 & 0.5& 0& 1 & 0 & 0 & -6\\
 && small $\rho_{v,x}$& 0.20& 0.71& 0 & 0.5& 0& 1 & 1 & 2 & -1\\
 && binary covariate V& 0.19& 0.71& 0 & 0.5& 0& 0 & 0 & 0 & -2\\
 && $n_{MS}$ = 5,000 vs $n_{MS}$ = 10,000 & 0.37& 0.71& 0 & 0.5& 0& 0 & 1 & 2 & -7\\
 && $n_{VS}$ = 150 vs $n_{VS}$ = 400& 0.37& 0.71& 0 & 0.5& 0& 2 & 2 & 2 & -5\\
 & DAG 8& base case & 0.37& 0.71& 0.18& 0.5& 0& 1 & 2 & 2 & -4\\
 && small $\rho_{x,z|v}$& 0.37& 0.37& 0.18& 0.5& 0& 4 & 4 & 3 & -16 \\
 && small effect $\beta_x$& 0.37& 0.71& 0.18& 0.1& 0& -3& 2 & 3 & -8\\
 && negative $\rho_{v,x}$ & -0.37& 0.71& 0.18& 0.5& 0& 1 & 1 & 0 & -5\\
 && small $\rho_{v,x}$& 0.20& 0.71& 0.19& 0.5& 0& 1 & 2 & 2 & 1 \\
 && large $\rho_{v,z|x}$ & 0.37& 0.71& 0.97& 0.5& 0& 1 & 3 & 2 & 293 \\
 && binary covariate V& 0.19& 0.71& 0.1 & 0.5& 0& 0 & 0 & 0 & -1\\
 && $n_{MS}$ = 5,000 vs $n_{MS}$ = 10,000 & 0.37& 0.71& 0.18& 0.5 & 0  & 0 & 1 & 2 & -5\\
 && $n_{VS}$ = 150 vs $n_{VS}$ = 400& 0.37& 0.71& 0.18& 0.5& 0& 2 & 2 & 2 & -3 \\
\hline
\end{longtable}

\newpage
\section{All Simulations Results: Empirical Relative Efficiency as well as Variance for Valid Estimators}
\begin{longtable}{|| ll| lrrrrr | c c c c ||}
\caption{Empirical Relative Efficiency ($ERE (\hat{\beta}_{(\cdot)}) = \frac{Var(\hat{\beta}_{(OM)})}{Var(\hat{\beta}_{(\cdot)})}$) and Empirical Variance for Continuous Outcome} \\
\hline
\textbf{Y Type} & \textbf{DAG} & \textbf{Scenario} & $\rho_{v,x}$ & $\rho_{x,z|v}$ & $\rho_{v,z|x}$ & $\beta_x$ & $\beta_v$ & \multicolumn{4}{c||}{ $\widehat{ARE}$ ($\hat{\beta}_{(\cdot)}$) ($\widehat{Var}(\hat{\beta}_{(\cdot)}) \times 10^{3})$} \\
 &  &  &  &  &  &  &  & \textbf{$\hat{\beta}_{(OM)}$} & \textbf{$\hat{\beta}_{(--)}$} & \textbf{$\hat{\beta}_{(-M)}$} & \textbf{$\hat{\beta}_{(O-)}$} \\
\hline
Cont' & DAG 1& base case & 0 & 0.71& 0 & 0.5& 0.8& 1 (1.14) & 0.79 (1.44)& 0.24 (4.81)& 1 (1.14) \\
&& small $\rho_{x,z|v}$& 0 & 0.37& 0 & 0.5& 0.8& 1 (6.35) & 0.87 (7.29)& 0.07 (90.77) & 1.01 (6.32)\\
&& small effect $\beta_x$ & 0 & 0.71& 0 & 0.1& 0.8& 1 (0.45) & 0.6 (0.74) & 0.11 (4.1) & 1 (0.45) \\
&& weak risk factor V, small $\beta_v$& 0 & 0.71& 0 & 0.5& 0.2& 1 (1.14) & 0.98 (1.16)& 0.84 (1.36)& 1 (1.14) \\
&& binary covariate V& 0 & 0.71& 0 & 0.5& 0.8& 1 (1.14) & 0.96 (1.19)& 0.63 (1.81)& 1 (1.14) \\
&& $n_{MS}$ = 2,000 vs $n_{MS}$ = 5,000& 0 & 0.71& 0 & 0.5& 0.8& 1 (1.92) & 0.69 (2.76)& 0.29 (6.63)& 1 (1.91) \\
&& $n_{VS}$ = 150 vs $n_{VS}$ = 400& 0 & 0.71& 0 & 0.5& 0.8& 1 (2.25) & 0.9 (2.5)& 0.19 (11.79) & 1.01 (2.23)\\
& DAG 2& base case & 0 & 0.71& 0.2 & 0.5& 0.8& 1 (1.14) && 0.24 (4.81)&\\
&& small $\rho_{x,z|v}$& 0 & 0.37& 0.2 & 0.5& 0.8& 1 (6.35) && 0.07 (90.77) &\\
&& small effect $\beta_x$ & 0 & 0.71& 0.2 & 0.1& 0.8& 1 (0.45) && 0.11 (4.1) &\\
&& large $\rho_{v,z|x}$ & 0 & 0.71& 0.97& 0.5& 0.8& 1 (1.14) && 0.24 (4.81)&\\
&& weak risk factor V, small $\beta_v$& 0 & 0.71& 0.2 & 0.5& 0.2& 1 (1.14) && 0.84 (1.36)&\\
&& binary covariate V& 0 & 0.71& 0.1 & 0.5& 0.8& 1 (1.14) && 0.63 (1.81)&\\
&& $n_{MS}$ = 2,000 vs $n_{MS}$ = 5,000& 0 & 0.71& 0.2 & 0.5& 0.8& 1 (1.92) && 0.29 (6.63)&\\
&& $n_{VS}$ = 150 vs $n_{VS}$ = 400& 0 & 0.71& 0.2 & 0.5& 0.8& 1 (2.25) && 0.19 (11.79) &\\
& DAG 5& base case & 0 & 0.71& 0 & 0.5& 0& 1 (1.14) & 1 (1.14) & 1.01 (1.13)& 1 (1.14) \\
&& small $\rho_{x,z|v}$& 0 & 0.37& 0 & 0.5& 0& 1 (6.35) & 1 (6.32) & 1.05 (6.03)& 1.01 (6.32)\\
&& small effect $\beta_x$ & 0 & 0.71& 0 & 0.1& 0& 1 (0.45) & 1 (0.45) & 1.01 (0.45)& 1 (0.45) \\
&& binary covariate V& 0 & 0.71& 0 & 0.5& 0& 1 (1.14) & 1 (1.14) & 1.01 (1.13)& 1 (1.14) \\
&& $n_{MS}$ = 2,000 & 0 & 0.71& 0 & 0.5& 0& 1 (1.92) & 1.01 (1.91)& 1.01 (1.91)& 1 (1.91) \\
&& $n_{VS}$ = 150 vs $n_{VS}$ = 400& 0 & 0.71& 0 & 0.5& 0& 1 (2.25) & 1.01 (2.23)& 1.04 (2.16)& 1.01 (2.23)\\
& DAG 6& base case & 0 & 0.71& 0.2 & 0.5& 0& 1 (1.14) & 0.97 (1.18)& 1.01 (1.13)&\\
&& small $\rho_{x,z|v}$& 0 & 0.37& 0.2 & 0.5& 0& 1 (6.35) & 0.97 (6.53)& 1.05 (6.03)&\\
&& small effect $\beta_x$ & 0 & 0.71& 0.2 & 0.1& 0& 1 (0.45) & 0.98 (0.46)& 1.01 (0.45)&\\
&& large $\rho_{v,z|x}$ & 0 & 0.71& 0.97& 0.5& 0& 1 (1.14) & 0.05 (23.68) & 1.01 (1.13)&\\
&& binary covariate V& 0 & 0.71& 0.1 & 0.5& 0& 1 (1.14) & 0.99 (1.15)& 1.01 (1.13)&\\
&& $n_{MS}$ = 2,000 vs $n_{MS}$ = 5,000& 0 & 0.71& 0.2 & 0.5& 0& 1 (1.92) & 0.98 (1.97)& 1.01 (1.91)&\\
&& $n_{VS}$ = 150 vs $n_{VS}$ = 400& 0 & 0.71& 0.2 & 0.5& 0& 1 (2.25) & 0.97 (2.31)& 1.04 (2.16)&\\
& DAG 7& base case & 0.37& 0.71& 0 & 0.5& 0& 1 (1.14) & 1.19 (0.96)& 1.31 (0.87)&\\
&& small $\rho_{x,z|v}$& 0.37& 0.37& 0 & 0.5& 0& 1 (6.35) & 1.25 (5.1) & 2.28 (2.79)&\\
&& small effect $\beta_x$ & 0.37& 0.71& 0 & 0.1& 0& 1 (0.45) & 1.22 (0.37)& 1.28 (0.35)&\\
&& negative $\rho_{v,x}$ & -0.37 & 0.71& 0 & 0.5& 0& 1 (1.14) & 1.16 (0.99)& 1.3 (0.88) &\\
&& small $\rho_{v,x}$& 0.20& 0.71& 0 & 0.5& 0& 1 (1.14) & 1.05 (1.09)& 1.09 (1.05)&\\
&& binary covariate V& 0.19& 0.71& 0 & 0.5& 0& 1 (1.14) & 1.04 (1.09)& 1.06 (1.08)&\\
&& $n_{MS}$ = 2,000 vs $n_{MS}$ = 5,000& 0.37& 0.71& 0 & 0.5& 0& 1 (1.92) & 1.21 (1.59)& 1.29 (1.48)&\\
&& $n_{VS}$ = 150 vs $n_{VS}$ = 400& 0.37& 0.71& 0 & 0.5& 0& 1 (2.25) & 1.23 (1.82)& 1.41 (1.6) &\\
& DAG 8& base case & 0.37& 0.71& 0.18& 0.5& 0& 1 (1.14) & 1.28 (0.89)& 1.31 (0.87)&\\
&& small $\rho_{x,z|v}$& 0.37& 0.37& 0.18& 0.5& 0& 1 (6.35) & 1.66 (3.83)& 2.28 (2.79)&\\
&& small effect $\beta_x$ & 0.37& 0.71& 0.18& 0.1& 0& 1 (0.45) & 1.27 (0.35)& 1.28 (0.35)&\\
&& negative $\rho_{v,x}$ & -0.37 & 0.71& 0.18& 0.5& 0& 1 (1.14) & 1 (1.14) & 1.3 (0.88) &\\
&& small $\rho_{v,x}$& 0.20& 0.71& 0.19& 0.5& 0& 1 (1.14) & 1.08 (1.05)& 1.09 (1.05)&\\
&& large $\rho_{v,z|x}$ & 0.37& 0.71& 0.97& 0.5& 0& 1 (1.14) & 0.46 (2.45)& 1.31 (0.87)&\\
&& binary covariate V& 0.19& 0.71& 0.1 & 0.5& 0& 1 (1.14) & 1.06 (1.08)& 1.06 (1.08)&\\
&& $n_{MS}$ = 2,000 vs $n_{MS}$ = 5,000& 0.37& 0.71& 0.18& 0.5& 0& 1 (1.92) & 1.28 (1.5) & 1.29 (1.48)&\\
&& $n_{VS}$ = 150 vs $n_{VS}$ = 400& 0.37& 0.71& 0.18& 0.5& 0& 1 (2.25) & 1.35 (1.67)& 1.41 (1.6) &\\
\hline
\end{longtable}

\newpage
\begin{longtable}{|| ll| lrrrrr | c c c c ||}
\caption{Empirical Relative Efficiency ($ERE (\hat{\beta}_{(\cdot)}) = \frac{Var(\hat{\beta}_{(OM)})}{Var(\hat{\beta}_{(\cdot)})}$) and Empirical Variance for Binary Outcome}\\
\hline
\textbf{Y Type} & \textbf{DAG} & \textbf{Scenario} & $\rho_{v,x}$ & $\rho_{x,z|v}$ & $\rho_{v,z|x}$ & $\beta_x$ & $\beta_v$ & \multicolumn{4}{c||}{ $ERE (\hat{\beta}_{(\cdot)}) (Var( \hat{\beta}_{(\cdot)}) \times 10^{3})$}\\
   &  &  &  &  &  &  &  & \textbf{$\hat{\beta}_{(OM)}$} & \textbf{$\hat{\beta}_{(--)}$} & \textbf{$\hat{\beta}_{(-M)}$} & \textbf{$\hat{\beta}_{(O-)}$} \\
  \hline
  \endhead
Binary& DAG 1& base case & 0 & 0.71& 0 & 0.5& 0.8& 1 (2.11) & 1.01 (2.29) & 0.92 (2.29)& 1 (2.11) \\
&& small $\rho_{x,z|v}$& 0 & 0.37& 0 & 0.5& 0.8& 1 (8.22) & 1.01 (8.10) & 0.54 (15.16) & 1 (8.22) \\
&& small effect $\beta_x$ & 0 & 0.71& 0 & 0.1& 0.8& 1 (2.19) & 1.01 (2.87) & 0.89 (2.46)& 1 (2.19) \\
&& weak risk factor V, small $\beta_v$& 0 & 0.71& 0 & 0.5& 0.2& 1 (2.85) & 1 (3)& 1.01 (2.83)& 1 (2.85) \\
&& binary covariate V& 0 & 0.71& 0 & 0.5& 0.8& 1 (1.65) & 1 (1.65)& 0.99 (1.67)& 1 (1.66) \\
&& $n_{MS}$ = 5,000 vs $n_{MS}$ = 10,000& 0 & 0.71& 0 & 0.5& 0.8& 1 (4.58) & 1.02 (4.58) & 0.96 (4.79)& 1 (4.58) \\
&& $n_{VS}$ = 150 vs $n_{VS}$ = 400& 0 & 0.71& 0 & 0.5& 0.8& 1 (2.16) & 1.01 (2.16) & 0.73 (2.94)& 1 (2.16) \\
& DAG 2& base case & 0 & 0.71& 0.2 & 0.5& 0.8& 1 (2.11) && 0.92 (2.29)&\\
&& small $\rho_{x,z|v}$& 0 & 0.37& 0.2 & 0.5& 0.8& 1 (8.22) && 0.54 (15.16) &\\
&& small effect $\beta_x$ & 0 & 0.71& 0.2 & 0.1& 0.8& 1 (2.19) && 0.89 (2.46)&\\
&& large $\rho_{v,z|x}$ & 0 & 0.71& 0.97& 0.5& 0.8& 1 (2.11) && 0.92 (2.29)&\\
&& weak risk factor V, small $\beta_v$& 0 & 0.71& 0.2 & 0.5& 0.2& 1 (2.84) && 1.01 (2.83)&\\
&& binary covariate V& 0 & 0.71& 0.1 & 0.5& 0.8& 1 (1.65) && 0.99 (1.67)&\\
&& $n_{MS}$ = 5,000 vs $n_{MS}$ = 10,000& 0 & 0.71& 0.2 & 0.5& 0.8& 1 (4.58) && 0.96 (4.79)&\\
&& $n_{VS}$ = 150 vs $n_{VS}$ = 400& 0 & 0.71& 0.2 & 0.5& 0.8& 1 (2.16) && 0.73 (2.94)&\\
& DAG 5& base case & 0 & 0.71& 0 & 0.5& 0& 1 (2.9)& 1 (2.91)& 1 (2.89) & 1 (2.91) \\
&& small $\rho_{x,z|v}$& 0 & 0.37& 0 & 0.5& 0& 1 (11.14)& 1 (11.14) & 1.02 (10.94) & 1 (11.14)\\
&& small effect $\beta_x$ & 0 & 0.71& 0 & 0.1& 0& 1 (3.23) & 1 (3.23)& 1 (3.22) & 1 (3.23) \\
&& binary covariate V& 0 & 0.71& 0 & 0.5& 0& 1 (2.91) & 1 (2.91)& 1.01 (2.89)& 1 (2.91) \\
&& $n_{MS}$ = 2,000 vs $n_{MS}$ = 5,000& 0 & 0.71& 0 & 0.5& 0& 1 (6.26) & 1 (6.26)& 1 (6.25) & 1 (6.26) \\
&& $n_{VS}$ = 150 vs $n_{VS}$ = 400& 0 & 0.71& 0 & 0.5& 0& 1 (3.00)& 1 (3.00)& 1 (2.99) & 1 (3.00)\\
& DAG 6& base case & 0 & 0.71& 0.2 & 0.5& 0& 1 (2.90)& 0.97 (3.00) & 1 (2.89) &\\
&& small $\rho_{x,z|v}$& 0 & 0.37& 0.2 & 0.5& 0& 1 (11.14)& 0.96 (11.62)& 1.02 (10.94) &\\
&& small effect $\beta_x$ & 0 & 0.71& 0.2 & 0.1& 0& 1 (3.23) & 0.98 (3.29) & 1 (3.22) &\\
&& large $\rho_{v,z|x}$ & 0 & 0.71& 0.97& 0.5& 0& 1 (2.9)& 0.09 (32.25)& 1 (2.89) &\\
&& binary covariate V& 0 & 0.71& 0.1 & 0.5& 0& 1 (2.91) & 1 (2.92)& 1.01 (2.89)&\\
&& $n_{MS}$ = 5,000 vs $n_{MS}$ = 10,000& 0 & 0.71& 0.2 & 0.5& 0& 1 (6.26) & 0.97 (6.45) & 1 (6.25) &\\
&& $n_{VS}$ = 150 vs $n_{VS}$ = 400& 0 & 0.71& 0.2 & 0.5& 0& 1 (3.00)& 0.97 (3.09) & 1 (2.99) &\\
& DAG 7& base case & 0.37& 0.71& 0 & 0.5& 0& 1 (2.84) & 1.25 (2.28) & 1.34 (2.13)&\\
&& small $\rho_{x,z|v}$& 0.37& 0.37& 0 & 0.5& 0& 1 (10.76)& 1.33 (8.09) & 2.28 (4.72)&\\
&& small effect $\beta_x$ & 0.37& 0.71& 0 & 0.1& 0& 1 (3.21) & 1.25 (2.56) & 1.34 (2.4) &\\
&& negative $\rho_{v,x}$ & -0.37 & 0.71& 0 & 0.5& 0& 1 (2.78) & 1.24 (2.24) & 1.31 (2.12)&\\
&& small $\rho_{v,x}$& 0.20& 0.71& 0 & 0.5& 0& 1 (2.89) & 1.06 (2.74) & 1.08 (2.68)&\\
&& binary covariate V& 0.19& 0.71& 0 & 0.5& 0& 1 (2.63) & 1.05 (2.52) & 1.06 (2.48)&\\
&& $n_{MS}$ = 5,000 vs $n_{MS}$ = 10,000& 0.37& 0.71& 0 & 0.5& 0& 1 (6.26) & 1.22 (5.13) & 1.3 (4.83) &\\
&& $n_{VS}$ = 150 vs $n_{VS}$ = 400& 0.37& 0.71& 0 & 0.5& 0& 1 (2.89) & 1.24 (2.34) & 1.32 (2.18)&\\
& DAG 8& base case & 0.37& 0.71& 0.18& 0.5& 0& 1 (2.84) & 1.31 (2.17) & 1.34 (2.13)&\\
&& small $\rho_{x,z|v}$& 0.37& 0.37& 0.18& 0.5& 0& 1 (10.76)& 1.71 (6.29) & 2.28 (4.72)&\\
&& small effect $\beta_x$ & 0.37& 0.71& 0.18& 0.1& 0& 1 (3.21) & 1.32 (2.44) & 1.34 (2.4) &\\
&& negative $\rho_{v,x}$ & -0.37 & 0.71& 0.18& 0.5& 0& 1 (2.78) & 1.14 (2.44) & 1.31 (2.12)&\\
&& small $\rho_{v,x}$& 0.20& 0.71& 0.19& 0.5& 0& 1 (2.89) & 1.07 (2.69) & 1.08 (2.68)&\\
&& large $\rho_{v,z|x}$ & 0.37& 0.71& 0.97& 0.5& 0& 1 (2.84) & 0.71 (4.00) & 1.34 (2.13)&\\
&& binary covariate V& 0.19& 0.71& 0.1 & 0.5& 0& 1 (2.63) & 1.06 (2.49) & 1.06 (2.48)&\\
&& $n_{MS}$ = 5,000 vs $n_{MS}$ = 10,000& 0.37& 0.71& 0.18& 0.5& 0& 1 (6.26) & 1.27 (4.93) & 1.3 (4.83) &\\
&& $n_{VS}$ = 150 vs $n_{VS}$ = 400& 0.37& 0.71& 0.18& 0.5& 0& 1 (2.89) & 1.3 (2.22)& 1.32 (2.18)&\\
\hline
\end{longtable}

\newpage
\section{Additional Description of Real Data Example}
For this example, we created a cohort consisting of participants who completed the sleep duration question in 1987 and sunscreen use question in 1992, as these two questions were asked among a subset of HPFS cohort along with the 4-year interval waves of FFQs. Among 22,577 eligible cohort participants, 198 were missing marital status, smoking status, BMI or physical activity. Since the percentage of missingness was so small, we conducted the complete case analysis assuming missing at random\cite{rubin1976inference,tsiatis2006semiparametric}.

For figure 2 of the real data example, Furthermore, by considering sleep duration and depression as $V_{2(-ZY)}$, we are assuming that conditional on $V_{3(X-Y)}$ and $V_{4(XZY)}$, these variables are independent of the unmeasured covariate $U$, $V_{7(X-{}-)}$, and the true exposure $X$. Note that we also allow correlation between covariate sets through the unmeasured common cause $U$ on the DAG. 

We also empirically assessed how each covariate, conditional on the other covariates, was associated with validated fiber intake $X$, measurement error thus mismeasured FFQ fiber intake $Z$ (conditional on true intake) and CVD outcome $Y$, to ensure that the associations are in line with our a priori knowledge. However, we note that such evaluation might be underpowered in validation study and is only suggestive evidence of presence or absence of arrows for the DAG. The result of the assessment is presented in the table below. 

\begin{threeparttable}
\caption{Partial Correlation Coefficient, Odds Ratio (OR) and P values}
\begin{tabular}{|l l | c c c |}
\hline
$V_j$ & Covariate $V=\overset{\cdot}{\cup} V_i$ \tnote{\textdagger} & $\rho_{_{X,V_i| V \setminus V_i}}$ (p value) \tnote{\textdaggerdbl} & $\rho_{_{Z,V_i|X, V \setminus V_i}}$ (p value) \tnote{\textdaggerdbl} &  Y $\sim$V: OR(p value) \\
\hline
\textbf{$V_{_{1(-{}-{}Y)}}$} & No Family history of MI vs Yes  & -0.069(0.161) & 0.023(0.645)  & 0.792(0.001) \\
& Not married vs married & -0.001(0.977) & 0.04(0.415)& 0.968(0.709) \\
\hline 
$V_{_{2(-{}Z{}Y)}}$ & sleep duration $\le$ 6 vs 6-8 hrs  & 0.01(0.842)& 0.087(0.08)& 0.913(0.122) \\
 & sleep duration $\ge$ 8 vs 6-8 hrs  & -0.017(0.729) & -0.009(0.85)  & 0.955(0.657) \\
 & Not depressed vs depressed& -0.06(0.229)  & -0.067(0.179) & 0.866(0.395) \\
\hline
$V_{_{3(X{}-{}Y)}}$ & Energy intake (calories) & 0.053(0.281)  & 0.023(0.647)  & 1(0.122)  \\
 & High cholesterol& -0.033(0.507) & 0.042(0.4) & 0.853(0.002) \\
 & Diabetes& -0.009(0.85)  & 0.037(0.456)  & 0.582(0)  \\
 & No high blood pressure & 0.028(0.576)  & 0.037(0.457)  & 0.753(0)  \\
 & Former vs current smoker  & -0.009(0.854) & -0.086(0.083) & 0.789(0.009) \\
 & Never vs current smoker& 0.054(0.273)  & 0.115(0.02)& 0.727(0.001) \\
 \hline
$V_{_{4(X{}Z{}Y)}}$ & Metabolic equivalent task& 0.204(0)& 0.232(0)& 0.998(0.09)  \\
 & Baseline age  & 0.031(0.534)  & 0.131(0.008)  & 1.054(0)  \\
 & BMI  & -0.159(0.001) & -0.096(0.053) & 1.054(0)  \\
 & Taking multivitamin vs not& -0.001(0.976) & 0.086(0.083)  & 0.975(0.599) \\
 & Alcohol intake $\ge$ 45 vs 0-5 g/d & 0.045(0.361)  & -0.088(0.075) & 0.979(0.865) \\
 & Alcohol intake 5 - 45 vs 0-5 g/d& -0.153(0.002) & -0.163(0.001) & 0.877(0.008) \\
 \hline
$V_{_{7(X{}-{}-)}}$ & Sunscreen use 100\%  vs 0-25\%  & 0.062(0.21)& 0.04(0.414)& 0.938(0.289) \\
 & Sunscreen use 25-50\%  vs 0-25\%& -0.092(0.062) & -0.072(0.146) & 0.901(0.115) \\
 & Sunscreen use Not in the sun vs 0-25\%& 0(0.997)& 0.089(0.072)  & 0.992(0.912)\\
 \hline
\end{tabular}
    \begin{tablenotes}
        \item[\textdagger] $V_i$ is each of the covariate in this column under consideration and belongs to one of the four covariate sets $V_{_{1(-{}-{}Y)}}$, $V_{_{2(-{}Z{}Y)}}$, $V_{_{3(X{}-{}Y)}}$,  $V_{_{4(X{}Z{}Y)}}$ and $V_{_{7(-{}-{}Y)}}$. 
        \item[\textdaggerdbl] $\rho_{\cdot}$ is the partial correlation coefficient where $V \setminus V_i$ indicates the covaraite sets that are conditioned on, i.e. the full set of covariate $V$ minus each covaraite $V_i$ under consideration. 
    \end{tablenotes}
\end{threeparttable}

\newpage
\section{LASSO Regression}
\normalsize
We used glmnet package in R to perform the LASSO regression where covariates are standardized, with hyperparameter $\alpha=1$ (i.e. LASSO regresson) and candidate $log_{10} \lambda$ values ranging from $-3$ to $3$ with interval of 0.25. With 10-fold cross validation we obtained the following model, with best $log_{10} \lambda$ at $-0.5$:

\begin{threeparttable}
  \caption{Final Model Coefficients Obtained through LASSO Procedure}
    \begin{tabular}{||l | l |r||}
        \hline
        Covariate Set & Covariate &  LASSO Coefficient \\[1ex] 
        \hline
        $Z$ & Daily total fiber intake (FFQ) & 0.36196 \\
        
        $V_{1(-{}-{}Y)}$ & Family history (yes vs. no) & -1.04779 \\
                         & Marital status (married vs. others) & 0\\
        
        $V_{2(-{}Z{}Y)}$ & Sleep duration ($\le$ 6 versus (6,8] hours) & 0  \\
                         & Sleep duration ($>$8 versus (6,8] hours) & 0 \\
                         & Self-reported depression (yes vs. no) & 0 \\
        
        $V_{3(X{}-{}Y)}$ & Calories intake & 0.00003\\
                         & Baseline hypertension (yes vs. no) & 0 \\
                         & Baseline diabetes (yes vs. no) & 0\\
                         & Baseline hypercholesterolemia (yes vs. no) & -0.15392\\
                         & Smoking status (former vs. current smoker) & 0.05163\\
                         & Smoking status (never vs. current smoker) & 0\\
        $V_{4(X{}Z{}Y)}$ & Age at baseline & 0 \\
                         & Physical activity (metabolic equivalent hours) & 0.02347\\
                         & Body mass index & -0.20923\\
                         & Use of multivitamin supplement (yes vs no) & -0.01406\\
                         & Alcohol drinking ($>$45 g/d vs. $<=$5 g/d) & 0.97227\\
                         & Alcohol drinking ((5,45] g/d vs. $<=$5 g/d ) & -0.94688 \\
        $V_{7(X{}-{}-)}$ & Frequency of sunscreen use (100 vs. [50,100) \% of the time) & 0\\
                         & Frequency of sunscreen use ([25, 50) vs. [50,100) \% of the time) & -0.43524\\
                         & Frequency of sunscreen use ($<$25 vs. [50,100) \% of the time) & 0\\
        \hline
    \end{tabular}
\end{threeparttable}

\newpage
\section{Effect Modification by Covariate $V$} 

To understand whether the validity results in Table 1 of the manuscript under the four different covariate adjustment strategies still apply in the presence of effect modification, we first extend the original CRS estimators (i.e. estimating equations) in section 1, where the conditional mean models of $X$ and $Y$, $\beta(V)$ and $g(V)$ are all parametrically specified, to the less restricted semiparametric estimators motivated by the following equations: 
\begin{equation} \label{eq11}
E[Y|Z,V] = \kappa_1 + \beta_{_{(OM)}} (V) E[X|Z,V] + g(V),
\end{equation}
\begin{equation} \label{eq12}
E[Y|Z] = \kappa_2 + \beta_{_{(--)}} (V) E[X|Z],
\end{equation}
\begin{equation} \label{eq13}
E[Y|Z,V] = \kappa_3 + \beta_{_{(-M)}} (V) E[X|Z,V], \text{ and}
\end{equation}
\begin{equation} \label{eq14}
E[Y|Z,V] = \kappa_4 + \beta_{_{(O-)}} (V) E[X|Z] + g(V), 
\end{equation}where $E[X|Z=z, V]$, $E[X|Z=z]$, $\beta_{_{(\cdot)}} (V)$ and $g(V)$ can be modeled parametrically or semiparametrically.

We see that $\beta_{_{(OM)}} (V)$ in \eqref{eq11} is equal to $\beta (V)$ in equation \eqref{identify_formula_V} and as long as $\hat{E}[X|Z, V]\overset{p}{\to} E[X|Z, V]$ and $\hat{g} (V)\overset{p}{\to} g(V)$ under conditions such as prescribed in Gilvenko-Cantelli theorem \cite{vaart1996weak}, then $\hat{\beta}_{_{(OM)}}(V) \overset{p}{\to} \beta_{_{(OM)}} (V) = \beta (V)$ is valid for all eight DAGs.

Under DAG 1, $X \perp V|Z$ thus, $E[X|Z,V] = E[X|Z]$, $\kappa_1 = \kappa_4$ and $\beta_{_{(OM)}} (V) = \beta_{_{(O-)}} (V) = \beta (V)$. And under DAG 5, the true effect does not depend on $V$ and $X \perp V|Z$ and $Y \perp V|Z$ and thus the trivial result that $\beta_{_{(OM)}}$, $\beta_{_{(--)}}$, $\beta_{_{(-M)}}$ and $\beta_{_{(O-)}}$ equal to the true effect $\beta$ in equation \eqref{identify_formula_V_noEMM}. 

Under DAG 3 and 4, we have shown in section 1 that even under parametric model for $E[X|Z,V]$, $\beta(V)$ and $g(V)$ assuming no effect modification, $\beta_{_{(--)}}$, $\beta_{_{(-M)}}$ and $\beta_{_{(O-)}}$ are not equal to $\beta$, therefore $\beta_{_{(--)}} (V)$, $\beta_{_{(-M)}} (V)$ and $\beta_{_{(O-)}} (V)$ in general are not equal to $\beta(V)$. Therefore, under DAG 3 and 4, only $\beta_{_{(OM)}} (V) = \beta (V)$ identifies the conditional average treatment effect. 

Under DAG 6, 7 and 8, true effect $\beta$ cannot depend on covaraite $V$ \cite{webster2021directed} and we have shown that $\beta$ in equation \eqref{identify_formula_noV} or equivalently $\beta_{_{(--)}}$ in equation \eqref{eq12} identifies the true effect under DAG 6, 7 and 8. We also note that in section 1 that even under parametric model for $E[X|Z]$, $\beta(V)$ and $g(V)$, $\beta_{_{(O-)}}$ is not equal to $\beta$. Now we show that under DAG 6, 7 and 8, $\beta_{_{(-M)}}$ in \eqref{eq13} is euqal to $\beta$ in \eqref{identify_formula_noV}. Particularly, we have the following transformations:

\begin{align*}
 E[Y^x] & = \kappa + \beta x& \text{(constant linear treatment effect)}\\
        & = E[Y^x|X=x] & \text{(exchangeability condition $Y^x \perp X$)} \\
        & = E[Y|X=x] & \text{(consistency)}\\
        & = E[Y|X=x, Z] & \text{(surrogacy assumption: $Y \perp Z|X$)}\\
        & = E[Y|X=x, Z, V] & \text{(independence under DAG 6, 7 and 8: $Y \perp V|X, Z$)}\\  
\end{align*}

The resulting equation $E[Y|X=x,Z, V]= \kappa + \beta x$ holds for all realized values $x$ of observed random variable $X$. This allows us to treat $X$ as a random variable and rewrite the equation as $E[Y|X,Z,V] = \kappa + \beta X$. Taking expectation with respect to $X|Z, V$ on both sides gives us the following equation: 

\begin{align*} \tag{G1}  \label{G1}
    E[Y|Z,V] = \kappa + \beta E[X|Z,V],
\end{align*} where now $\kappa_3 = \kappa$ and $\beta_{_{(-M)}} = \beta$ in \eqref{G1} and we can estimate $\beta_{_{(-M)}}$ possibly through estimating equations. 

The proof so far does not depend on additional modeling assumptions other than what's required for consistent estimation of mean model of $X$ (i.e., $E[X|Z,V]$ or $E[X|Z]$) and $g(V)$. We now show that under additional statistical assumptions, $\beta_{_{(OM)}} = \beta_{_{(-M)}}$ under DAG 1 and 2. Suppose the following statistical model is appropriate for the data: 
\begin{align*}  \tag{G2}  \label{G2}
    Y & = \kappa_1 + \beta_{_{(OM)}} (V) E[X|Z,V] + g(V) + \epsilon_Y \text{ (statistical assumption I)}
\end{align*} where $E[\epsilon_Y|E[X|Z,V],V]=0$ and $Cov(\epsilon_Y,E[X|Z,V])=Cov(\epsilon_Y,g(V)) =0$. This statistical model can give us the mean model $E[Y|Z,V] = \kappa_1 + \beta_{_{(OM)}} (V) E[X|Z,V] + g(V)$ in equation $\eqref{eq11}$. If $E[V|E[X|Z,V]]$ is a linear function of $E[X|Z,V]$ (statistical assumption II), then equation \eqref{eq11} can be re-expressed as equation \eqref{eq13}.

Now suppose $E[Y|Z,V] = \kappa_3 + \beta_{_{(-M)}} (V) E[X|Z,V]$ in equation \eqref{eq13} is derived from the more restrictive statistical model, i.e.:
\begin{align*}  \tag{G3}  \label{G3}
    Y & = \kappa_3 + \beta_{_{(-M)}} (V) E[X|Z,V] + \tilde{\epsilon}_Y \text{ (statistical assumption III)},
\end{align*} where $E[\tilde{\epsilon}_Y|E[X|Z,V]]=0$ and $Cov(\tilde{\epsilon}_Y,E[X|Z,V])=0$.

From \eqref{G2} and \eqref{G3} we can respectively have:
\begin{align*} \tag{G4} \label{G4}
    \beta_{_{(OM)}} (V) = \frac{Cov(Y, E[X|Z,V])}{Var(E[X|Z,V])} {{-}} \frac{Cov(g(V), E[X|Z,V])}{Var(E[X|Z,V])}
\end{align*} and
\begin{align*} \tag{G5} \label{G5}
    \beta_{_{(-M)}} (V) = \frac{Cov(Y, E[X|Z,V])}{Var(E[X|Z,V])}    
\end{align*}

In general $\beta_{_{(OM)}} (V) \ne \beta_{_{(-M)}} (V)$, but if true exposure $X$ can be expressed as conditional mean $E[X|Z,V]$ plus  some error term that is independent of $V$ (and thus any functions of $V$):
\begin{align*} \tag{G6} \label{G6}
X = \tau + E[X|Z,V] + \epsilon_x \text{ where $\tau$ is a constant and $Cov(\epsilon_x,g(V))=0$ (statistical assumption IV)},
\end{align*} then under DAG 1 and 2, {{since $X\perp V$, we have $Cov(X, g(V))=0$, and since}} $Cov(X,g(V)) = Cov(E[X|Z,V],g(V)) + Cov(\epsilon_x,g(V)) = Cov(E[X|Z,V],g(V)) = 0$ and thus $\beta_{_{(OM)}} (V)=\beta_{_{(-M)}}(V)$. As a side note, we note that the so-called Berkson measurement error model similarly assumes $X = E[X|Z] + \epsilon_x$ where $E[\epsilon_x]=0$ \cite{CRS_book}. Under DAG 1, we further have $\beta_{_{(-M)}} (V) = \beta_{_{(--)}}(V)$ (recall $X\perp V|Z$) and therefore $\beta_{_{(OM)}} (V)$, $\beta_{_{(--)}} (V)$, $\beta_{_{(-M)}} (V)$ and $\beta_{_{(O-)}} (V)$ all equal to $\beta(V)$. Last, we note that statistical conditions (I) through (IV) are satisfied when for example $(V,X,Z,Y)$ are jointly normal.  We summarize the validity results under semiparametric models in table \ref{tab:validity_semiparametric}.

\begin{table}
\centering
  \begin{threeparttable}
  \caption{Validity of Semiparametric CRS Estimators}
  \label{tab:validity_semiparametric}
    \begin{tabular}{|c || c  c  c  c||}
        \hline
        $V_j$ as in \tnote{a} & $\hat{\beta}_{(OM)} (V)$ & $\hat{\beta}_{(--)} (V)$ & $\hat{\beta}_{(-M)} (V)$ & $\hat{\beta}_{(O-)} (V)$ \\[1ex] 
        \hline
        DAG 1, $V_{1(-{}-{}Y)}$ & valid & valid \tnote{b} & valid \tnote{b} & valid \\ 
        DAG 2, $V_{2(-{}Z{}Y)}$ & valid  & biased & valid \tnote{b} & biased \\
        DAG 3, $V_{3(X{}-{}Y)}$ & valid & biased & biased & biased \\
        DAG 4, $V_{4(X{}Z{}Y)}$ & valid & biased & biased & biased \\        
        \hline
        $V_j$ as in \tnote{a} & $\hat{\beta}_{(OM)}$ & $\hat{\beta}_{(--)}$ & $\hat{\beta}_{(-M)}$ & $\hat{\beta}_{(O-)}$ \\[1ex] 
        \hline
        DAG 5, $V_{5(-{}-{}-)}$ & valid & valid & valid & valid \\ 
        DAG 6, $V_{6(-{}Z{}-)}$ & valid & valid & valid & biased\\ 
        DAG 7, $V_{7(X{}-{}-)}$ & valid & valid & valid & biased\\ 
        DAG 8, $V_{8(X{}Z{}-)}$ & valid & valid & valid & biased\\ 
        \hline
    \end{tabular}
    
    \begin{tablenotes}
        \item[a] The subscript such as $(-{}-{}Y)$ emphasizes how the given covariate relate to $X, Z$ and $Y$. For example, DAG 2 describes a situation where covariate $V_{2(-ZY)}$ systematically affects measurement error and is a risk factor for the outcome.
        \item[b] These estimators are generally not valid but if additional statistical assumptions (I) through (IV) listed in text are satisfied then these estimators are valid. 
    \end{tablenotes}
 \end{threeparttable}
\end{table}

\newpage
\section{Bibliography}
%\nocite{*}% Show all bib entries - both cited and uncited; comment this line to view only cited bib entries;
\bibliography{WileyNJD-AMA}
%\begin{enumerate}[1]
%\item Use \verb"\bibliography{wileyNJD-AMA}" BST file for AMA reference style
%\item Use \verb"\bibliography{wileyNJD-APA}" BST file for APA reference style
%\item Use \verb"\bibliography{wileyNJD-AMS}" BST file for AMS reference style
%\item Use \verb"\bibliography{wileyNJD-VANCOUVER}" BST file for Vancouver reference style
%\item Use \verb"\bibliography{wileyNJD-ACS}" BST file for Chemistry reference style
%\end{enumerate}

%The normal commands for producing the reference list are:

%\begin{verbatim}
%\begin{thebibliography}{99}
%\bibitem{<x-ref label>}
%         <Reference details>

%\end{thebibliography}
%\end{verbatim}